\documentclass[10pt]{iopart}

\expandafter\let\csname equation*\endcsname\relax
\expandafter\let\csname endequation*\endcsname\relax
\usepackage{amsmath}
\usepackage{algorithmic}
\usepackage{subcaption}
\usepackage{threeparttable}
\usepackage[toc,page]{appendix}

\usepackage{array}
\usepackage{graphicx}
\usepackage{ifpdf}
\usepackage{times}
\usepackage{comment}
\usepackage{xcolor}
\usepackage{soul}
\usepackage[utf8]{inputenc}
\usepackage{amsfonts}
\usepackage{epsfig}
\usepackage{amsopn}
\usepackage{float}
\usepackage{epstopdf}

\def\bf{\textbf}
\def\it{\textit}

\usepackage{url}
\usepackage{booktabs} 
\usepackage{colortbl}
\usepackage[ruled]{algorithm2e} 

\usepackage{caption}

\usepackage{multirow}
\usepackage{tabularx}
\usepackage{mathrsfs}

\usepackage{balance}
\usepackage{bm}
\usepackage{rotating}
\usepackage{lipsum}
\usepackage{float}
\usepackage{footnote}
\newcommand{\ie}{\textit{i.e.}}
\newcommand{\eg}{\textit{e.g.}}

\usepackage[bottom]{footmisc}

\usepackage{pdflscape}
\usepackage{afterpage}
\usepackage{capt-of}

\begin{document}

\title[A Survey on Deep Learning-based Non-Invasive Brain Signals]{A Survey on Deep Learning-based Non-Invasive Brain Signals: Recent Advances and New Frontiers}

\author{Xiang Zhang$^{1,5}$, Lina Yao$^1$, Xianzhi Wang$^2$, Jessica Monaghan$^3$, David McAlpine$^3$, Yu Zhang$^4$}
\address{$^1$University of New South Wales, Australia\\
$^2$University of Technology Sydney, Australia\\
$^3$Macquarie university, Australia\\
$^4$Lehigh University, USA\\
$^5$Harvard University, USA} 
\ead{xiang\_zhang@hms.harvard.edu, \\lina.yao@unsw.edu.au, xianzhi.wang@uts.edu.au, \{jessica.monaghan,david.mcalpine\}@mq.edu.au, yuzi20@lehigh.edu}

\vspace{10pt}
\begin{indented}
\item[Accepted]October 2020
\end{indented}

\begin{abstract}
Brain signals refer to the biometric information collected from the human brain.
The research on brain signals aims to discover the underlying neurological or physical status of the individuals by signal decoding. The emerging deep learning techniques have improved the study of brain signals significantly in recent years. 
In this work, we first present a taxonomy of non-invasive brain signals and the basics of deep learning algorithms.
Then, we provide the frontiers of applying deep learning for non-invasive brain signals analysis, by summarizing a large number of recent publications. Moreover, upon the deep learning-powered brain signal studies, we report the potential real-world applications which benefit not only disabled people but also normal individuals. Finally, we discuss the opening challenges and future directions.
\end{abstract}

%
\submitto{\JNE}
%
\maketitle
%
\ioptwocol

\section{Introduction}
\label{sec:introduction}

Brain signals measure the instinct biometric information from the human brain, which reflects the user’s passive or active mental state. Through precise brain signal decoding, we can recognize the underlying psychological and physical status of the user and further improve his/her life quality. Based on the signal collection, brain signals contain invasive signals and non-invasive signals. The former are acquired by electrodes deployed under the scalp while the latter are collected upon human scalp without electrodes being inserted. In this survey, we mainly consider non-invasive brain signals\footnote{Without specification, the brain signals mentioned in this work refer to non-invasive signals.}. 

\subsection{General Workflow} 
\label{sub:general_workflow}

\begin{figure}[t]
\centering
  \includegraphics[width=\linewidth]{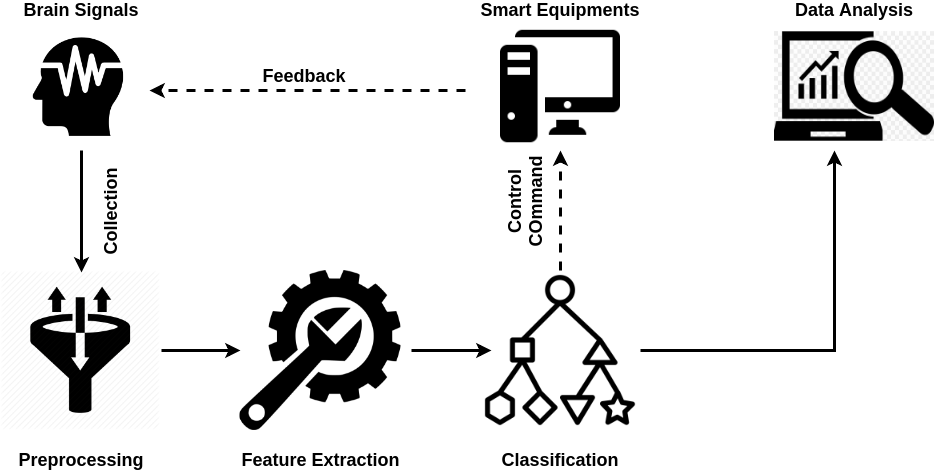}
  \caption{Generally workflow of brain signal analysis. It is named as a Brain-Computer Interface if the classified signal are used to control smart equipment (dashed lines).}
  \label{fig:bci_system}
\end{figure}

Figure~\ref{fig:bci_system} shows the general paradigm of brain signal decoding, which receives brain signals and produces the user's latent informatics. The workflow includes several key components: brain signal collection, signal preprocessing, feature extraction, classification, and data analysis. The brain signals are collected from humans and sent to the preprocessing component for denoising and enhancement. Then, the discriminating features are extracted from the processed signals and sent to the classifier for further analysis.

The collection methods differ from signal to signal. For example, EEG signals measure the voltage fluctuation resulting from ionic current within the neurons of the brain. Collecting EEG signals requires placing a series of electrodes on the scalp of the human head to record the electrical activity of the brain. Since the ionic current generated within the brain is measured at the scalp, obstacles (\eg, skull) greatly decrease the signal quality---the fidelity of the collected EEG signals, measured as Signal-to-Noise Ratio (SNR), is only approximately 5\% of that of original brain signals \cite{ball2009signal}. The collection methods of more non-invasive signals can be found in Appendix~\ref{sec:brain_signals}.  

Therefore, brain signals are usually preprocessed before feature extraction to increase the SNR. The preprocessing component contains multiple steps such as signal cleaning (smoothing the noisy signals or resolving the inconsistencies), signal normalization (normalizing each channel of the signals along time-axis), signal enhancement (removing direct current), and signal reduction (presenting a reduced representation of the signal). 

Feature extraction refers to the process of extracting discriminating features from the input signals through domain knowledge. Traditional features are extracted from time-domain (\eg, variance, mean value, kurtosis), frequency-domain (\eg, fast Fourier transform), and time-frequency domains (\eg, discrete wavelet transform). They will enrich distinguishable information regarding user intention. Feature extraction is highly dependent on the domain knowledge. For example, neuroscience knowledge is required to extract distinctive features from motor imagery EEG signals. Manual feature extraction is also time-consuming and difficult. Recently, deep learning provides a better option to automatically extract distinguishable features.

The classification component refers to the machine learning algorithms that classify the extracted features into logical control signals recognizable by external devices. Deep learning algorithms are shown to be more powerful than traditional classifiers
\cite{zhang2018internet,an2014deep,tabar2016novel}.

The classification results reflect the user's psychological or physical status and can inspire further information analysis. This is widely used in real-world applications such as neurological disorder diagnosis, emotion measuring, and driving fatigue detection. Appropriate treatment, therapy, and precaution could be conducted based on the analysis results.

In specific, the system is called a Brain-Computer Interface (BCI) while the decoded brain signals are converted into digital commands to control
the smart equipment and react with the user (dashed lines in Figure~\ref{fig:bci_system}). BCI\footnote{Apart from BCI, there are a number of similar terms to define the system that machines are directly controlled by human brain signals, like Brain-Machine Interface (BMI), Brain Interface (BI), Direct Brain Interface (DBI), Adaptive Brain Interface (ABI), and so on.} systems interpret the human brain patterns into messages or commands to communicate with the outer world \cite{lotte2018review}. BCI is generally a closed-loop system with an external device (\eg, wheelchair and robotic arm), which can directly serve the user. In contrast, brain signal analysis doesn't require a specific device as long as the analysis results can benefit society and individuals.

In this survey, we summarize the state-of-the-art studies which adopt deep learning models: 1) for feature extraction only; 2) for classification only; 3) for both feature extraction and classification. The details will be introduced in Section~\ref{sec:state_of_the_art_dl_techniques_for_bci}.
Brain signal underpins many novel applications that are important to people's daily life.
For example, the brain signal-based user identification system, with high fake-resistance, allows normal people to enjoy enhanced entertainment and security \cite{zhang2018mindid}; for people with psychological/physical deceases or disabilities, brain signals enable them to control smart device such as wheelchairs, home appliances, and robots. We present a wide range of deep learning-based brain signal applications in Section~\ref{sec:bci_applications}.




\subsection{Why Deep Learning?} 
\label{sub:why_dl_is_important_in_bci_system_}

Although traditional brain signal system has made tremendous progress \cite{abdulkader2015brain,bashashati2007survey}, it still faces significant challenges. 
First, brain signals are easily corrupted by various biological (\eg, eye blinks, muscle artifacts, fatigue, and the concentration level) and environmental artifacts (\eg, noises) \cite{abdulkader2015brain}. Therefore, it is crucial to distill informative data from corrupted brain signals and build a robust system that works in different situations.
Second, it faces the low SNR of non-stationary electrophysiological brain signals \cite{samek2012brain}. The low SNR cannot be easily addressed by traditional preprocessing or feature extraction methods due to the time complexity of those method and the risk of information loss \cite{zhang2018converting}.
Third, feature extraction highly depends on human expertise in the specific domain. For example, it requires the basic biological knowledge to investigate sleep state through Electroencephalogram (EEG) signals. Human experience may help on certain aspects but fall insufficient in more general circumstances. An automatic feature extraction method is highly desirable.
Moreover, most existing machine learning research focuses on static data and therefore, cannot classify rapidly changing brain signals accurately. For instance, the state-of-the-art classification accuracy for multi-class motor imagery EEG is generally below 80\% \cite{lotte2007review}. It requires novel learning methods to deal with dynamical data streams in brain signal systems.

Until now, deep learning has been applied extensively in brain signal applications and shown success in addressing the above challenges \cite{cecotti2014single,mahmud2018applications}. Deep learning has two advantages. First, it works directly on raw brain signals, thus avoiding the time-consuming preprocessing and feature extraction. Second, deep neural networks can capture both representative high-level features and latent dependencies through deep structures. 

\begin{table*}[t]
\centering
\caption{The existing survey on brain signals in the last decade. The column `Comprehensiveness' indicates whether the survey covers all subcategories of non-invasive brain signals or not. MI EEG refers to Motor Imagery EEG signals.}
\label{tab:survey}
\resizebox{\textwidth}{!}{
\small
\begin{tabular}{lllllll}
\hline
\textbf{No.} & \textbf{Reference} & \textbf{Comprehensiveness} & \textbf{Signal} & \textbf{Deep Learning} & \textbf{\begin{tabular}[c]{@{}l@{}}Publication\\ Time\end{tabular}} & \textbf{Area} \\ \hline
2 & \cite{wen2018deeplearning} & No & fMRI & Yes & 2018 & \begin{tabular}[c]{@{}l@{}}Mental Disease \\ Diagnosis\end{tabular} \\
3 & \cite{lotte2007review} & Partial & EEG (MI EEG, P300) & No & 2007 & Classification \\
4 & \cite{lotte2018review} & Partial & EEG (MI EEG, P300) & Partial & 2018 & Classification \\
5 & \cite{mason2007comprehensive} & Partial & EEG (ERD, P300, SSVEP, VEP, AEP) & No & 2007 &  \\
6 & \cite{litjens2017survey} & No &MRI, CT  & Partial & 2017 & \begin{tabular}[c]{@{}l@{}}Medical Image \\ Analysis\end{tabular} \\
7 & \cite{yannick2019deep} & No & EEG & \begin{tabular}[c]{@{}l@{}}Yes\end{tabular} & 2019 &  \\
8 & \cite{bashashati2007survey} & No & EEG & No & 2007 & Signal Processing \\
9 & \cite{wang2016survey} & Partial & EEG & No & 2016 & BCI Applications \\
10 & \cite{abdulkader2015brain} & Yes &  & No & 2015 &  \\
11 & \cite{movahedi2018deep} & No & EEG & Partial & 2018 &  \\
12 & \cite{soekadar2015brain} & No & EEG, fMRI & No & 2015 & \begin{tabular}[c]{@{}l@{}}Neurorehabilitation\\  of Stroke\end{tabular} \\
13 & \cite{ahn2015performance} & No & MI EEG & No & 2015 &  \\
14 & \cite{ruiz2014real} & No & fMRI & No & 2014 &  \\
15 &\cite{haider2017application}    &No&    ERP (P300)    &No    &2017    &    \begin{tabular}[c]{@{}l@{}}Applications\\  of ERP"\end{tabular}\\
16 &\cite{liu2018applications}    &No    &fMRI    &Yes&    2018    &    \begin{tabular}[c]{@{}l@{}}Applications\\  of fMRI\end{tabular}\\
17 &\cite{tsinalis2016automatic}    &No    &ERP&    No&    2017    &    Classification \\
18 & \cite{Gui2019survey}    &Partial    &EEG    &No    &2019 & Brain Biometrics \\
19 & \cite{abiri2019comprehensive}    &Partial    &EEG    &No    &2018 & BCI Paradigms \\
20 & Current Study & Yes & \begin{tabular}[c]{@{}l@{}}EEG and the subcategories,\\ fNIRS, fMRI, MEG\end{tabular} & Yes &  &  \\ \hline
\end{tabular}
}
\end{table*}

\subsection{Why this Survey is Necessary?} 
\label{sub:why_this_survey_}

We conduct this survey for three reasons. 
First, there lacks a comprehensive survey on the non-invasive brain signals. Table~\ref{tab:survey} shows a summary of the existing survey on brain signals. As our best knowledge, the limited existing surveys \cite{wen2018deeplearning,liu2018applications,abdulkader2015brain,lotte2007review,lotte2018review,bashashati2007survey,mason2007comprehensive} only focus on partial EEG signals. 
For example, Lotte et al. \cite{lotte2007review} and Wang et al. \cite{wang2016survey} focus on general EEG without analyzing EEG subtypes;
Cecotti et al. \cite{cecotti2017best} focus on Event-Related Potentials (ERP);
Haseer et al. \cite{naseer2015fnirs} focus on functional near-infrared spectroscopy (fNIRS);
Mason et al. \cite{mason2007comprehensive} brief the neurological phenomenons like event-related desynchronization (ERD), P300, SSVEP, Visual Evoked Potentials (VEP), Auditory Evoked Potentials (AEP) but have not organized them systematically;
Abdulkader et al. \cite{abdulkader2015brain} present a topology of brain signals but have not mentioned spontaneous EEG and Rapid Serial Visual Presentation (RSVP);
Lotte et al. \cite{lotte2018review} have not considered ERD and RSVP;
VEP should be a subtype of ERP in \cite{bashashati2007survey}.
Ahn et al. \cite{ahn2015performance} review the performance variation in MI-EEG based BCI systems.
Roy et al. \cite{yannick2019deep} list some deep learning-based EEG studies but present little technical inspirations and have less analysis on deep learning algorithms, they also failed to investigate other non-invasive brain signals beyond EEG. In particular, compared to \cite{yannick2019deep}, this work provides a better introduction of deep learning including the basic concepts, algorithms, and popular models (Section~\ref{sec:deep_learning_models} and Appendix~\ref{sec:deep_learning_models_appendix}). Moreover, this paper discusses the high-level guidelines in brain signal analysis in terms of the brain signal paradigms, the suitable deep learning frameworks and the promising real-world applications (Section~\ref{sec:analysis}).

Second, few research has investigated the association between deep learning (\cite{schmidhuber2015deep,deng2014tutorial}) and brain signals (\cite{fatourechi2007emg,abdulkader2015brain,lotte2007review,lotte2018review,bashashati2007survey,mason2007comprehensive}). To the best of our knowledge, this paper is in the first batch of comprehensive survey on recent advances on deep learning-based brain signals. We also point out frontiers and promising directions in this area.

Lastly, the existing surveys focus on specific areas or applications and lack an overview of broad scenarios. For example, Litjens et al. \cite{litjens2017survey} summarize several deep neural network concepts aiming at medical image analysis; Soekadar et al. \cite{soekadar2015brain} review the BCI systems and machine learning methods for stroke-related motor paralysis based on Sensori-Motor Rhythms (SMR); Vieira et al. \cite{vieira2017using} investigate the application of brain signals on the neurological disorder and psychiatric.

\subsection{Our Contributions} 
\label{sub:our_contributions}
This survey can mainly benefit: 1) the researchers with computer science background who are interested in the brain signal research; 2) the biomedical/medical/neuroscience experts who want to adopt deep learning techniques to solve problems in basic science. 

To our best knowledge, this survey is the first comprehensive survey of the recent advances and frontiers of deep learning-based brain signal analysis. To this end, we have summarized over 200 contributions, most of which were published in the last five years.
We make several key contributions in this survey: 
\begin{itemize}
  \item We review brain signals and deep learning techniques to help readers gain a comprehensive understanding of this area of research. 
  \item We discuss the popular deep learning techniques and state-of-the-art models for brain signals, providing practical guidelines for choosing the suitable deep learning models given a specific subtype of signal.
  \item We review the applications of deep learning-based brain signal analysis and highlight some promising topics for future research.
\end{itemize}

The rest of this survey is structured as followed. 
Section~\ref{sec:signal_recording} briefly introduces an taxonomy of brain signals in order to help the reader build a big picture in this field. Section~\ref{sec:deep_learning_models} overviews the commonly used deep learning models to present the basic knowledge for researchers (\eg, neurological and biomedical scholars ) who are not familiar with deep learning.
Section~\ref{sec:state_of_the_art_dl_techniques_for_bci} presents the state-of-the-art deep learning techniques for brain signals and Section~\ref{sec:bci_applications} discusses the applications related to brain signals. 
Section~\ref{sec:analysis} provides a detailed analysis and gives guidelines for choosing appropriate deep learning models based on the specific brain signal.
Section~\ref{sec:future_directions} points out the opening challenges and future directions. Finally, Section~\ref{sec:conclusion} gives the concluding remarks.

\section{Brain Imaging Techniques}
\label{sec:signal_recording}

\begin{figure*}[t]
\centering
\includegraphics[width=0.9\textwidth]{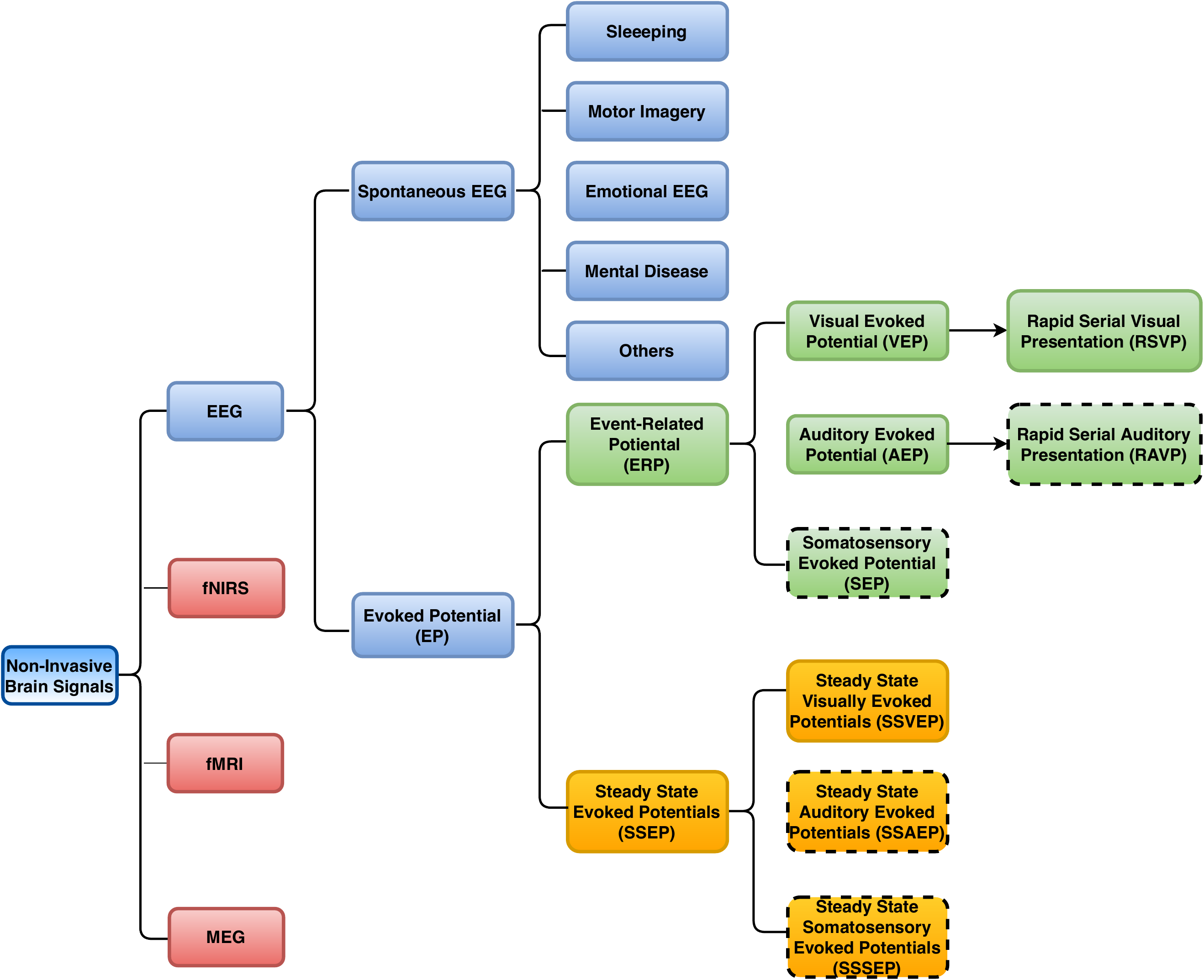}
\caption{The taxonomy of non-invasive brain signals. The dashed quadrilaterals (RAVP, SEP, SSAEP, and SSSEP) are not included in this survey because there is no existing work focusing on them involving deep learning algorithms. P300, which is a positive potential recorded approximately 300 ms after the onset of presented stimuli, is not listed in this signal tree because it is included by ERP (which refers to all the potentials after the presented stimuli). In this classification, other brain imaging technique beyond EEG (\eg, MEG and fNIRS) could also include visual/auditory tasks theoretically, but we omitted them since there is no existing work adopting deep learning on these tasks.
 }
\label{fig:BCI_signals}
\end{figure*}

In this section, we present a brief introduction of typical non-invasive brain imaging techniques. \textit{More fundamental details about non-invasive brain signal (\eg, concepts, characteristics, advantages, and drawbacks) are provided in Appendix~\ref{sec:brain_signals}.}

Figure~\ref{fig:BCI_signals} shows a taxonomy of non-invasive brain signals based on the signal collection method. 
Non-invasive signals divides into Electroencephalogram (EEG), Functional near-infrared spectroscopy (fNIRS), Functional magnetic resonance imaging (fMRI), and Magnetoencephalography (MEG) \cite{arico2018passive}. 
Table~\ref{tab:signal_comparison} summarizes the characteristics of various brain signals.
In this survey, we mainly focus on EEG signals and its subcategories because they dominate the non-invasive signals. EEG monitors the voltage fluctuations generated by an electrical current within human neurons. The electrodes attached on scalp can measure various types of EEG signals, including spontaneous EEG \cite{pfurtscheller2001motor} and evoked potentials (EP) \cite{huang2012electroencephalography}. Depending on the scenario, spontaneous EEG further diverges into sleep EEG, motor imagery EEG, emotional EEG, mental disease EEG, and others. Similarly, EP divides into event-related potentials (ERP) \cite{cecotti2017best} and steady-state evoked potentials (SSEP) \cite{regan1977steady} according to the frequency of external stimuli. Each potential contains visual-, auditory-, and somatosensory-potentials based on the external stimuli types. 

Regarding the other non-invasive techniques, fNIRS produces functional neuroimages by employing near-infrared (NIR) light to measure the aggregation degree of oxygenated hemoglobin (Hb) and deoxygenated-hemoglobin (deoxy-Hb), both of which have higher absorbers of light than other head components such as skull and scalp \cite{naseer2016analysis};
fMRI monitors brain activities by detecting the blood flow changes in brain areas \cite{wen2018deeplearning};
MEG reflects brain activities via magnetic changes \cite{Singh2018study}.

\begin{table*}[]
\centering
\caption{Summary of non-invasive brain signals' characteristics.}
\label{tab:signal_comparison}
\begin{tabular}{l|llll}
\hline
 \textbf{Signals}&  \textbf{EEG} & \textbf{fNIRS} & \textbf{fMRI}  & \textbf{MEG} \\ \hline
\textbf{Spatial resolution} & Low & Intermediate & High & Intermediate \\
\textbf{Temporal resolution} & High & Low & Low  & High \\
\textbf{Signal-to-Noise Ratio} & Low & Low & Intermediate & Low \\
\textbf{Portability}  & High & High & Low  & Low \\
\textbf{Cost} & Low & Low & High  & High \\
\textbf{Characteristic}& Electrical & Metabolic & Metabolic & Magnetic \\ \hline
\end{tabular}
\end{table*}

\section{Overview on Deep Learning Models}
\label{sec:deep_learning_models}

\begin{figure*}
\centering
  \includegraphics[width=\linewidth]{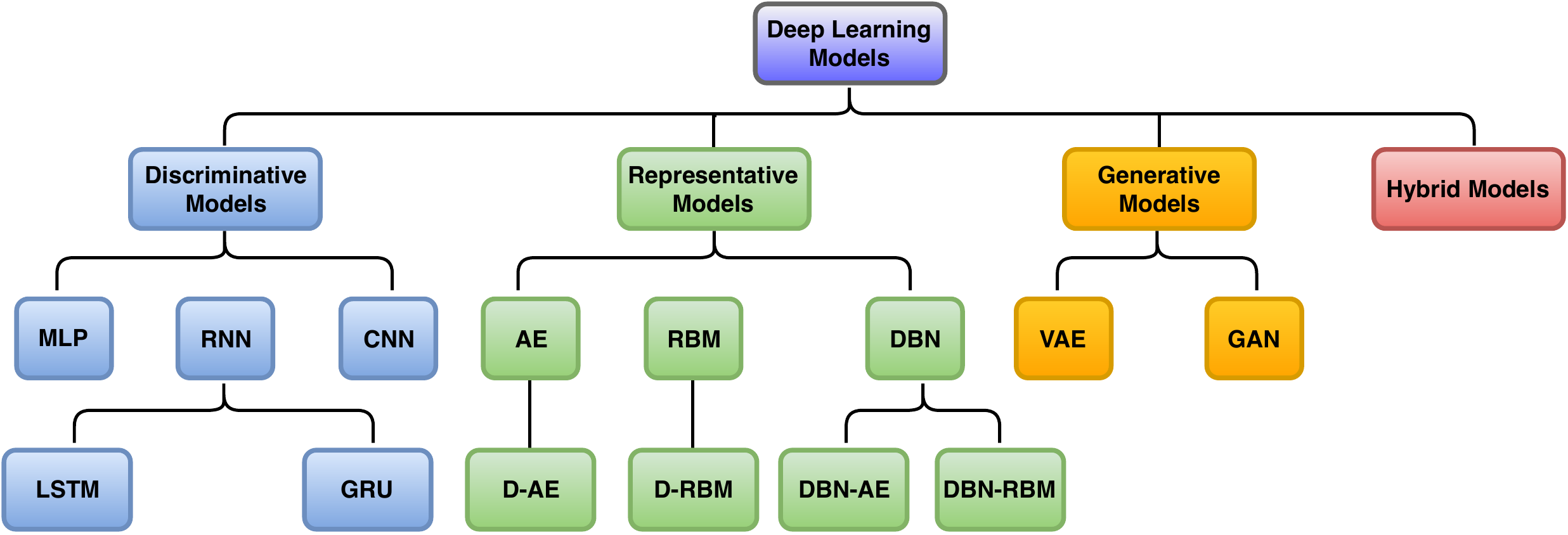}
  \caption{Deep learning models. They can be divided into discriminative, representative, generative and hybrid models based on the algorithm functions.
Discriminative models (Appendix~\ref{sub:discriminative_deep_learning_models}) mainly include Multi-Layer Perceptron (MLP),  Recurrent Neural Networks(RNN), and Convolutional Neural Networks (CNN). The two mainstreams of RNN are Long Short-Term Memory (LSTM) and Gated Recurrent Unit (GRU). Representative models (Appendix~\ref{sub:representative_deep_learning_models}) can be divided into Authoencoder (AE), Restricted Boltzmann Machine (RBM), and Deep Belief Networks (DBN). D-AE denotes Deep-Autoencoder which refers to the Autoencoder with multiple hidden layers. Likewise, D-RBM denotes Deep-Restricted Boltzmann Machine with multiple hidden layers. Deep Belief Network can be composed of AE or RBM, therefore, we divided DBN into DBN-AE and DBN-RBM. Generative models (Appendix~\ref{sub:generative_deep_learning_models}) that are commonly used in non-invasive brain signal analysis include Variational Autoencoder (VAE) and Generative Adversarial Networks (GAN). }
  \label{fig:dl_models}
\end{figure*}

\begin{table*}[tb]
\centering
\caption{Summary of deep learning model types}
\label{tab:model_type}
\begin{tabular}{lllll}
\hline
\textbf{Deep Learning } & \textbf{Input} & \textbf{Output} & \textbf{Function} & \textbf{Training method} \\ \hline
\textbf{Discriminative} & Input data & Label & Feature extraction, Classification & Supervised \\
\textbf{Representative} & Input data & Representation & Feature extraction & Unsupervised \\
\textbf{Generative} & Input data & New Sample & Generation, Reconstruction & Unsupervised \\
\textbf{Hybrid} & Input data & -- & -- & --  \\ \hline
\end{tabular}
\end{table*}

In this section, we formally introduce the deep learning models including concepts, architectures, and techniques that are commonly used in the field of brain signal researches.
Deep learning is a class of machine learning techniques that uses many layers of information-processing stages in hierarchical architectures for pattern classification and feature/representation learning \cite{deng2014tutorial}. 
\textit{More detailed information about the deep learning techniques which are common-used in brain signal analysis can be find in Appendix \ref{sec:deep_learning_models_appendix}.}


Deep learning algorithms contain several subcategories based on the aim of the techniques (Figure~\ref{fig:dl_models}):
\begin{itemize}
  \item Discriminative deep learning models, which classify the input data into a pre-known label based on the adaptively learned discriminative features. Discriminative algorithms are able to learn distinctive features by non-linear transformation, and classification through probabilistic prediction\footnote{The classification function is achieved by the combination of a softmax layer and one-hot label encoding. The one-hot label encoding refers to encoding the label by the one-hot method, which is a group of bits among which the only valid combinations of values are those with a single high (1) bit and all the others low (0) bits. For instance, a set of labels {0, 1, 2, 3} can be encoded as {(1, 0, 0, 0), (0, 1, 0, 0), (0, 0, 1, 0), (0, 0, 0, 1)}.}. Thus these algorithms can play the role of both feature extraction and classification (corresponding to Figure~\ref{fig:bci_system}). 
  Discriminative architectures mainly include Multi-Layer Perceptron (MLP) \cite{pal1992multilayer}, Recurrent Neural Networks (RNN) \cite{mikolov2010recurrent}, Convolutional Neural Networks (CNN) \cite{sainath2013deep}, along with their variations. 
  \item Representative deep learning models, which learn the pure and representative features from the input data. These algorithms only have the function of feature extraction (Figure~\ref{fig:bci_system}) but cannot make classification. Commonly used deep learning algorithms for representation are Autoencoder (AE) \cite{kramer1991nonlinear}, Restricted Boltzmann Machine (RBM) \cite{hinton2006reducing}, Deep Belief Networks (DBN) \cite{hinton2009deep}, along with their variations. 
  \item Generative deep learning models, which learn the joint probability distribution of the input data and the target label. In the brain signal scope, generative algorithms are mostly used to generate a batch of brain signals samples to enhance the training set. Generative models commonly used in brain signal analysis include Variational Autoencoder (VAE)\footnote{VAE is a variation of AE. However, they are working on different aspects. Therefore, we separately introduce AE and VAE.} \cite{kingma2013auto}, Generative Adversarial Networks (GANs) \cite{goodfellow2014generative}, etc.
  \item Hybrid deep learning models, which combine more than two deep learning models. For example, the typical hybrid deep learning model employs a representation algorithm for feature extraction and discriminative algorithms for classification. 
\end{itemize}

The summary of the characteristics of each deep learning subcategories are listed in Table~\ref{tab:model_type}. Almost all the classification functions in neural networks are implemented by a softmax layer, which will not be regarded as an algorithmic component in this survey. For instance, a model combining a DBN and a softmax layer will still be regarded as a representative model instead of a hybrid model.

\section{State-of-The-Art DL Techniques for Brain Signals} 
\label{sec:state_of_the_art_dl_techniques_for_bci}
In this section, we thoroughly summarize the advanced studies on deep learning-based brain signals (Table~\ref{tab:summary}).
The hybrid models are divided into three parts: the combination of RNN and CNN, the combination of representative and discriminative models (denoted as `Repre + Discri'), and others hybrid models.

\subsection{EEG} 
\label{sub:eeg_}
Due to the advantages of high portability and low price, EEG signals have attracted much attention. Most of the latest publications on non-invasive brain signals are related to EEG.
In this section, we summarize two aspects of EEG signals: spontaneous EEG and evoked potentials. As implied by the name, the former are spontaneous and the latter requires outside stimuli.

\subsubsection{Spontaneous EEG} 
\label{sub:eeg_oscillatory}

We present the deep learning models for spontaneous EEG according to the application scenarios as follows.

(1) \bf{Sleep EEG.}
Sleep EEG is mainly used for recognizing the sleep stage and diagnosing sleep disorders or cultivating the healthy habit \cite{chambon2018deep,zhang2016automatic}.
According to Rechtschaffen and Kales (R\&K) rules, the sleep stage includes wakefulness, non-REM (rapid eye movement) 1, non-REM 2, non-REM 3, non-REM 4, and REM. The American Academy of Sleep Medicine (AASM) recommends segmentation of sleep in five stages: wakefulness, non-REM 1, non-REM 2, slow wave sleep (SWS), and REM. The non-REM 3 and non-REM 4 are combined into SWS since there is no clear distinction between them \cite{zhang2016automatic}.  Generally, in the sleep stage analysis, the EEG signals are preprocessed by a filter which has various passband in different papers, but all notched at 50 Hz. The EEG signals are usually segmented into 30s windows.

(i) \it{Discriminative models.} CNN are frequently used for sleep stage classification on single-channel EEG \cite{tsinalis2016automatic,sors2018convolutional}. For example, Viamala et al. \cite{vilamala2017neural} manually extracted the time-frequency features and achieved a classification accuracy of 86\%.
Others used RNN \cite{biswal2017sleepnet} and LSTM \cite{tsiouris2018long} based on various features from the frequency domain, correlation, and graph theoretical features.

(ii) \it{Representative models.} Tan et al. \cite{tan2015sleep} adopted a DBN-RBM algorithm to detect sleep spindle based on Power Spectral Density (PSD) features extracted from sleep EEG signals and achieved an F-1 of 92.78\% on a local dataset. Zhang et al. \cite{zhang2016automatic} further combined DBN-RBM with three RBMs for sleep feature extraction. 

(iii) \it{Hybrid models.} Manzano et al. \cite{manzano2017combination} presented a multi-view algorithm in order to predict sleep stage by combining CNN and MLP. The CNN was employed to receive the raw time-domain EEG oscillations  while the MLP received the spectrum singles processed by the Short-Time Fourier Transform (STFT) among 0.5-32 Hz. Fraiwan et al. \cite{fraiwan2017neonatal} combined DBN with MLP for neonatal sleep state identification.
Supratak et al. \cite{supratak2017deepsleepnet} proposed a model by combing a multi-view CNN and LSTM for automatic sleep stage scoring, in which the former was adopted to discover time-invariant dependencies while the latter (a bidirectional LSTM) was adopted the temporal features during the sleep.
Dong et al. \cite{dong2018mixed} proposed a hybrid deep learning model aiming at temporal sleep stage classification and took advantage of MLP for detecting hierarchical features along with LSTM for sequential information learning.

(2) \bf{MI EEG.}
Deep learning models have shown the superior on the classification of Motor-Imagery (MI) EEG and real-motor EEG \cite{hartmann2018hierarchical,nurse2016decoding}.

(i) \it{Discriminative models.} Such models mostly use CNN to recognize MI EEG \cite{zhang2017deepkey}. Some are based on manually extracted features \cite{yang2015use,jingwei2015deep}.
For instance, Lee et al. \cite{lee2018convolution} and Zhang et al. \cite{zhang2017intent} employed CNN and 2-D CNN, respectively, for classification;
Zhang et al. \cite{zhang2017intent} learned affective information from EEG signals to built a modified LSTM control smart home appliances. 
Others also used CNN for feature extraction \cite{tang2017single}.
For example, Wang et al. \cite{wang2017multi} first used CNN to capture latent connections from MI-EEG signals and then applied
weak classifiers to choose important features for the final classification;
Hartmann et al. \cite{hartmann2018hierarchical} investigated how CNN represented spectral features through the sequence of the MI EEG samples.
MLP has also been applied for MI EEG recognition \cite{sturm2016interpretable}, which showed higher sensitivity to EEG phase features at earlier stages and higher sensitivity to EEG amplitude features at later stages.

    \clearpage
    \thispagestyle{empty}
    \begin{landscape}
    \begin{table}
        \centering 
        \captionof{table}{A summary of non-invasive brain signal studies based on deep learning models}
        \label{tab:summary}
        \resizebox{1.28\textwidth}{!}{
\begin{tabular}{l|l|l|l|l|lll|llll|ll|lll}
\hline
\multicolumn{5}{c|}{\multirow{4}{*}{\textbf{brain signals}}} & \multicolumn{12}{c}{\textbf{Deep Learning Models}} \\ \cline{6-17}
\multicolumn{5}{c|}{} & \multicolumn{3}{c|}{\textbf{Discriminative Models}} & \multicolumn{4}{c}{\textbf{Representative Models}} & \multicolumn{2}{|c|}{\textbf{Generative Models}} & \multicolumn{3}{c}{\textbf{Hybrid Models}} \\ \cline{6-17}
\multicolumn{5}{c|}{} & \multicolumn{1}{c}{\multirow{2}{*}{\textbf{MLP}}} & \multicolumn{1}{c}{\multirow{2}{*}{\textbf{RNN}}} & \multicolumn{1}{c|}{\multirow{2}{*}{\textbf{CNN}}} & \multicolumn{1}{c}{\multirow{2}{*}{\textbf{AE (D-AE)}}} & \multicolumn{1}{c}{\multirow{2}{*}{\textbf{RBM (D-RBM)}}} & \multicolumn{2}{|c|}{\textbf{DBN}} & \multicolumn{1}{c}{\multirow{2}{*}{\textbf{VAE}}} & \multicolumn{1}{c|}{\multirow{2}{*}{\textbf{GAN}}} & \multicolumn{1}{c}{\multirow{2}{*}{\textbf{LSTM+CNN}}} & \multicolumn{1}{c}{\multirow{2}{*}{\textbf{Repre + Discri}}} & \multicolumn{1}{c}{\multirow{2}{*}{\textbf{Others}}} \\  \cline{11-12}
\multicolumn{5}{c|}{} & \multicolumn{1}{c}{} & \multicolumn{1}{c}{} & \multicolumn{1}{c|}{} & \multicolumn{1}{c}{} & \multicolumn{1}{c}{} & \multicolumn{1}{|c}{\textbf{DBN-AE}} & \multicolumn{1}{c|}{\textbf{DBN-RBM}} & \multicolumn{1}{c}{} & \multicolumn{1}{c|}{} & \multicolumn{1}{c}{} & \multicolumn{1}{c}{} & \multicolumn{1}{c}{} \\ \hline
\multirow{6}*{\textbf{\begin{tabular}[c]{@{}l@{}}Non-\\ invasive \\ Signals\end{tabular}}} & \multirow{6}{*}{\textbf{EEG}} & \multirow{6}{*}{\textbf{\begin{tabular}[c]{@{}l@{}} \\ \\ Spont- \\ aneous  \\EEG \end{tabular}}} & \multicolumn{2}{l|}{\textbf{Sleep EEG}} & \cite{shahin2017deep,biswal2017sleepnet} & \cite{tsiouris2018long,biswal2017sleepnet} & \begin{tabular}[c]{@{}l@{}}\cite{vilamala2017neural},\cite{chambon2018deep},\\ \cite{tsinalis2016automatic,sors2018convolutional},\\ \cite{fernandez2018sleep,biswal2017sleepnet}\end{tabular} &  &  &  & \cite{zhang2016automatic,tan2015sleep} &  &  & \begin{tabular}[c]{@{}l@{}}\cite{supratak2017deepsleepnet}\\ \cite{biswal2017sleepnet}\end{tabular} & \cite{fraiwan2017neonatal} & \begin{tabular}[c]{@{}l@{}}\cite{manzano2017combination},\\ \cite{dong2018mixed}\end{tabular} \\
 &  &  & \multicolumn{2}{l|}{\textbf{MI EEG}} & \cite{chiarelli2018deep},\cite{sturm2016interpretable} & \begin{tabular}[c]{@{}l@{}}\cite{zhang2018mindid},\\ \cite{zhang2017deepkey,zhang2017intent}\end{tabular} & \begin{tabular}[c]{@{}l@{}}\cite{lee2018convolution}, \cite{uktveris2017application},\\ \cite{nurse2016decoding},\\ \cite{jingwei2015deep},\cite{lawhern2018eegnet},\\ \cite{hartmann2018hierarchical,yang2015use}\\ \cite{tang2017single}\end{tabular} & \begin{tabular}[c]{@{}l@{}}\cite{li2015feature,redkar2015using}\\ \cite{zhang2017multi}\end{tabular} &  & \cite{li2014deeplearningof} & \begin{tabular}[c]{@{}l@{}}\cite{ren2014convolutional,kumar2016deep},\\ \cite{lu2017deep}\end{tabular} & \cite{dai2019eeg} &  & \cite{tan2017multimodal,zhang2018converting} & \cite{tabar2016novel},\cite{duan2016classification} & \begin{tabular}[c]{@{}l@{}}\cite{nurse2015generalizable,zhang2018multimodality}, \\ \cite{wang2017multi,sakhavi2015parallel}\\ \cite{zhang2018internet}\end{tabular} \\
 &  &  & \multicolumn{2}{l|}{\textbf{\begin{tabular}[c]{@{}l@{}}Emotional \\ EEG\end{tabular}}} & \cite{frydenlund2015emotional} & \cite{zhang2018spatial} & \begin{tabular}[c]{@{}l@{}}\cite{li2016implementation},\cite{liu2017analyze},\\ \cite{wang2018data},\\ \cite{li2017hierarchical,wang2016research}\end{tabular} & \begin{tabular}[c]{@{}l@{}}\cite{chai2016unsupervised},\\ \cite{liu2016emotion}\end{tabular} & \begin{tabular}[c]{@{}l@{}}\cite{zheng2015investigating, zheng2015revealing}\\ \cite{jia2014novel}\end{tabular} & \cite{xu2016affective} & \begin{tabular}[c]{@{}l@{}}\cite{jia2014novel},\cite{xu2016affective},\\ \cite{li2015eeg},\cite{xu2016eeg},\\ \cite{zheng2014eeg,li2013affective}\end{tabular} &  &  & \cite{mioranda2018multi} & \begin{tabular}[c]{@{}l@{}}\cite{kawde2017deep,gao2015deep}\\ \cite{yin2017recognition}\end{tabular} & \cite{alhagry2017emotion} \\
 &  &  & \multicolumn{2}{l|}{\textbf{\begin{tabular}[c]{@{}l@{}}Mental Disease\\  EEG\end{tabular}}} & \cite{yuan2018novel} & \cite{talathi2017deep},\cite{ruffini2016eeg} & \begin{tabular}[c]{@{}l@{}}\cite{ullah2018automated},\cite{acharya2018automated},\\ \cite{acharya2018deep},\cite{morabito2016deep},\\ \cite{schirrmeister2017deep},\cite{hosseini2017deep},\\ \cite{johansen2016epileptiform,ansari2018neonatal}\\ \cite{hosseini2017optimized}\end{tabular} & \begin{tabular}[c]{@{}l@{}}\cite{yuan2017novel},\cite{lin2016classification},\\ \cite{morabito2017deep},\\ \cite{wen2018deepconvolution}\end{tabular} &  & \cite{page2014comparing} & \cite{zhao2014deep,turner2014deep} &  &  & \cite{shah2017optimizing} & \begin{tabular}[c]{@{}l@{}}\cite{hosseini2017cloud,hosseini2017optimized},\\ \cite{golmohammadi2017deep,al2015multichannel}\end{tabular} &  \\
 &  &  & \multicolumn{2}{l|}{\textbf{\begin{tabular}[c]{@{}l@{}}Data \\ Augmentation\end{tabular}}} &  &  &  &  &  &  &  &  & \begin{tabular}[c]{@{}l@{}}\cite{abdelfattah2018augmenting,dai2019eeg},\\ \cite{palazzo2017generative}\\ \cite{Kavasidis2017brain}\end{tabular} &  &  &  \\
 &  &  & \multicolumn{2}{l|}{\textbf{Others}} & \begin{tabular}[c]{@{}l@{}}\cite{teo2017deep,reddy2016online}\\ \cite{yepes2017improving}\end{tabular} & \cite{shang2017cognitive} & \begin{tabular}[c]{@{}l@{}}\cite{behncke2018signature},  \cite{hung2017brain},\\ \cite{baltatzis2017bullying},\cite{stober2014classifying},\\ \cite{shang2017cognitive},\cite{volker2018deep},\cite{hernandez2018eeg},\\ \cite{almogbel2018eeg,putten2018predicting}\\ \cite{hajinoroozi2015prediction,hajinoroozi2015prediction}\\ \cite{sternin2015tempo,chu2017individual}\end{tabular} & \cite{yin2017cross},\cite{du2017detecting} &  & \cite{narejo2016eeg} & \begin{tabular}[c]{@{}l@{}}\cite{hajinoroozi2015feature},\cite{narejo2016eeg},\\ \cite{san2016eeg,li2016single}\end{tabular} &  &  & \cite{zhang2018learning,hajinoroozi2015prediction} & \cite{stober2015deep},\cite{chai2017improving,bashivan2015single} & \cite{zhang2018brain2object} \\ \cline{3-17}
\multirow{9}{*}{\textbf{\begin{tabular}[c]{@{}l@{}} \end{tabular}}} & \multirow{5}{*}{} & \multirow{4}{*}{\textbf{EP}} & \multirow{3}{*}{\textbf{ERP}} & \textbf{VEP} & \cite{koike2016high,kawasaki2015visualizing}, & \cite{spampinato2017deep,Kavasidis2017brain} & \begin{tabular}[c]{@{}l@{}}\cite{spampinato2017deep},\cite{lawhern2018eegnet}\\ \cite{liu2018deep,hajinoroozi2015prediction},\\ \cite{hajinoroozi2015prediction,sarkar2016wearable}\\ \cite{cecotti2011convolutional}\end{tabular} & \cite{gao2015multi} &  &  & \begin{tabular}[c]{@{}l@{}}\cite{liu2017deep,ma2017extraction},\\ \cite{sarkar2016wearable}\end{tabular} &  &  & \begin{tabular}[c]{@{}l@{}}\cite{maddula2017deep},\cite{bashivan2016learning},\\ \cite{bashivan2016mental}\end{tabular} & \begin{tabular}[c]{@{}l@{}}\cite{zheng2015revealing,shanbhag2017p300}\\ \cite{zheng2015investigating}\end{tabular} &  \\
 &  &  &  & \textbf{RSVP} & \cite{mao2014classification,mao2016deep}, &  & \begin{tabular}[c]{@{}l@{}}\cite{manor2015convolutional},\cite{cecotti2017convolutional},\\ \cite{solon2017deep},\cite{hajinoroozi2017deep},\\ \cite{mao2017eeg,manor2016multimodal},\\ \cite{lin2017method,mao2016deep}\\ \cite{gordon2017real,yoon2018spatial}\\ \cite{shamwell2016single,cecotti2014single}\end{tabular} &  &  & \cite{vavreka2017stacked} &  &  &  &  & \cite{manor2016multimodal,mao2016deep} & \cite{cecotti2014single} \\
 &  &  &  & \textbf{AEP} &  &  & \begin{tabular}[c]{@{}l@{}}\cite{carabez2017identifying,sarkar2016wearable},\\ \cite{cecotti2011convolutional,stober2014using}\end{tabular} &  &  &  & \cite{sarkar2016wearable} &  &  &  &  &  \\ \cline{4-5}
 &  &  & \textbf{SSEP} & \textbf{SSVEP} & \cite{hachem2014effect} & \cite{thomas2017deep} & \begin{tabular}[c]{@{}l@{}}\cite{kwak2017convolutional}, \cite{waytowich2018compact},\\ \cite{thomas2017deep,aznan2018classification}\\ \cite{tu2018relating}\end{tabular} &  &  & \cite{kulasingham2016deep} & \cite{kulasingham2016deep} &  &  & \cite{attia2018time} & \cite{perez2018development} &  \\ \cline{2-5}
 & \multicolumn{4}{c|}{\textbf{fNIRS}} & \begin{tabular}[c]{@{}l@{}}\cite{naseer2016analysis},\cite{huve2017brain},\\ \cite{huve2018brain},\cite{chiarelli2018deep},\\ \cite{hennrich2015investigating}\end{tabular} &  & \cite{huve2017brain} &  &  &  &  &  &  &  & \cite{hiroyasu2014gender} &  \\ \cline{2-5}
 & \multicolumn{4}{c|}{\textbf{fMRI}} & \cite{koyamada2015deep,shen2019deep} &  & \begin{tabular}[c]{@{}l@{}}\cite{cichy2016comparison},\cite{jingwei2015deep},\\ \cite{havaei2017brain},\cite{shreyas2017deep},\\ \cite{hosseini2017deep},\cite{sarraf2016deep},\\ \cite{li2014deeplearningbased,tu2018relating}\end{tabular} & \cite{suk2016state} &  & \cite{suk2015deep}, & \begin{tabular}[c]{@{}l@{}}\cite{plis2014deep},\cite{suk2015deep},\cite{ortiz2016ensembles,suhaimi2015studies}\end{tabular} &  & \begin{tabular}[c]{@{}l@{}}\cite{seeliger2018generative,han2018gan},\\ \cite{zhang2018multi,shen2019deep}\end{tabular} &  & \cite{hu2016clinical}&  \\ \cline{2-5}
 & \multicolumn{4}{c|}{\textbf{MEG}} &  &  & \cite{garg2017automatic},\cite{cichy2016comparison} & \cite{shu2013sparse} &  &  &  &  &  &  & \cite{hasasneh2018deep} &  \\ \hline
\end{tabular}
}
\end{table}
    \end{landscape}

(ii) \it{Representative models.}
DBN is widely used as a basis for MI EEG classification for its high representative ability \cite{lu2017deep,kumar2016deep}.
For example, Ren et al. \cite{ren2014convolutional} applied a convolutional DBN based on RBM components, showing better feature representation than hand-crafted features.
Li et al. \cite{li2014deeplearningof} processed EEG signals with discrete wavelet transformation and then applied a DBN-AE based on denoising AE.
Other models include the combination of AE model (for feature extraction) and a KNN classifier \cite{redkar2015using},
%
the combination of Genetic Algorithm (for hyper-parameter tuning) and MLP (for classification) \cite{nurse2015generalizable},
the combination AE and XGBoost for multi-person scenarios \cite{zhang2017multi},
and the combination of LSTM and reinforcement learning for multi-modality signal classification \cite{zhang2018multimodality,zhang2018internet}.

(iii) \it{Hybrid models.} Several studies proposed hybrid models for the recognition of MI EEG \cite{dai2019eeg}.
For example, 
Tabar et al. \cite{tabar2016novel} extracted high-level representations from the time, frequency domain and location information of EEG signals using CNN and then used a DBN-AE with seven AEs as the classifier;
Tan et al. \cite{tan2017multimodal} used a denoising AE for dimensional reduction, a multi-view CNN combined with RNN for discovering latent temporal and spatial information, and finally achieved
an average accuracy of 72.22\% on a public dataset.

(3) \bf{Emotional EEG.}
The emotion of an individual can be evaluated in three aspects: valence, arousal, and dominance.
The combination of the three aspects form emotions such as fear, sadness, and anger, which can be revealed by EEG signals.

(i) \it{Discriminative models.}
MLP are traditionally used \cite{yepes2017improving,frydenlund2015emotional} while CNN and RNN are increasingly popular in EEG based emotion prediction \cite{li2016implementation,liu2017analyze}.
%
Typical CNN-based work in this category includes hierarchical CNN \cite{li2016implementation,li2017hierarchical} and augmenting the training set for CNN \cite{wang2018data}.
Li et al. \cite{li2016implementation} were the first to propose capturing the spatial dependencies among EEG channels via converting multi-channel EEG signals into a 2-D matrix.
%
Besides, Talathi \cite{talathi2017deep} used a discriminative deep learning model composed of GRU cells.
Zhang et al. \cite{zhang2018spatial} proposed a spatial-temporal recurrent neural network, which employs a multi-directional RNN layer to discover long-range contextual cues and a bi-directional RNN layer to capture sequential features produced by the previous spatial RNN.

(ii) \it{Representative models.} DBN, especially DBN-RBM, is widely used for the unsupervised representation ability in emotion recognition \cite{li2015eeg,gao2015deep,li2013affective}.
For instance, Xu et al. \cite{xu2016affective,xu2016eeg} proposed a DBN-RBM algorithm with three RBMs and an RBM-AE to predict affective state;
Zhao et al. \cite{zhao2014deep} and Zheng et al. \cite{zheng2014eeg} cobmined DBN-RBM with SVM and Hidden Markov Model (HMM), respectively, addressing the same problem;
%
Zheng et al. \cite{zheng2015investigating, zheng2015revealing} introduced a D-RBM with five hidden RBM layers to search the important frequency patterns and informative channels in affection recognition;
Jia et al. \cite{jia2014novel} eliminated channels with high errors and then used D-RBM for affective state recognition based on representative features of the residual channels.

The emotion is affected by many subjective and environmental factors (\eg, gender and fatigue). Yan et al. \cite{liu2016emotion} investigated the discrepancy of emotional patterns between men and women by proposing a novel model called Bimodal Deep AutoEncoder (BDAE) which received both EEG and eye movement features and shared the information in a fusion layer which connected with an SVM classifier. 
The results showed that the females have higher EEG signal diversity on the fearful emotion while males on sad emotion.
Moreover, for women, the inter-subject differences in fear is more significant then other emotions \cite{liu2016emotion}.
To overcome the mismatched distribution among the samples collected from different subjects or different experimental sessions, Chai et al. \cite{chai2016unsupervised} proposed an unsupervised domain adaptation technology which is called subspace alignment autoencoder (SAAE) by combing an AE and a subspace alignment solution. The proposed approach obtained a mean accuracy of 77.88\% in person independent scenario.

(iii) \it{Hybrid models.} One common-used hybrid model is a combination of RNN and MLP. For example, Alhagry et al. \cite{alhagry2017emotion} employed an LSTM architecture for feature extraction from emotional EEG signals and the features are forwarded into an MLP for classification.
Furthermore, Yin et al. \cite{yin2017recognition} proposed a multi-view ensemble classifier to recognize individual emotions using multimodal physiological signals. The ensemble classifier contains several D-AEs with three hidden layers and a fusion structure. Each D-AE receives one physiological signal (\eg, EEG) and then sends the outputs of D-AE to a fusion structure which is composed of another D-AE. At last, an MLP classifier makes the prediction based on the mixed features. 
Kawde et al. \cite{kawde2017deep} implemented an affect recognition system by combining a DBN-RBM for effective feature extraction and an MLP for classification.

(4) \bf{Mental Disease EEG.}
A large number of researchers exploited EEG signals to diagnose neurological disorders, especially epileptic seizure \cite{yuan2018novel}. 

(i) \it{Discriminative models.} The CNN is widely used in the automatic detection of epileptic seizure \cite{ullah2018automated,acharya2018deep,schirrmeister2017deep,wang2016research}. For example, Johansen et al. \cite{johansen2016epileptiform} adopted CNN to work on the high-passed (1 Hz) EEG signals of epileptic spike and achieved an AUC of 94.7\%. Acharya et al. \cite{acharya2018automated} employed a CNN model with 13 layers on depression detection, which was evaluated on a local dataset with 30 subjects
 and achieved the accuracies of 93.5\% and 96.0\% based on the left- and right- hemisphere EEG signals, respectively. 
Morabito et al. \cite{morabito2016deep} tried to exploit a CNN structure to extract suitable features of multi-channel EEG signals to classify Alzheimer's Disease from the patients with Mild Cognitive Impairment and healthy control group. The EEG signals are filtered in bandpass ($0.1\sim30$ Hz) and achieved an accuracy of around 82\% for three-class classification. 
Eapid Eye Movement Behavior Disorder (RBD) may cause many mental disorder diseases like Parkinson's disease (PD).  Ruffini et al. \cite{ruffini2016eeg} described an Echo State Networks (ESNs) model, a particular class of RNN, to distinguish RBD from healthy individuals. 
%
In some research, the discriminative model is only employed for feature extraction. For example, Ansari et al. \cite{ansari2018neonatal} used CNN to extract the latent features and fed into a Random Forest classifier for the final seizure detection of neonatal babies. Chu et al. \cite{chu2017individual} combined CNN and a traditional classifier for schizophrenia recognition.

(ii) \it{Representative models.} For disease detection, one commonly used method is adopting a representative model (\eg, DBN) followed by a softmax layer for classification \cite{turner2014deep}. Page et al. \cite{page2014comparing} adopted DBN-AE to extract informative features from seizure EEG signals. The extracted features were fed into a traditional logistic regression classifier for seizure detection.
Al et al. \cite{al2015multichannel} proposed a multi-view DBN-RBM structure to analyze EEG signals from depression patients. The proposed approach contains multiple input pathways, composed of two RBMs, while each corresponded to one EEG channel. All the input pathways would merge into a shared structure which is composed of another RBMs.
%
Some papers would like to preprocess the EEG signals through dimensionality reduction methods such as PCA \cite{hosseini2017cloud} while others prefer to directly fed the raw signals to the representative model \cite{lin2016classification}. Lin et al. \cite{lin2016classification} proposed a sparse D-AE with three hidden layers to extract the representative features from epileptic EEG signals while Hosseini et al. \cite{hosseini2017cloud} adopted a similar sparse D-AE with two hidden layers.

(iii) \it{Hybrid models.} A popular hybrid method is a combination of RNN and CNN. Shah et al. \cite{shah2017optimizing} investigated the performance of CNN-LSTM on seizure detection after channel selection and the sensitivities range from 33\% to 37\% while false alarms ranges from 38\% to 50\%.
Golmohammadi et al. \cite{golmohammadi2017deep} proposed a hybrid architecture for automatic interpretation of EEG by integrating both the temporal and spatial information.
2D and 1D CNNs capture the spatial features while LSTM networks capture the temporal features. The authors claimed a sensitivity of 30.83\% and a specificity of 96.86\% on the well-known TUH EEG seizure corpus. 
In the detection of early-stage Creutzfeldt-Jakob Disease (SJD), Morabito et al. \cite{morabito2017deep} combined D-AE and MLP together. The EEG signals of SJD were first filtered by bandpass (0.5$ \sim $70 Hz) and then fed into a D-AE with two hidden layers for feature representation. At last, the MLP classifier obtained the accuracy of 81$\sim$ 83\% in a local dataset. 
Convolutional autoencoder, replacing the fully-connected layers in a standard AE by convolutional and de-convolutional layers, is applied to extract the seizure features in an unsupervised manner \cite{wen2018deepconvolution}.

(5) \bf{Data augmentation.}
The generative models such as GAN could be used for data augmentation in brain signal classification \cite{abdelfattah2018augmenting}.
Palazzo et al. \cite{palazzo2017generative} first demonstrated that the information contained in brainwaves are empowered to distinguish the visual object and then extracted more robust and distinguishable representations of EEG data using RNN. At last, they employed the GAN paradigm to train an image generator conditioned by the learned EEG representations, which could convert the EEG signals into images \cite{palazzo2017generative}. 
Kavasidis et al. \cite{Kavasidis2017brain} aiming at converting EEG signals into images. The EEG signals were collected when the subjects were observing images on a screen. An LSTM layer was employed to extract the latent features from the EEG signals, and the extracted features were regarded as the input of a GAN structure. The generator and the discriminator of the GAN were both composed of convolutional layers. The generator was supposed to generate an image based on the input EEG signals after the pre-training. 
Abdelfattach et al. \cite{abdelfattah2018augmenting} adopted a GAN on seizure data augmentation. The generator and discriminator are both composed of fully-connected layers. The authors demonstrated that GAN outperforms other generative models such as AE and VAE. After the augmentation, the classification accuracy increased dramatically from 48\% to 82\%.

(6) \bf{Others.}
Some researches have explored a wide range of exciting topics. 
The first one is how EEG signals are affected by audio/visual stimuli. This differs from the potentials evoked by audio/visual stimulations because the stimuli in this phenomenon always exist instead of flicking in a particular frequency. Stober et al. \cite{stober2014using,stober2014classifying} claimed that the rhythm-evoked EEG signals are informative enough to distinguish the rhythm stimuli.
The authors conducted an experiment where 13 participants were stimulated by 23 rhythmic stimuli, including 12 East African and 12 Western stimuli. For the 24-category classification, the proposed CNN achieved a mean accuracy of 24.4\%. After that, the authors exploited convolutional AE for representation learning and CNN for recognition and achieved an accuracy of 27\% for 12-class classification \cite{stober2015deep}. Sternin et al. \cite{sternin2015tempo} adopted CNN to capture discriminative features from the EEG oscillations to distinguish whether the subject was listening or imaging music.
Similarly, Sarkar et al. \cite{sarkar2016wearable} designed two deep learning models to recognize the EEG signals aroused by audio or visual stimuli. For this binary classification task, the proposed CNN and DBN-RBM with three RBMs achieved the accuracy of 91.63\% and 91.75\%, respectively. Furthermore, the spontaneous EEG could be used to distinguish the user's mental state (logical versus emotional) \cite{bashivan2016mental}.

Moreover, some researchers focus on the impact on EEG of cognitive load \cite{shang2017cognitive} or physical workload \cite{gordienko2017deep}. Bashivan et al. \cite{bashivan2015single} first extract informative features through wavelet entropy and band-specific power, which would be fed into a DBN-RBM for further refining. At last, an MLP is employed for cognitive load level recognition. The authors, in another work \cite{bashivan2016learning}, also denoted to find the general features which are constant in inter-/intra- subjects scenarios under various mental load.
Yin et al. \cite{yin2017cross} collected the EEG signals from different mental workload levels (\eg, high and low) for binary classification. The EEG signals are filtered by a low-pass filter, transformed to the frequency domain and be calculated the power spectral density (PSD). The extracted PSD features were fed into a denoising D-AE structure for future refining. They finally got an accuracy of 95.48\%.  
Li et al. \cite{li2016single} worked on the recognition of mental fatigue level, including alert, slight fatigue, and severe fatigue. 

In addition, EEG based driver fatigue detection is an attractive area \cite{chai2017improving,du2017detecting,hajinoroozi2015prediction}. Huang et al. \cite{hung2017brain} designed a 3D CNN to predict the reaction time in drowsiness driving. This is meaningful to reduce traffic accident. Hajinoroozi et al. \cite{hajinoroozi2015feature} adopted a DBN-RBM to handle the EEG signals which were processed by ICA. They achieved an accuracy of around 85\% in binary classification (`drowsy' or `alert'). The strength of this paper is that it evaluated the DBN-RBM on three levels: time samples, channel epochs, and windowed samples. The experiments illustrated that the channel epoch level outperformed the other two levels. San et al. \cite{san2016eeg} combined deep learning models with a traditional classifier to detect driver fatigue. The model contains a DBN-RBM structure followed by an SVM classifier, which achieved the detection accuracy of 73.29\%. Almogbel et al. \cite{almogbel2018eeg} investigated the drivers' mental state under different low workload levels. A proposed CNN is claimed to detect the driving workload directly based on the raw EEG signals. 

The research of the detection of eye state has shown exceeding accuracy. Narejo et al. \cite{narejo2016eeg} explored the detection of eye state (closed or open) based on EEG signals. They tried a DBN-RBM with three RBMs and a DBN-AE with three AEs and achieved a high accuracy of 98.9\%. Reddy et al. \cite{reddy2016online} tried a simpler structure, MLP, and got a slightly lower accuracy of 97.5\%.


Furthermore, to make this survey more complete, we provide a brief introduction of Event-related desynchronization/synchronization (ERD/ERS).
ERD/ERS refers to the phenomena that the magnitude and frequency distribution of the EEG signal power changes during a specific brain state \cite{huang2012electroencephalography}. In particular, ERD denotes the power decrease of ongoing EEG signals while ERS represents the power increase of EEG signals. This characteristic of ERD/ERS of brain signals can be used to detect the event which caused the EEG fluctuation. For example, \cite{pfurtscheller1999event} presents the ERD/ERS phenomena in motor cortex recorded during a motor-imagery task.

ERD/ERS mainly appears in sensory, cognitive and motor procedures, which is not widely used in brain research due to the drawbacks like unstable accuracy cross subjects \cite{huang2012electroencephalography}. 
In most of the situations, the ERD/ERS is regarded as a specific feature of EEG powers for further analysis \cite{dai2019eeg,tabar2016novel}. The task causes an ERD in the mu band (8-13 Hz) of EEG and an ERS in the beta band (13-30 Hz). In particular, the ERD/ERS were calculated as relative changes in power concerning baseline:
$ERD/ERS = (P_e - P_b)/P_b$,
where $P_e$ denotes the signals power over one-second segment when the event occurring and $P_b$ denotes the signal power in a one-second segment during baseline which is before the event \cite{chiarelli2018deep}. Generally, the baseline refers to the rest state. For example, Sakhavi et al. calculated the ERD/ERS map and analyzed the different patterns among different tasks. The analysis demonstrated that the dynamic of energy should be considered because the static energy does not contains enough information \cite{sakhavi2015parallel}.

There are several overlooked yet promising areas. Baltatzis et al. \cite{baltatzis2017bullying} adopted CNN to detect school bullying through the EEG when watching the specific video. They achieved 93.7\% and 88.58\% for binary and four-class classification. Khurana et al. \cite{khurana2016class} proposed deep dictionary learning that outperformed several deep learning methods. Volker et al. \cite{volker2018deep} evaluated the use of Deep CNN in flanker task, which achieved an averaging accuracy of 84.1\% on the seen subject and 81.7 on the unseen subject. Zhang et al. \cite{zhang2018brain2object} combined CNN and graph network to discover the latent information from the EEG signal.

Miranda-Correa et al. \cite{mioranda2018multi} proposed a cascaded framework by combing RNN and CNN to predict individuals' affective level and personal factors (Big-five personality traits, mood, and social context).
An experiment conducted by Putten et al. \cite{putten2018predicting} attempted to identify the user's gender based on their EEG signals. They employed a standard CNN algorithm and achieved the binary classification accuracy of 81\% over a local dataset. The detection of emergency braking intention could help to reduce the responses time. Hernandez et al. \cite{hernandez2018eeg} demonstrated that the driver's EEG signals could distinguish braking intention and normal driving state. They combined a CNN algorithm which achieved the accuracy of 71.8\% in binary classification. Behncke et al. \cite{behncke2018signature} applied deep learning, a CNN model, in the context of robot assistive devices. They attempted to use CNN to improve the accuracy of decoding robot errors from EEG while the subject was watching the robot both during an object grasping and a pouring task.

Teo et al. \cite{teo2017deep} tried to combine the brain signal and recommender system, which predicted the user's preference by EEG signals. There were sixteen participants took the experiments which collected the EEG signals when the subject was presented 60 bracelet-like objects as rotating visual stimuli (a 3D object).
Then, an MLP algorithm was adopted to classify the user like or dislike the object. This exploration got the prediction accuracy of 63.99\%.
Some researchers have tried to explore a common framework which can be used for various brain signal paradigms. Lawhern et al. \cite{lawhern2018eegnet} introduced EEGNet based on a compact CNN and
evaluated its robustness in various brain signal contexts \cite{lawhern2018eegnet}.

\subsubsection{Evoked Potential} 
\label{sub:evoked_potential_}
Next, we introduce the latest researches on evoked potentials including ERP and SSEP.

(1) \bf{ERP.}
In most situations, the ERP signals are analyzed through P300 phenomena. Meanwhile, almost all the studies on P300 are based on the scenario of ERP. Therefore, in this section, a majority of the P300 related publications are introduced in the subsection of VEP/AEP according to the scenario.

(i) \it{VEP.}
VEP is one of the most popular subcategories of ERP \cite{haider2017application,yin2015hybrid,spampinato2017deep}.
Ma et al. \cite{min2017deep} worked on motion-onset VEP (mVEP) by extracting representative features through deep learning and adopted genetic algorithm combined with a multi-level sensing structure to compress the raw signals.
The compressed signals were sent to a DBN-RBM algorithm to capture the more abstract high-level features. 
Maddula et al. \cite{maddula2017deep} filtered the P300 signals with visual stimuli by a bandpass filter ($2 \sim 35$ Hz) and then fed into a proposed hybrid deep learning model for further analysis. The model includes a 2D CNN structure to capture the spatial features followed by an LSTM layer for temporal feature extraction. Liu et al. \cite{liu2017deep} combined a DBN-RBM representative model with an SVM classifier for concealed information test and achieved a high accuracy of 97.3\% over a local dataset. Gao et al. \cite{gao2015multi} employed an AE model for feature extraction followed by an SVM classifier. In the experiment, each segment contains 150 points, which were divided into five time-steps, and each step had 30 points. This model achieved an accuracy of 88.1\% over a local dataset. 
A wide range of P300 related studies is based on P300 speller \cite{shanbhag2017p300}, which allows the user to write characters. Cecotti et al. \cite{cecotti2017convolutional} tried to increase the P300 detection accuracy for more precise word-spelling. A new model was presented based on CNN, which including five low-level CNN classifiers with the different feature set, and the final high-level results are voted by the low-level classifiers. The highest accuracy reached 95.5\% over the dataset II from the third BCI competition. Liu et al. \cite{liu2018deep} proposed a Batch Normalized Neural Network (BN$^3$) which is a variant of CNN in P300 speller. The proposed method consists of six layers, and the batch normalization was operated in each batch. Kawasaki et al. \cite{kawasaki2015visualizing} employed an MLP model to detect P300 segments from non-P300 segments and achieved the accuracy of 90.8\%. 

(ii) \it{AEP.}
A few works focused on the recognition of AEP. For example, Carabez et al. \cite{carabez2017identifying} proposed and tested 18 CNN structures to classify single-trial AEP signals. In the experiment, the volunteers were required to wear on an earphone which produces auditory stimulus designed based on the oddball paradigm. The experimental analysis demonstrated that the CNN frameworks, regardless of the number of convolutional layers, were effective to extract the temporal and spatial features and provided competitive results.
The AEP signals are filtered by $0.1\sim 8$ Hz and downsampled from 256 Hz to 25 Hz. The experimental results showed that the downsampled data work better. 

(iii) \it{RSVP.}
Among various VEP diagrams, RSVP has attracted much attention \cite{gordon2017real}. In the analysis of RSVP, a number of discriminative deep learning models (e,g., CNN \cite{cecotti2017convolutional,solon2017deep,lin2017method} and MLP \cite{mao2014classification}) has achieved a big success. A common preprocessing method used in RSVP signals is frequency filtering. The pass bands are generally ranged from $ 0.1 \sim 50$ Hz \cite{manor2015convolutional,shamwell2016single}. 
Cecotti et al. \cite{cecotti2014single} worked on the classification of ERP signals in RSVP scenario and proposed a modified CNN model for the detection of the specific target in RSVP. In the experiment, the images of faces and cars were regarded as target or non-target, respectively. The image presenting frequency is 2 Hz.
 In each session, the target probability was 10\%. The proposed model offered an AUC of 86.1\%. 
Hajinoroozi et al. \cite{hajinoroozi2017deep} adopted a CNN model targeting the inter-subject and inter-task detection of RSVP. 
The experimental results showed that CNN worked good in cross-task but failed to get satisfying performance in the cross-subject scenario. Mao et al. \cite{mao2016deep} compared three different deep neural network algorithms in the prediction of whether the subject had seen the target or not. The MLP, CNN, and DBN models obtained the AUC of 81.7\%, 79.6\%, and 81.6\%, respectively. The author also applied a CNN model to analyze the RSVP signals for person identification \cite{mao2017eeg}. 

The representative deep learning models are also applied in RSVP. Vareka et al. \cite{vavreka2017stacked} verified if deep learning performs well for single trial P300 classification. They conducted an RSVP experiment while the subjects were asked to recognize the target from non-target and distracters. Then a DBN-AE was implemented and compared with some non-deep learning algorithms. The DBN-AE was composed of five AEs while the hidden layer of the last AE only has two nodes which can be used for classification through softmax function. Finally, the proposed model achieved the accuracy of 69.2\%. Manor et al. \cite{manor2016multimodal} applied two deep neural networks to deal with the RSVP signals after lowpass filtering ($0\sim 51$ Hz). Discriminative CNN achieved the accuracy of 85.06\%. Meanwhile, the representative convolutional D-AE achieved the accuracy of 80.68\%. 

(2) \bf{SSEP.}
Most of deep learning-based studies in SSEP area focus on SSVEP like \cite{kwak2017convolutional}. SSVEP refers to brain oscillations evoked by the flickering visual stimuli, which generally produced from the parietal and occipital regions \cite{waytowich2018compact}. 
Attia et al. \cite{attia2018time} aimed at finding an intermediate representation of SSVEP. A hybrid method combined CNN and RNN was proposed to capture the meaningful features from the time domain directly, which achieved the accuracy of 93.59\%. Waytowich et al. \cite{waytowich2018compact} applied a compact CNN model to directly work on the raw SSVEP signals without any hand-crafted features. The reported cross subject mean accuracy was approximately 80\%. Thomas et al. \cite{thomas2017deep} first filter the raw SSVEP signals through a bandpass filter ($5 \sim 48$ Hz) and then operated discrete  FFT on consecutive 512 points. The processed data were classified by a CNN (69.03\%) and an LSTM (66.89\%) independently. 

Perez et al. \cite{perez2018development} adopted a representative model, a sparse AE, to extract the distinct features from the SSVEP from multi-frequency visual stimuli. The proposed model employed a softmax layer for the final classification and achieved the accuracy of 97.78\%. Kulasingham et al. \cite{kulasingham2016deep} classified SSVEP signals in the context of guilty knowledge test. The authors applied DBN-RBM and DBN-AE independently and achieved the accuracy of 86.9\% and 86.01\%, respectively.
Hachem et al. \cite{hachem2014effect} investigated the influence of fatigue on SSVEP through an MLP model during wheelchair navigation. The goal of this study was to seek the key parameters to switch between manual, semi-autonomous, and autonomous wheelchair command. Aznan et al. \cite{aznan2018classification} explored the SSVEP classification, where the signals were collected through dry electrodes. The dry signals were more challenging for the lower SNR than standard EEG signals. This study applied a CNN discriminative model and achieved the highest accuracy of 96\% over a local dataset.

\subsection{fNIRS} 
\label{sub:fnirs}
Up to now, only a few of researchers paid attention on deep learning-based fNIRS. 
Naseer et al. \cite{naseer2016analysis} analyzed the difference between two mental tasks (mental arithmetic and rest) based on fNIRS signals. The authors manually extracted six features from the prefrontal cortex fNIRS and compared six different classifiers. The results demonstrated that the MLP with the accuracy of 96.3\% outperformed all the traditional classifiers, including SVM, KNN, naive Bayes, etc. Huve et al. \cite{huve2017brain} classified the fNIRS signals, which were collected from the subjects during three mental states, including substractions, word generation, and rest. The employed MLP model achieved the accuracy of 66.48\% based on the hand-crafted features (\eg, the concentration of OxyHb/DeoxyHb). After that, the authors study the mobile robot control through fNIRS signals and got the binary classification accuracy of 82\% (offline) and 66\% (online) \cite{huve2018brain}. Chiarelli et al. \cite{chiarelli2018deep} exploited the combination of fNIRS and EEG for left/right MI EEG classification. Sixteen features extracted from fNIRS signals (eight from OxyHb and eight from DeoxyHb) were fed into an MLP classifier with four hidden layers. 

On the other hand, Hiroyasu et al. \cite{hiroyasu2014gender} attempted to detect the gender of the subject through their fNIRS signals. The authors employed a denoising D-AE with three hidden layers to extract distinctive features to be fed into an MLP classifier for gender detection. The model was evaluated over a local dataset and gained the average accuracy of 81\%. 
In this study, the authors also pointed out that, compared with Positron Emission Tomography (PET) and fMRI, fNIRS has higher time resolution and more affordable \cite{hiroyasu2014gender}.

~
\subsection{fMRI} 
\label{sub:fmri}
Recently, several deep learning methods have been applied to fMRI analysis, especially on the diagnosis of cognitive impairment \cite{wen2018deeplearning,vieira2017using}.

(1) \textbf{Discriminative models.}
Among the discriminative models, CNN is a promising model to analyze fMRI \cite{shreyas2017deep}. For example, Havaei et al. built a segmentation approach for brain tumor based on fMRI with a novel CNN algorithm which can capture both the global features and the local features simultaneously \cite{havaei2017brain}. The convolutional filters have different size. Thus, the small-size and large-size filter could exploit the local and global features, independently. 
Sarraf et al. \cite{sarraf2016deepad,sarraf2016deep} applied deep CNN to recognize Alzheimer's Disease based on fMRI and MRI data. Morenolopez et al. \cite{Morenolopez2017deep} employed a CNN model to deal with fMRI of brain tumor patients for three-class recognition (normal, edema, or active tumor). The model was evaluated over BRATS dataset and obtained the F1 score of 88\%. Hosseini et al. \cite{hosseini2017deep} employed CNN for feature extraction. The extracted features were classified by SVM for the detection of an epileptic seizure.

Furthermore, Li et al. proposed a data completion method based on CNN. In particular, utilizing the information from fMRI data to complete PET, then train the classifier based on both fMRI and PET \cite{li2014deeplearningbased}. In the model, the input data of the proposed CNN is the fMRI patch, and the output is a PET patch. There are two convolutional layers with ten filters mapping the fMRI to PET. The experiments illustrated that the classifier trained by the combination of fMRI and PET (92.87\%) outperformed the one trained by solo fMRI (91.92\%)
Moreover, Koyamada et al. used a nonlinear MLP to extract common features from different subjects. The model is evaluated over a dataset from the Human Connectome Project (HCP) \cite{koyamada2015deep}.

(2) \textbf{Representative models.}
A wide range of publications demonstrated the effectiveness of representative models in recognition of fMRI data \cite{suhaimi2015studies}. Hu et al. \cite{hu2016clinical} used demonstrated that deep learning outperforms other machine learning methods in the diagnosis of neurological disorders such as Alzheimer's disease.
Firstly, the fMRI images were converted to a matrix to represent the activity of 90 brain regions. Secondly, a correlation matrix is obtained by calculating the correlation between each pair of brain regions to represent the functional connectivity between different brain regions. Furthermore, a targeted AE is built to classify the correlation matrix, which is sensitive to AD. The proposed approach achieved an accuracy of 87.5\%. 
Plis et al. \cite{plis2014deep} employed a DBN-RBM with three RBM components to extract the distinctive features from ICA processed fMRI and finally achieved an average F1 measure of above 90\% over four public datasets. Suk et al. compared the effectiveness of DBN-RBM and DBN-AE on Alzheimer's disease detection and the experimental results showed that the former obtained the accuracy of 95.4\%, which is slightly lower than the latter (97.9\%) \cite{suk2015deep}. Suk et al. \cite{suk2016state} applied a D-AE model to extract latent features from the resting-state fMRI data on the diagnosis of Mild Cognitive Impairment (MCI). The latent features are fed into a SVM classifier which achieved the accuracy of 72.58\%. Ortiz et al. \cite{ortiz2016ensembles} proposed a multi-view DBN-RBM to receives the information of MRI and PET simultaneously. The learned representations were sent to several simple SVM classifiers which were ensembled to form a high-level stronger classifier by voting. 

(3) \textbf{Generative models.}
The reconstruction of natural image (\eg, fMRI) has been attracted lots of attention \cite{han2018gan,zhang2018spatial,shen2019deep}. Seeliger et al. \cite{seeliger2018generative} proposed a deep convolutional GAN for reconstructing visual stimuli from fMRI, which aimed at training a generator to create an image similar to the visual stimuli. 
The generator contains four convolutional layers in order to convert the input fMRI to a natural image. Han et al. \cite{han2018gan} focused on the generation of synthetic multi-sequence fMRI using GAN. The generated image can be used for data augmentation for better diagnostic accuracy or physician training to help better understand various diseases. The authors applied the existing Deep Convolutional GAN (DCGAN) \cite{radford2015unsupervised} and Wasserstein GAN (WGAN) \cite{arjovsky2017wasserstein} and found that the former works better. Shen et al. \cite{shen2019deep} presented another image recovery approach by minimizing the distance between the real image and the image generated based on real fMRI.

\subsection{MEG} 
\label{sub:meg}

Garg et al. \cite{garg2017automatic} worked on the refining of MEG signals by removing the artifacts like eye-blinks and cardiac activity. The MEG singles were decomposed by ICA first and then classified by a 1-D CNN model. At last, the proposed approach achieved the sensitivity of 85\% and specificity of 97\% over a local dataset. Hasasneh et al. \cite{hasasneh2018deep} also focused on artifacts detection (cardiac and ocular artifacts). The proposed approach uses CNN  to capture temporal features and MLP to extract spatial information. Shu et al. \cite{shu2013sparse} employed a sparse AE to learn the latent dependencies of MEG signals in the task of single word decoding. The results demonstrated that the proposed approach is advantageous for some subjects, although it did not produce an overall increase in decoding accuracy.
Cichy et al. \cite{cichy2016comparison} applied a CNN model to recognize visual object based on MEG and fMRI signals.

\section{Brain Signal-based Applications} 
\label{sec:bci_applications}
Deep learning models have contributed to various of brain signal applications as summarized in Table~\ref{tab:application}.
The papers focused on signal classification without application background are not listed in this table. 
Therefore, the publication amounts in this table are less than in Table~\ref{tab:summary}.

\subsection{Health Care} 
\label{sub:health_care}
In the health care area, the deep learning-based brain signal systems mainly works on the detection and diagnosis of mental diseases such as sleep disorders, Alzheimer's Disease, epileptic seizure, and other disorders. In the first place, for the sleep disorder detection, most studies are focused on the sleep stage detection based on sleep spontaneous EEG. In this situation, the researchers do not need to recruit patients with sleep disorder because the sleep EEG signals can be easily collected from healthy individuals. In terms of the algorithm, it can be observed from Table~\ref{tab:application} that the DBN-RBM and CNN are widely adopted for feature selection and classification. Ruffini et al. \cite{ruffini2016eeg} walk one step further by detecting the REM Behavior Disorder (RBD), which may cause neurodegenerative diseases such as Parkinson's disease. They achieved an average accuracy of 85\% in recognition of the RBD from healthy controls. 

Moreover, fMRI is widely applied in the diagnosis of Alzheimer's Disease. By taking advantage of the high spatial resolution of fMRI, the diagnosis achieved the accuracy of above 90\% in several studies. Another reason that contributes to competitive performance is the binary classification scenario. Apart from that, there are several publications diagnose the AD based on spontaneous EEG \cite{morabito2016deep,zhao2014deep}. 

Besides, the diagnosis of epileptic seizure attracted much attention. The seizure detection mainly based on spontaneous EEG.
The popular deep learning models in this scenario contain the independent CNN and RNN, along with hybrid models combined RNN and CNN. Some models integrated the deep learning models for feature extraction and traditional classifier for detection \cite{turner2014deep,page2014comparing}. For example, Yuan et al. \cite{yuan2017novel} applied a D-AE in feature extraction followed by SVM for seizure diagnosis. Ullah et al. \cite{ullah2018automated} adopted the voting for post-processing, which proposed several different CNN classifiers and predicted the final result by voting. 

Furthermore, there are a lot of other healthcare issues can be solved by brain signal research. The cardiac artifacts in MEG can be automatically detected by deep learning models\cite{garg2017automatic,hasasneh2018deep}. Several modified CNN structures are proposed to detect brain tumor based on fMRI from the public BRATS dataset \cite{havaei2017brain,shreyas2017deep}. Researchers have demonstrated the effectiveness of deep learning models in the detection of a wide number of mental diseases such as depression \cite{acharya2018automated}, Interictal Epileptic Discharge (IED) \cite{antoniades2016deep}, schizophrenia \cite{plis2014deep}, Creutzfeldt-Jakob Disease (CJD) \cite{morabito2017deep}, and Mild Cognitive Impairment (MCI) \cite{suk2016state}.

\subsection{Smart Environment} 
\label{sub:smart_environment}
The smart environment is a promising application scenario for brain signals in the future. With the development of Internet of Things (IoT), an increasing number smart environment can be connected to brain signals. For example, the assisting robot can be used in smart home \cite{zhang2017intent,zhang2018internet}, in which the robot can be controlled by brain signals of the individuals.
Moreover, Behncke et al. \cite{behncke2018signature} and Huve et al. \cite{huve2018brain} investigated the robot control problem based on the visual stimulated spontaneous EEG and fNIRS signals. The brain signal controlled exoskeleton could help the disabilities who damaged the motor system in sub-limb in walking and daily activities \cite{kwak2017convolutional}. In the future, the research on brain-controlled appliances may be beneficial to the elders or disabilities in smart home and smart hospital. 

\subsection{Communication} 
\label{sub:brain_communication}
One of the biggest advantages of brain signals, compared to other human-machine interface techniques, is that brain signal enables the patient who lost most motor abilities like speaking to communicate with the outer world. The deep learning technology improved the efficiency of brain signal based communications. One typical diagram which enables individual typing without any motor system is P300 speller, which can convert the user's intent into text \cite{kawasaki2015visualizing}. The powerful deep learning models empower the brain signal systems to recognize the P300 segment from the non-P300 segment while the former contains the communication information of the user \cite{cecotti2011convolutional}. In a higher level, the representative deep learning models can help to detect what character the user is focusing on and print it on the screen to chat with others \cite{cecotti2011convolutional,maddula2017deep,liu2018deep}. 
Additionally, Zhang et al. \cite{zhang2018converting} proposed a hybrid model that combined RNN, CNN, and AE to extract the informative features from MI EEG to recognize what letter the user wants to speak. 

\begin{figure}[t]
    \centering
    \begin{subfigure}[t]{0.25\textwidth}
        \centering
        \includegraphics[width=\textwidth]{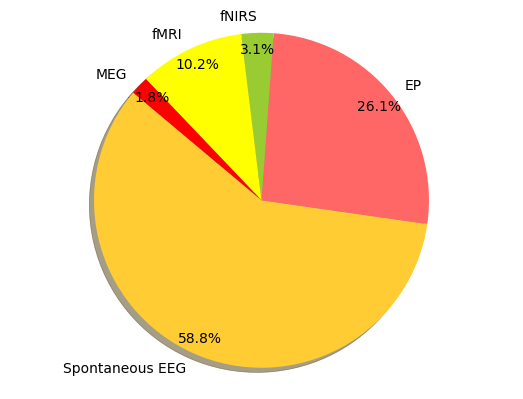}
        \caption{Brain signals}
    \end{subfigure}%
    ~
    \begin{subfigure}[t]{0.25\textwidth}
        \centering
        \includegraphics[width=\textwidth]{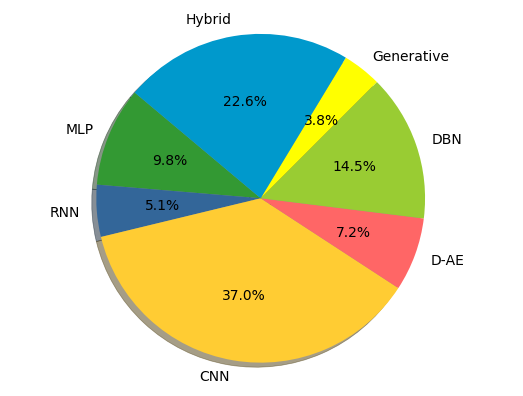}
        \caption{Deep learning models}
    \end{subfigure}%
 \caption{Illustration of the publications proportion for crucial brain signals and deep learning models.}
\label{fig:bar}
\end{figure}

\begin{table*}[]
\centering
\caption{Summary of deep learning-based brain signal applications. The `local' dataset refers to private or not available dataset. The public datasets (along with download links) will be introduced in Section~\ref{sec:datasets}. In the signals, S-EEG, MD EEG, and E-EEG separately denote sleep EEG, mental disease EEG, and emotional EEG. The single `EEG' refers to the other subcategory of spontaneous EEG. In the models, RF and LR denote to random forest and logistic regression algorithms, respectively. In the performance column, `N/A', `sen', `spe', 'aro', `val', `dom', and `like' denote not-found, sensitivity, specificity, arousal, valence, dominance, and liking, respectively. For each application scenario, the literature are sorted out by signal types and deep learning models. 
}
\label{tab:application}
\resizebox{\textwidth}{!}{
\begin{tabular}{l|l|lllll}
\hline
\multicolumn{2}{c|}{\textbf{Brain Signal Applications}} & \textbf{Reference} & \textbf{Signals} & \textbf{\begin{tabular}[c]{@{}l@{}}Deep Learning \\ Models\end{tabular}} & \textbf{Dataset} & \textbf{Performance} \\ \hline
\multirow{40}{*}{\textbf{\begin{tabular}[c]{@{}l@{}}Health \\ Care\end{tabular}}} & \multirow{15}{*}{\textbf{\begin{tabular}[c]{@{}l@{}}Sleep \\ Quality \\ Evaluation\end{tabular}}}  & Shahin et al.  \cite{shahin2017deep} & S-EEG & MLP & \begin{tabular}[c]{@{}l@{}}University\\  Hospital\\  in Berlin\end{tabular} & 0.9 \\
&  & Biswai et al. \cite{biswal2017sleepnet} & S-EEG & RNN & Local & 0.8576 \\
&  & Ruffini et al. \cite{ruffini2016eeg} & S-EEG & RNN & Local & 0.85 \\
&  & Vilamala et al. \cite{vilamala2017neural} & S-EEG & CNN & Sleep-EDF & 0.86 \\
&  & Tsinalis et al. \cite{tsinalis2016automatic} & S-EEG & CNN & Sleep-EDF & 0.82 \\
&  & Sors et al. \cite{sors2018convolutional} & S-EEG & CNN & SHHS & 0.87 \\
&  & Chambon et al. \cite{chambon2018deep} & S-EEG & Multi-view CNN & MASS session 3 & N/A \\
&  & Manzano et al. \cite{manzano2017combination} & S-EEG & CNN + MLP & Sleep-EDF & 0.732 \\
&  & Fraiwan et al. \cite{fraiwan2017neonatal} & S-EEG & DBN-AE + MLP & Local & 0.804 \\
&  & Tan et al. \cite{tan2015sleep} & S-EEG & DBN-RBM & Local & 0.9278 (F1) \\
&  & Zhang et al. \cite{zhang2016automatic} & S-EEG & DBN + voting & UCD & 0.9131 \\
&  & Fernandez et al. \cite{fernandez2018sleep} & S-EEG & CNN & SHHS & 0.9 (F1) \\
&  & Supratak et al. \cite{supratak2017deepsleepnet} & S-EEG & CNN + LSTM & \begin{tabular}[c]{@{}l@{}}MASS/\\ Sleep-EDF\end{tabular} & 0.862/0.82 \\  \cline{2-7}
& \multirow{7}{*}{\textbf{\begin{tabular}[c]{@{}l@{}}AD \\ Detection\end{tabular}}} & Morabito et al. \cite{morabito2016deep} & MD EEG & CNN & Local & 0.82 \\
&  & Zhao et al. \cite{zhao2014deep} & MD EEG & DBN-RBM & Local & 0.92 \\
&  & Suk et al. \cite{suk2015deep} & fMRI & \begin{tabular}[c]{@{}l@{}}DBN-AE; \\ DBN-RBM\end{tabular} & ADNI & \begin{tabular}[c]{@{}l@{}}0.979; \\ 0.954\end{tabular} \\
&  & Sarraf et al. \cite{sarraf2016deep} & fMRI & CNN & ADNI & 0.9685 \\
&  & Li et al. \cite{li2014deeplearningbased} & fMRI & CNN + LR & ADNI & 0.9192 \\  
& & Hu et al. \cite{hu2016clinical} & fMRI & D-AE + MLP & ADNI & 0.875 \\ 
&  & Ortiz et al. \cite{ortiz2016ensembles} & fMRI, PET & \begin{tabular}[c]{@{}l@{}}DBN-RBM\\  + SVM\end{tabular} & ADNI & 0.9 \\ \cline{2-7}
 & \multirow{17}{*}{\textbf{\begin{tabular}[c]{@{}l@{}}Seizure \\ Detection\end{tabular}}}  & Hosseini et al. \cite{hosseini2017optimized} & EEG & CNN & Local & 0.96 \\ 
&  & Yuan et al. \cite{yuan2018novel} & MD EEG & Attention-MLP & CHB-MIT & 0.9661 \\
&  & Tsiouris et al. \cite{tsiouris2018long} & MD EEG & LSTM & CHB-MIT & $>$0.99 \\
&  & Talathi et al. \cite{talathi2017deep} & MD EEG & GRU & BUD & 0.996 \\
&  & Acharya et al. \cite{acharya2018deep} & MD EEG & CNN & UBD & 0.8867 \\
&  & Schirmeister et al. \cite{schirrmeister2017deep} & MD EEG & CNN & TUH & 0.854 \\
&  & Hosseini et al. \cite{hosseini2017deep} & MD EEG & CNN & Local & N/A \\
&  & Johansen et al. \cite{johansen2016epileptiform} & MD EEG & CNN & Local & 0.947 (AUC) \\
&  & Ansari et al. \cite{ansari2018neonatal} & MD EEG & CNN + RF & Local & 0.77 \\ 
&  & Ullah et al. \cite{ullah2018automated} & MD EEG & CNN + voting & UBD & 0.954 \\
&  & Wen et al. \cite{wen2018deepconvolution} & MD EEG & AE & Local & 0.92 \\
&  & Lin et al.\cite{lin2016classification} & MD EEG & D-AE & UBD & 0.96 \\
&  & Yuan et al. \cite{yuan2017novel} & MD EEG & D-AE + SVM & CHB-MIT & 0.95 \\
&  & Page et al. \cite{page2014comparing} & MD EEG & DBN-AE + LR & N/A & 0.8 $\sim$ 0.9 \\
& & Turner et al. \cite{turner2014deep} & MD EEG & \begin{tabular}[c]{@{}l@{}}DBN-RBM \\ + LR\end{tabular} & Local & N/A \\
&  & Hosseini et al. \cite{hosseini2017cloud} & MD EEG & D-AE + MLP & Local & 0.94 \\
&  & Golmohammadi et al. \cite{golmohammadi2017deep} & MD EEG & RNN+CNN & TUH & \begin{tabular}[c]{@{}l@{}}Sen: 0.3083; \\ Spe: 0.9686\end{tabular} \\
& & Shah et al. \cite{shah2017optimizing} & MD EEG & CNN+ LSTM & TUH & \begin{tabular}[c]{@{}l@{}}Sen: 0.39; \\ Spe:  0.9037\end{tabular} \\ \cline{2-7}
 \hline
\end{tabular}
}
\end{table*}

\begin{table*}[]
\centering
\renewcommand\thetable{\ref{tab:application}}
\caption*{Table~\ref{tab:application}. Summary of deep learning-based brain signal applications (Continued). IEF and CJD refer to Interictal Epileptic Discharge and Creutzfeldt-Jakob Disease, respectively. }
\label{tab:application_2}
\resizebox{\textwidth}{!}{
\begin{tabular}{ll|lllll}
\hline
\multicolumn{2}{c|}{\textbf{Brain Signal Applications}} & \textbf{Reference} & \textbf{Signals} & \textbf{\begin{tabular}[c]{@{}l@{}}Deep Learning \\ Models\end{tabular}} & \textbf{Dataset} & \textbf{Performance} \\ \hline
\multirow{18}{*}{\textbf{\begin{tabular}[c]{@{}l@{}}Health\\ Care\end{tabular}}} 
& \textbf{Others:} &  &  &  &  &  \\
& \multirow{1}{*}{\textbf{IED}} & Antoniades et al. \cite{antoniades2018deep} & EEG & AE + CNN & Local & 0.68 \\
& \multirow{1}{*}{\bf{CJD}} & Morabito et al. \cite{morabito2017deep} & MD EEG & D-AE & Local & 0.81 $\sim$ 0.83 \\
& \multirow{2}{*}{\textbf{Depression}} & Acharya et al. \cite{acharya2018automated} & MD EEG & CNN & Local & 0.935 $\sim$ 0.9596 \\
&  & Al et al. \cite{al2015multichannel} & MD EEG & \begin{tabular}[c]{@{}l@{}}DBN-RBM \\ + MLP\end{tabular} & Local & 0.695 \\
& \multirow{3}{*}{\textbf{Brain Tumor}} & Morenolopez et al. \cite{Morenolopez2017deep} & fMRI & CNN & BRATS & 0.88 (F1) \\
&  & Shreyas et al. \cite{shreyas2017deep} & fMRI & CNN & BRATS & 0.83 \\
&  & Havaei et al.  \cite{havaei2017brain} & fMRI & Muli-scale CNN & BRATS & 0.88 (F1) \\
& \multirow{2}{*}{\textbf{Schizophrenia}} & Plils et al. \cite{plis2014deep} & fMRI & DBN-RBM & Combined & 0.9 (F1) \\
&  & Chu et al. \cite{chu2017individual} &  & \begin{tabular}[c]{@{}l@{}}CNN + RF\\  + Voting\end{tabular} & Local & 0.816, 0.967, 0.992 \\
& \textbf{\begin{tabular}[c]{@{}l@{}}Mild Cognitive \\ Impairment (MCI)\end{tabular}} & Suk et al. \cite{suk2016state} & fMRI & AE + SVM & ADNI2 & 0.7258 \\ 
& \multirow{2}{*}{\textbf{\begin{tabular}[c]{@{}l@{}}Cardiac \\ Detection\end{tabular}}} & Garg \cite{garg2017automatic} & MEG & CNN & Local & \begin{tabular}[c]{@{}l@{}}Sen: 0.85,\\ Spe: 0.97\end{tabular} \\
&  & Hasasneh et al. \cite{hasasneh2018deep} & MEG & CNN + MLP & Local & 0.944 \\ \cline{1-7}
\multirow{4}{*}{\textbf{\begin{tabular}[c]{@{}l@{}}Smart \\ Environment\end{tabular}}} & \textbf{Robot Control} & Behncke et al. \cite{behncke2018signature} & EEG & CNN & Local & 0.75 \\
 & \textbf{\begin{tabular}[c]{@{}l@{}}Smart \\ Home\end{tabular}} & Zhang et al. \cite{zhang2017intent} & MI EEG & RNN & EEGMMI & 0.9553 \\
  & \textbf{\begin{tabular}[c]{@{}l@{}}Exoskeleton \\ Control\end{tabular}} & Kwak et al. \cite{kwak2017convolutional} & SSVEP & CNN & Local & 0.9403 \\ 
 &  & Huve et al. \cite{huve2018brain} & fNIRS & MLP & Local & 0.82 \\   \hline
\multicolumn{2}{c|}{\multirow{6}{*}{\textbf{\begin{tabular}[c]{@{}l@{}} Communication\end{tabular}}}}  & Zhang et al. \cite{zhang2018converting} & MI EEG & \begin{tabular}[c]{@{}l@{}}LSTM+CNN\\  +AE\end{tabular} & Local & 0.9452 \\
 & & Kawasaki et al. \cite{kawasaki2015visualizing} & VEP & MLP & Local & 0.908 \\
 \multicolumn{2}{l|}{} & Cecotti et al. \cite{cecotti2011convolutional} & VEP & CNN & \begin{tabular}[c]{@{}l@{}}The third BCI \\ competition, \\ Dataset II\end{tabular} & 0.945 \\
 \multicolumn{2}{l|}{} & Liu et al. \cite{liu2018deep} & VEP & CNN & \begin{tabular}[c]{@{}l@{}}The third BCI \\ competition, \\ Dataset II\end{tabular} & 0.92 $\sim$ 0.96 \\ 
\multicolumn{2}{l|}{} & Cecotti et al. \cite{cecotti2011convolutional} & VEP & CNN + Voting & \begin{tabular}[c]{@{}l@{}}The third BCI \\ competition, \\ Dataset II\end{tabular} & 0.955 \\
\multicolumn{2}{l|}{} & Maddula et al. \cite{maddula2017deep} & VEP & RCNN & Local & 0.65$\sim$0.76 \\ \hline
\multirow{4}{*}{\textbf{Security}} & \multirow{3}{*}{\textbf{Identification}} & Zhang et al. \cite{zhang2018mindid} & MI-EEG & \begin{tabular}[c]{@{}l@{}}Attention-based\\  RNN\end{tabular} & EEGMMI + local & 0.9882 \\
 &  & Koike et al. \cite{koike2016high} & VEP & MLP & Local & 0.976 \\
 &  & Mao et al. \cite{mao2017eeg} & RSVP & CNN & Local & 0.97 \\
 & \textbf{Authentication} & Zhang et al. \cite{zhang2017deepkey} & MI EEG & Hybrid & EEGMMI + local & 0.984 \\ \hline
\multicolumn{2}{c|}{\multirow{12}{*}{\textbf{\begin{tabular}[c]{@{}l@{}}Affective  Computing\end{tabular}}}} & Frydenlund et al. \cite{frydenlund2015emotional} & E-EEG & MLP & DEAP & N/A \\
\multicolumn{2}{l|}{} & Zhang et al. \cite{zhang2018spatial} & E-EEG & RNN & SEED & 0.895 \\
\multicolumn{2}{l|}{} & Li et al. \cite{hiroyasu2014gender} & E-EEG & CNN & SEED & 0.882 \\
\multicolumn{2}{l|}{} & Liu et al. \cite{liu2017analyze} & E-EEG & CNN & Local & 0.82 \\
\multicolumn{2}{l|}{} & Li et al. \cite{li2016implementation} & E-EEG & \begin{tabular}[c]{@{}l@{}}Hierarchical \\ CNN\end{tabular} & SEED & 0.882 \\
\multicolumn{2}{l|}{} & Chai et al. \cite{chai2016unsupervised} & E-EEG & AE & SEED & 0.818 \\ 
\multicolumn{2}{l|}{} & Xu et al. \cite{xu2016affective} & E-EEG & \begin{tabular}[c]{@{}l@{}}DBN-AE, \\ DBN-RBM\end{tabular} & DEAP & $>$0.86 (F1) \\
\multicolumn{2}{l|}{} & Jia et al. \cite{jia2014novel} & E-EEG & DBN-RBM & DEAP & \begin{tabular}[c]{@{}l@{}}0.8 $\sim$\\  0.85 (AUC)\end{tabular} \\
\multicolumn{2}{l|}{} & Li et al. \cite{li2015eeg} & E-EEG & DBN-RBM & DEAP & \begin{tabular}[c]{@{}l@{}}Aro:0.642, \\ Val:0.584, \\ Dom 0.658\end{tabular}  \\
\multicolumn{2}{l|}{} & Xu et al. \cite{xu2016eeg} & E-EEG & DBN-RBM & DEAP & \begin{tabular}[c]{@{}l@{}}Aro:0.6984,\\  Val:0.6688, \\ Lik: 0.7539\end{tabular} \\ \hline
\end{tabular}
}
\end{table*}

\begin{table*}[]
\centering
\renewcommand\thetable{\ref{tab:application}}
\caption*{Table~\ref{tab:application}. Summary of deep learning-based brain signal applications (Continued). }
\label{tab:application_3}
\resizebox{\textwidth}{!}{
\begin{tabular}{ll|lllll}
\hline
\multicolumn{2}{c|}{\textbf{Brain Signal Applications}} & \textbf{Reference} & \textbf{Signals} & \textbf{\begin{tabular}[c]{@{}l@{}}Deep Learning \\ Models\end{tabular}} & \textbf{Dataset} & \textbf{Performance} \\ \hline
\multicolumn{2}{c|}{\multirow{8}{*}{\textbf{Affective Computing}}} &  Zheng et al. \cite{zheng2014eeg} & E-EEG & \begin{tabular}[c]{@{}l@{}}DBN-RBM\\  + HMM\end{tabular} & Local & 0.8762 \\
\multicolumn{2}{l|}{} & Zhang et al. \cite{zheng2015investigating, zheng2015revealing} & E-EEG & \begin{tabular}[c]{@{}l@{}}DBN-RBM \\ + MLP\end{tabular} & SEED & 0.8608 \\
\multicolumn{2}{l|}{} & Gao et al. \cite{gao2015deep} & E-EEG & \begin{tabular}[c]{@{}l@{}}DBN-RBM\\  + MLP\end{tabular} & Local & 0.684 \\
\multicolumn{2}{l|}{} & Yin et al. \cite{yin2017recognition} & E-EEG & \begin{tabular}[c]{@{}l@{}}Multi-view D-AE\\  + MLP\end{tabular} & DEAP & \begin{tabular}[c]{@{}l@{}}Aro: 0.7719; \\ Val: 0.7617\end{tabular} \\
\multicolumn{2}{l|}{} & Mioranda et al. \cite{mioranda2018multi} & E-EEG & RNN + CNN & AMIGOS & <0.7 \\
\multicolumn{2}{l|}{} & Alhagry et al. \cite{alhagry2017emotion} & E-EEG & LSTM + MLP & DEAP & \begin{tabular}[c]{@{}l@{}}Aro:0.8565, \\ Val:0.8545, \\ Lik: 0.8799\end{tabular} \\
\multicolumn{2}{l|}{} & Liu et al. \cite{liu2016emotion} & \begin{tabular}[c]{@{}l@{}}EEG\end{tabular} & AE & \begin{tabular}[c]{@{}l@{}}SEED,\\  DEAP\end{tabular} & 0.9101, 0.8325 \\ 
\multicolumn{2}{l|}{} & Kawde et al. \cite{kawde2017deep} & EEG & DBN-RBM & DEAP & \begin{tabular}[c]{@{}l@{}}Aro: 0.7033;\\  Val: 0.7828; \\ Dom: 0.7016\end{tabular} \\\hline

\multicolumn{2}{c|}{\multirow{9}{*}{\textbf{Drive Fatigue Detection}}} & Hung et al. \cite{hung2017brain,hung2017brain} & EEG & CNN & Local & 0.572 (RMSE) \\
\multicolumn{2}{l|}{} & Hung et al. \cite{hung2017brain} & EEG & CNN & Local &  \\
\multicolumn{2}{l|}{} & Almogbel et al. \cite{almogbel2018eeg} & EEG & CNN & Local & 0.9531 \\
\multicolumn{2}{l|}{} & Hajinoroozi et al. \cite{hajinoroozi2015prediction,hajinoroozi2015prediction} & EEG & CNN & Local & 0.8294 \\ 
\multicolumn{2}{l|}{} & Hajinoroozi et al. \cite{hajinoroozi2015feature} & EEG & DBN-RBM & Local & 0.85 \\
\multicolumn{2}{l|}{} & San et al. \cite{san2016eeg} & EEG & DBN-RBM + SVM & Local & 0.7392 \\
\multicolumn{2}{l|}{} & Chai et al. \cite{chai2017improving} & EEG & DBN + MLP & Local & 0.931 \\
\multicolumn{2}{l|}{} & Du et al. \cite{du2017detecting} & \begin{tabular}[c]{@{}l@{}}EEG\end{tabular} & D-AE + SVM & Local & 0.094 (RMSE) \\
\multicolumn{2}{l|}{} & Hachem et al. \cite{hachem2014effect} & SSVEP & MLP & Local & 0.75 \\ \hline
\multicolumn{2}{c|}{\multirow{7}{*}{\textbf{Mental Load Measurement}}} & Yin et al. \cite{yin2017cross} & EEG & D-AE & Local & 0.9584 \\
\multicolumn{2}{l|}{} & Bashivan et al. \cite{bashivan2015single} & EEG & DBN-RBM & Local & 0.92 \\
\multicolumn{2}{l|}{} & Li et al. \cite{li2016single} & EEG & DBN-RBM & Local & 0.9886 \\
\multicolumn{2}{l|}{} & Bashivan et al. \cite{bashivan2016learning} & EEG & R-CNN & Local & 0.9111 \\
\multicolumn{2}{l|}{} & Bashivan et al. \cite{bashivan2016mental} & EEG & DBN + MLP & Local & N/A \\
\multicolumn{2}{l|}{} & Naseer et al. \cite{naseer2016analysis} & fNIRS & MLP & Local & 0.963 \\
\multicolumn{2}{l|}{} & Hennrich et al. \cite{hennrich2015investigating} & fNIRS & MLP & Local & 0.641 \\ \hline

\multirow{22}{*}{\textbf{\begin{tabular}[c]{@{}l@{}}Other \\ Appli- \\ -cations\end{tabular}}}  & 
\textbf{School Bullying} & 
Baltatzis et al. \cite{baltatzis2017bullying} & EEG & CNN & Local & 0.937 \\
 & \multirow{4}{*}{\textbf{Music Detection}} & Stober et al. \cite{stober2014classifying} & EEG & CNN & Local & 0.776 \\
 &  & Stober et al. \cite{stober2015deep} & EEG & AE + CNN & Open MIIR & 0.27 for 12-class \\
 &  & Stober et al. \cite{stober2014using} & EEG & CNN & Local & 0.244 \\
 &  & Sternin et al. \cite{sternin2015tempo} & \begin{tabular}[c]{@{}l@{}}EEG \end{tabular} & CNN & Local & 0.75 \\
 & \textbf{\begin{tabular}[c]{@{}l@{}}Number \\ Choosing\end{tabular}} & Waytowich et al. \cite{waytowich2018compact} & SSVEP & CNN & Local & 0.8 \\
 & \multirow{6}{*}{\textbf{\begin{tabular}[c]{@{}l@{}}Visual Object \\ Recognition\end{tabular}}} & Cichy et al. \cite{cichy2016comparison} & \begin{tabular}[c]{@{}l@{}}fMRI, MEG\end{tabular} & CNN & N/A & N/A \\
 &  & Manor et al. \cite{manor2015convolutional} & RSVP & CNN & Local & 0.75 \\
 &  & Cecotti et al. \cite{cecotti2017convolutional} & RSVP & CNN & Local & 0.897 (AUC) \\
 &  & Hajinoroozi et al. \cite{hajinoroozi2017deep} & RSVP & CNN & Local & 0.7242 (AUC) \\
 &  & Shamwell et al. \cite{shamwell2016single} & RSVP & CNN & Local & 0.7252 (AUC) \\
  &  & Perez et al. \cite{perez2018development} & SSVEP & AE & Local & 0.9778 \\
 & \textbf{\begin{tabular}[c]{@{}l@{}}Guilty \\ Knowledge \\ Test\end{tabular}} & Kulasingham et al. \cite{kulasingham2016deep} & SSVEP & \begin{tabular}[c]{@{}l@{}}DBN-RBM; \\ DBN-AE\end{tabular} & Local & \begin{tabular}[c]{@{}l@{}}0.869; \\  0.8601\end{tabular} \\
 & \textbf{\begin{tabular}[c]{@{}l@{}}Concealed\\  Information \\ Test\end{tabular}} & Liu et al. \cite{liu2017deep} & EEG & DBN-RBM & Local & 0.973 \\
 & \textbf{Flanker Task} & Volker et al. \cite{volker2018deep} & EEG & CNN & Local & 0.841 \\
 & \multirow{2}{*}{\textbf{Eye State}} & Narejo et al. \cite{narejo2016eeg} & EEG & DBN-RBM & UCI & 0.989 \\
 &  & Reddy et al. \cite{reddy2016online} & EEG & MLP & Local & 0.975 \\
 & \textbf{User Preference} & Teo et al. \cite{teo2017deep} & EEG & MLP & Local & 0.6399 \\
 & \textbf{\begin{tabular}[c]{@{}l@{}}Emergency \\ Braking\end{tabular}} & Hernandez et al. \cite{hernandez2018eeg} & EEG & CNN & Local & 0.718 \\
 & \multirow{2}{*}{\textbf{\begin{tabular}[c]{@{}l@{}}Gender \\ Detection\end{tabular}}} & Putten et al. \cite{putten2018predicting} & EEG & CNN & Local & 0.81  \\
 &  & Hiroyasu et al. \cite{hiroyasu2014gender} & fNIRS & D-AE + MLP & Local & 0.81 \\ \hline
\end{tabular}
}
\end{table*}

\subsection{Security} 
\label{sub:security}
Brain signals can be used in security scenarios such as identification (or recognition) and authentication (or verification). The former conducts multi-class classification to recognize a person's identity \cite{zhang2018mindid}. The latter conducts binary classification to decide whether a person is authorized \cite{zhang2017deepkey}. 

The majority of the existing biometric identification/authentication systems rely on individuals' intrinsic physiological features such as face, iris, retina, voice, and fingerprint \cite{zhang2018mindid}. They are vulnerable to various attacks based on anti-surveillance prosthetic masks, contact lenses, vocoder, and fingerprint films. EEG-based biometric person identification is a promising alternative given its highly resilient to spoofing attacks---individual’s EEG signals are virtually impossible for an imposter to mimic. Koike et al. \cite{koike2016high} have adopted deep neural networks to identify the user's ID based on the VEP signals; Mao et al. \cite{mao2017eeg} applied CNN for person identification based on RSVP signals; Zhang et al. \cite{zhang2018mindid} proposed an attention-based LSTM model and evaluated it over both public and local datasets. EEG signals are also combined with gait information in a hybrid deep learning model for a dual-authentication system \cite{zhang2017deepkey}.

\subsection{Affective Computing} 

Affective states of a user provide critical information for many applications such as personalized information (\eg, multimedia content) retrieval or intelligent human-computer interface design \cite{xu2016affective}. Recent research illustrated that deep learning models can enhance the performance in affective computing.
The most widely used circumplex model believe the emotions are distributed in two dimensions: arousal and valence. The arousal refers to the intensity of the emotional stimuli or how strong is the emotion. The valence refers to the relationship within the person who experiences the emotion. In some other models, the dominance and liking dimensions are deployed. 

Some research \cite{li2016implementation,liu2017analyze,wang2018data} attempts to classify users' emotional state into two (positive/negative) or three categories (positive, neutral, and negative) based on EEG signals using deep learning algorithms such as CNN and its variants \cite{frydenlund2015emotional}.
DBN-RBM is the most representative deep learning model to discover the concealed features from emotional spontaneous EEG \cite{xu2016affective,zheng2015investigating}. Xu et al. \cite{xu2016affective} applied DBN-RBM as feature extractors to classify affective states based on EEG.

Further, some researchers aim to recognize the positive/negative state of each specific emotional dimension. For example, Yin et al. \cite{yin2017recognition} employed an ensemble classifier of AE in order to recognize the user's affection. Each AE uses three hidden layers to filter out noises and to derive stable physiological feature representations. The proposed model was evaluated over the benchmark, DEAP, and achieved the arousal of 77.19\% and valence of 76.17\%. 

\subsection{Driver Fatigue Detection} 
\label{sub:drive_fatigue_detection}

Vehicle drivers' ability to keep alert and maintain optimal performance will dramatically affect the traffic safety \cite{almogbel2018eeg}. EEG signals have proven useful in evaluating the human's cognitive state in different context. Generally, a driver is regarded as in an alert state if the reaction time is lower than 0.7 seconds and in fatigue state if it is higher than 2.1 seconds.
Hajinoroozi et al. \cite{hajinoroozi2015feature} considered the detection of driver's fatigue from EEG signals by discovering the distinct features. They explored an approach based on DBN for dimension reduction. 

Detecting driver fatigue is crucial because the drowsiness of the driver may lead to disaster. Driver fatigue detection is feasible in practice. In the hardware aspect, the collection equipment of EEG singles is off-the-shelf and portable enough to be used in a car. Moreover, the price of an EEG headset is affordable for most people. In the algorithm aspect, deep learning models have enhanced the performance of fatigue detection. As we summarized, the EEG based driving drowsiness can be recognized with high accuracy (82\% $\sim$ 95\%). 

Future scope of drive fatigue detection is in the self-driving scenario. As we know, in the most situation of self-driving (\eg, Automation level 3\footnote{https://en.wikipedia.org/wiki/Self-driving\_car}), the human driver is expected to respond appropriately to a request to intervene, which indicates that the driver should keep alert state. Therefore, we believe the application of brain signal-based drive fatigue detection will benefit the development of the self-driving car.

\subsection{Mental Load Measurement} 
\label{sub:mental_load_measurement}
The EEG oscillations can be used to measure the mental workload level, which can sustain decision making and strategy development in the context of human-machine interaction \cite{yin2017cross}. Additionally, the appropriate mental workload is essential for maintaining human health and preventing accidents.
For example, the abnormal mental workload of the human operator may result in performance degradation which could cause catastrophic accidents \cite{parasuraman2012individual}.
Evaluation of operator Mental Workload levels via ongoing EEG is quite promising in Human-Machine collaborative task environment to alarm the temporal operator performance degradation. 

\begin{table*}[t]
\centering
\caption{The summary of public dataset for brain signal studies. The `\# Sub', `\# Cla', and S-Rate denote the number of subject, number of class, and sampling rate, respectively. FM denote finger movement while BCI-C denote the BCI Competition. The `\# channel` refers to the number of brain signal channels.}
\label{tab:dataset}
\small
\begin{tabular}{ll|lllll}
\hline
\multicolumn{2}{c|}{\textbf{Brain Signals}} & \textbf{Name Link} & \textbf{\# Sub} & \textbf{\# Cla} & \textbf{S-Rate} & \textbf{\# Channel} \\ \hline
\multirow{26}{*}{\textbf{EEG}} & \multirow{8}{*}{\textbf{\begin{tabular}[c]{@{}l@{}}Sleep \\ EEG\end{tabular}}} & Sleep-EDF\footnotemark: Telemetry & 22 & 6 & 100 & \begin{tabular}[c]{@{}l@{}}2 \end{tabular} \\
 &  & Sleep-EDF: Cassette & 78 & 6 & 100, 1 & \begin{tabular}[c]{@{}l@{}}2 \end{tabular} \\
 &  & MASS-1\footnotemark & 53 & 5 & 256 & \begin{tabular}[c]{@{}l@{}}17\end{tabular} \\
 &  & MASS-2 & 19 & 6 & 256 & \begin{tabular}[c]{@{}l@{}}19\end{tabular} \\
 &  & MASS-3 & 62 & 5 & 256 & \begin{tabular}[c]{@{}l@{}}20\end{tabular} \\
 &  & MASS-4 & 40 & 6 & 256 & \begin{tabular}[c]{@{}l@{}}4\end{tabular} \\
 &  & MASS-5 & 26 & 6 & 256 & \begin{tabular}[c]{@{}l@{}}20\end{tabular} \\
 &  & SHHS\footnotemark & 5804 & N/A & 125, 50 & \begin{tabular}[c]{@{}l@{}}2\end{tabular} \\ \cline{2-7}
 & \multirow{2}{*}{\textbf{\begin{tabular}[c]{@{}l@{}}Seizure \\ EEG\end{tabular}}} & CHB-MIT\footnotemark & 22 & 2 & 256 & 18 \\
 &  & TUH\footnotemark & 315 & 2 & 200 & 19 \\ \cline{2-7}
 & \multirow{10}{*}{\textbf{\begin{tabular}[c]{@{}l@{}}MI \\ EEG\end{tabular}}} & EEGMMI\footnotemark & 109 & 4 & 160 & 64 \\
 &  & BCI-C II\footnotemark, Dataset III & 1 & 2 & 128 & 3 \\
 &  & BCI-C III, Dataset III a & 3 & 4 & 250 & 60 \\
 &  & BCI-C III, Dataset III b & 3 & 2 & 125 & 2 \\
 &  & BCI-C III, Dataset IV a & 5 & 2 & 1000 & 118 \\
 &  & BCI-C III, Dataset IV b & 1 & 2 & 1001 & 119 \\
 &  & BCI-C III, Dataset IV c & 1 & 2 & 1002 & 120 \\
 &  & BCI-C IV, Dataset I & 7 & 2 & 1000 & 64 \\
 &  & BCI-C IV, Dataset II a & 9 & 4 & 250 & 22 \\
 &  & BCI-C IV, Dataset II b & 9 & 2 & 250 & 3 \\ \cline{2-7}
 & \multirow{3}{*}{\textbf{\begin{tabular}[c]{@{}l@{}}Emotional\\  EEG\end{tabular}}} & AMIGOS\footnotemark & 40 & 4 & 128 & 14 \\
 &  & SEED\footnotemark & 15 & 3 & 200 & 62 \\
 &  & DEAP\footnotemark & 32 & 4 & 512 & 32 \\ \cline{2-7}
 & \textbf{\begin{tabular}[c]{@{}l@{}}Others \\ EEG\end{tabular}} & Open MIIR\footnotemark & 10 & 12 & 512 & 64 \\ \cline{2-7}
 & \multirow{2}{*}{\textbf{VEP}} & BCI-C II, Dataset II b & 1 & 36 & 240 & 64 \\
 &  & BCI-C III, Dataset II & 2 & 26 & 240 & 64 \\ \hline
\multicolumn{2}{c|}{\multirow{2}{*}{\textbf{fMRI}}} & ADNI\footnotemark & 202 & 3 & N/A & N/A \\
\multicolumn{2}{l|}{} & BRATS\footnotemark 2013 & 65 & 4 & N/A & N/A \\ \hline
\multicolumn{2}{c|}{\textbf{MEG}} & BCI-C IV, Dataset III & 2 & 4 & 400 & 10 \\ \hline
\end{tabular}
\end{table*}
\addtocounter{footnote}{-12}
\footnotetext[6]{https://physionet.org/physiobank/database/sleep-edfx/}
\addtocounter{footnote}{1}
\footnotetext[7]{https://massdb.herokuapp.com/en/}
\addtocounter{footnote}{1}
\footnotetext[18]{https://physionet.org/pn3/shhpsgdb/}
\addtocounter{footnote}{1}
\footnotetext[19]{https://physionet.org/pn6/chbmit/}
\addtocounter{footnote}{1}
\footnotetext[20]{https://www.isip.piconepress.com/projects/tuh\_eeg/html/downloads.shtml}
\addtocounter{footnote}{1}
\footnotetext[21]{https://physionet.org/pn4/eegmmidb/}
\addtocounter{footnote}{1}
\footnotetext[22]{http://www.bbci.de/competition/ii/}
\addtocounter{footnote}{1}
\footnotetext[23]{http://www.eecs.qmul.ac.uk/mmv/datasets/amigos/readme.html}
\addtocounter{footnote}{1}
\footnotetext[24]{http://bcmi.sjtu.edu.cn/~seed/download.html}
\addtocounter{footnote}{1}
\footnotetext[25]{https://www.eecs.qmul.ac.uk/mmv/datasets/deap/}
\addtocounter{footnote}{1}
\footnotetext[26]{https://owenlab.uwo.ca/research/the\_openmiir\_dataset.html}
\addtocounter{footnote}{1}
\footnotetext[27]{http://adni.loni.usc.edu/data-samples/access-data/}
\addtocounter{footnote}{1}
\footnotetext[28]{https://www.med.upenn.edu/sbia/brats2018/data.html}

Several researchers have been paid attention to this topic. The mental workload can be measured from fNIRS signals or spontaneous EEG. Naseer et al. adopted a MLP algorithm for fNIRS-based binary mental task level classification (mental arithmetic and rest) \cite{naseer2016analysis}.
The experiment results showed that the MLP outperformed the traditional classifiers like SVM, KNN, and achieved the highest accuracy of 96.3\%. Bashivan et al. \cite{bashivan2015single} presented a statistical approach, a DBN model, for the recognition of mental workload level based on single-trial EEG. Before the DBN, the authors manually extracted the wavelet entropy and band-specific power from three frequency bands (theta, alpha, and beta). At last, the experiments demonstrated the recognition of mental workload achieved an overall accuracy of 92\%. Zhang et al. \cite{zhang2018learning} investigate the mental load measurement across multiple mental tasks via a recurrent-convolutional framework. The model simultaneously learns EEG features
from the spatial, spectral, and temporal dimensions, which results in the accuracy of 88.9\% in binary classification (high/low workload levels).

\subsection{Other Applications} 
\label{sub:other_applications}
There are plenty of interesting scenarios beyond the above where deep learning-based brain signals can apply, such as recommender system \cite{teo2017deep} and emergency braking \cite{hernandez2018eeg}. One possible topic is the recognition of a visual object, which may be used in guilty knowledge test \cite{kulasingham2016deep} and concealed information test \cite{liu2017deep}. The neurons of the participant will produce a pulse when he/she suddenly watch a similar object. Based on the theory, the visual target recognition is mainly used RSVP signals. Cecotti et al. \cite{cecotti2017convolutional} 
aimed to build a common model for target recognition, which can work for various subjects instead of a specific subject. 

Besides, researchers have investigated to distinguish the subject's gender by the fNIRS \cite{hiroyasu2014gender} and spontaneous EEG \cite{putten2018predicting}. Hiriyasu et al. \cite{hiroyasu2014gender} adopted deep learning to recognize the gender of the subject based on the cerebral blood flow. The experiment results suggested that the cerebral blood flow changes in different ways for male and female.
Putten et al. \cite{putten2018predicting} tried to discover the sex-specific information from the brain rhythms and adopted a CNN model to recognize the participant's gender. This paper illustrated that fast beta activity (20 $\sim$25 Hz) is one of the most distinctive attributes.

\subsection{Benchmark Datasets} 
\label{sec:datasets}
We have extensively explored the benchmark datasets usable for deep learning-based brain signals (Table~\ref{tab:dataset}). We provide a bunch of public datasets with download links, which cover most brain signal types. In particular, BCI competition IV (BCI-C IV) contains five datasets via the same link. For better understanding, we present the number of subjects, the number of class (how many categories), sampling rate, and the number of channels of each dataset. In the `\# Channel' column, the default channel is for EEG signals. Some datasets contain more biometric signals (\eg, ECG), but we only list the channels related to brain signals.

\section{Analysis and Guidelines} 
\label{sec:analysis}

In this section, we first analyze what is the most suitable deep learning models for each brain signal. Then, we summarize the popular deep learning models in brain signal research. At last, we investigate the brain signals in terms of application. We hope this survey could help our readers to select the most effective and efficient methods when dealing with brain signals. Please recall Table~\ref{tab:summary} where we summarize the brain signals and the corresponding deep learning models of the state-of-the-art papers. 
Figure~\ref{fig:bar} illustrated of the publications proportion for crucial brain signals and deep learning models.

\subsection{Brain Signal Acquisition} 
\label{sub:bci_signal_based_discussion}

Among the non-invasive signals, the studies on EEG is far more than the sum of all the other brain signal paradigms (fNIRS, fMRI, and MEG). Furthermore, there are about 70\% of the EEG papers pay attention to the spontaneous EEG (133 publications). For better understanding, we split the spontaneous EEG into several aspects: the sleep, the motor imagery, the emotional, the mental disease, the data augmentation, and others. 

First, the classification of the sleep EEG mainly depends on the discriminative and the hybrid models. Among the nineteen studies about sleep stage classification, there are six employed CNN and the modified CNN models independently while two papers adopted RNN models. There are three hybrid models built on the combination of CNN and RNN. 

Second, in terms of the research on MI EEG (30 publications), the independent CNN and CNN-based hybrid models are widely used. As for the representative models, DBN-RBM is often applied to capture the latent features from the MI EEG signals. 

Third, there are twenty-five publications related to spontaneous emotional EEG. More than half of them employed representative models (such as D-AE, D-RBM, especially DBN-RBM) for unsupervised feature learning. The most typical state recognition works recognize the user's emotion as positive, neutral, or negative. Some researchers take a further step to classify the valence, and the arouse rate, which is more complex and challenging. 

Fourth, the research on mental disease diagnosis is promising and attracting. The majority of the related research focuses on the detection of epileptic seizure and Alzheimer's Disease. Since the detection is a binary classification problem which is rather easier than multi-class classification, many studies can achieve a high accuracy like above 90\%. In this area, the standard CNN model and the D-AE are prevalent. One possible reason is that CNN and AE are the most well-known and effective deep learning models for classification and dimensionality reduction. 

Fifth, several publications pay attention to the GAN based data augmentation. At last, about thirty studies are investigating other spontaneous EEG such as driving fatigue, audio/visual stimuli impact, cognitive/mental load, and eye state detection. These studies extensively apply standard CNN models and variants. 

Moreover, apart from spontaneous EEG, evoked potentials also attracted much attention. On the one hand, in ERP, VEP and the subcategory RSVP has drawn lots of investigations because visual stimuli, compared to other stimuli, is easier to be conducted and more applicable in the real world (\eg, P300 speller can be used for brain typing). For VEP (twenty-one publications), there are elven studies applied discriminative models, and six works adopted hybrid models. In terms of RSVP, the sole CNN dominates the algorithms. Apart from them, five papers focused on the analysis of AEP signals.
On the other hand, among the steady-state related researches, only SSVEP has been studied by deep learning models. Most of them only applied discriminative models on the recognition of the target image. 

Furthermore, beyond the diverse EEG diagrams, a wide range of papers paid attention to fNIRS and fMRI. The fNIRS images are rarely studied by deep learning, and the major studies just employed the simple MLP models. We believe more attention should be paid to the research on fNIRS for the high portability and low cost. As for the fMRI, twenty-three papers proposed deep learning models to the classification. The CNN model is widely used for its outstanding performance in feature learning from images. There are also several papers interested in image reconstruction based on fMRI signals. One reason why fMRI is so hot is that several public datasets are available on the Internet, although the fMRI equipment is expensive. 
The MEG signals are mainly used in the medical area, which is insensitive to the deep learning algorithm. Thus, we only found very few studies on MEG. The sparse AE and CNN algorithms have a positive influence on the feature refining and classification of MEG.

\subsection{Selection Criteria for Deep Learning Models} 
\label{sub:deep_learning_model_based_discussion}

Our investigation shows that discriminative models are most frequent in the summarized publications. This is reasonable at a high level because a large proportion of brain signal issues can be regarded as a classification problem. Another observation is that CNN and its variants are adopted in more than 70\% of the discriminative models, for which we provide reasons as follows. 

First, the design of CNN is powerful enough to extract the latent discriminative features and spatial dependencies from the EEG signals for classification. As a result, CNN structures are adopted for classification in some studies while adopted for feature extraction in some other studies. 

Second, CNN has been achieved great success in some research areas (\eg, computer vision), which makes it extremely famous and feasible (public codes). Thus, the brain signal researchers have more chance to understand and apply CNN on their works. 

Third, some brain signal diagrams (\eg, fMRI) are naturally formed as two-dimension images that are conducive to be processedg by CNN. Meanwhile, other 1-D signals (\eg, EEG) could be converted into 2-D images for further analysis by CNN. Here, we provide several methods converting 1-D EEG signals (with multiple channels) to the 2-D matrix: 1) convert each time-point\footnote{Time-point represents one sampling point. For example, we can have 100 time-points if the sampling rate is 100 Hz.} to a 2-D image; 2) convert a segment into a 2-D matrix. In the first situation, suppose we have 32 channels, and we can collect 32 elements (each element corresponding to a channel) at each time-point. As described in \cite{li2016implementation}, the collected 32 elements could be converted into a 2-D image based on the spatial position. In the second situation, suppose we have 32 channels, and the segment contains 100 time-points. The collected data can be arranged as a matrix with the shape of $[32, 100]$ where each row and column refers to a specific channel and time-point, respectively. 

Fourth, there are a lot of variants of CNN which are suitable for a wide range of brain signal scenarios. For example, the single-channel EEG signals can be processed by 1-D CNN. In terms of RNN, only about 20\% of discriminative model-based papers adopted RNN, which is much less than we expected since RNN has demonstrated powerful in temporal feature learning. One possible reason for this phenomena is that processing a long sequence by RNN is time-consuming and the EEG signals are generally formed as a long sequence. For example, the sleep signals are usually sliced into segments with 30 seconds, which has 3000 time-points under 100 Hz sampling rate. For a sequence with 3000 elements, through our preliminary experiments, RNN takes more than 20 folds training time than CNN. Moreover, MLP is not popular due to its inferior effectiveness (\eg, non-linear ability) to the other algorithms its simple deep learning architecture. 

As for representative models, DBN, especially DBN-RBM, is the most popular model for feature extraction. DBN is widely used in brain signal for two reasons: 1) it learns the generative parameters that reveal the relationship of variables in neighboring layers efficiently;
2) it makes it straightforward to calculate the values of latent variables in each hidden layer \cite{deng2012three}.
%
However, most works that employed the DBN-RBM model were published before 2016. It can be inferred that the researchers prefer to use DBN for feature learning followed by a non-deep learning classifier before 2016; but recently, an increasing number of studies would like to adopt CNN or hybrid models for both feature learning and classification. 

Moreover, generative models are rarely employed independently. The GAN- and VAE-based data augmentation and image reconstruction are mainly focused on fMRI and EEG signals. It is demonstrated that the trained classifier will achieve more competitive performance after data augmentation. Therefore, this is a promising research prospect in the future. 

Last but not the least, there are fifty-three publications proposed hybrid models for brain signal studies. Among them, the combinations of RNN and CNN take about one-fifth proportion. Since RNN and CNN are illustrated having excellent temporal and spatial feature extraction ability, it is natural to combine them for both temporal and spatial feature learning. Another type of hybrid models is the combination of representative and discriminative models. This is easy to understand because the former is employed for feature refining, and the latter is employed for classification. 
There are twenty-eight publications which almost covered all the brain signals proposed this type of hybrid deep learning models. The adopted representative models are mostly AE or DBN-RBM; at the meanwhile, the adopted discriminative models are mostly CNN. Apart from that, there are twelve papers proposed other hybrid models such as two discriminative models. For example, several studies proposed the combination of CNN and MLP where a CNN structure is used for extract spatial features and an MLP is used for classification.

\subsection{Application Performance} 
\label{sub:application_performance}

In order to have a closer observation of the recent advances on deep learning-based brain signal analysis, we analyze the brain signal acquisition methods and the deep learning algorithms in terms of application performance. In some cases, various studies adopt the same deep architecture working on the same dataset but results in different performance, which maybe caused by the different pre-processing methods and hyper-parameter settings.

To begin with, the most appealing and hot field is that using brain signal analysis on health care area. For sleep quality evaluation, the dominate brain signals are spontaneous EEG which are measured while the patient is sleeping.
The single RNN or CNN models seem have a good discriminative feature learning ability and lead to a comprehensive performance. Generally, most of the deep learning algorithms can achieve the accuracy of above 85\% in the context of multiple sleep stage scenario. Upon this, the combined hybrid models (\eg, CNN integrates with LSTM) can only have incremental improvements. 

One key method to detect Alzheimer's Disease is brain signal analysis by measuring the functions of specific brain regions. In detail, the diagnosis can be conducted by spontaneous EEG signals or fMRI images. For MD EEG, DBN is supposed to outperform CNN since the EEG signals contains more temporal instead of spatial information. As for the fMRI pictures, CNN have great advantages in the grid-arranged spatial information learning, which makes it obtain a very comprehensive classification accuracy (above 90\%). As for epileptic seizure, the diagnosis are generally based on EEG signals. The single RNN classifier (\eg, LSTM or GRU) seems work better than its counterparts due to the excellent temporal dependency representing ability. Here, the complex hybrid models indeed outperform the single component. For example, \cite{golmohammadi2017deep} achieves a better specification than \cite{schirrmeister2017deep} on the same dataset because of combing with RNN. Most of the epileptic seizure detection models claim a rather high classification accuracy (above 95\%). One possible reason is that the binary recognition scenario is much easier than multi-class classification. 

The brain signal-controlled smart environment only appear in a small number of publications. Among them, the brain signals are collected through very different methods. This is an emerging but promising field because it is easy to integrate with smart home and smart hospital to benefit the individuals whether healthy or disable. Another advantage of brain signals is bridging people's inside and outer world by communication techniques. In this area, lots of investigations are focusing on the VEP signals because the visual evoked potential is obvious and easy to be detected. One important data source is from the third BCI competition. In addition, brain signal analysis can be widely implement in security systems since the brain signals are invisible and very hard to be mimicked. The characteristic of high fake-resistance enables brain signal a raising star in the identification/authentication in confidential scenarios. The drawbacks of brain signal-based security systems are the expensive equipment and inconvenient (\eg, the subject have to wear an EEG headset to monitor the brainwaves).   

Affective computing has drawn much attention in recent years. The EEG signals have high temporal resolution and able to capture the quick-varying emotions. Therefore, almost all the studies are based on spontaneous EEG signals. The signals are gathered when the subject is watching video which is supposed to arouse the subject's specific emotion.
Another reason for this phenomenon is that there are several open-source EEG-based affecting analysis datasets (\eg, DEAP and SEED) which greatly promote the investigation in this area. The EEG-based affective computing contains two mainstreams. One of them focuses on developing powerful discriminative classifiers (such as hierarchical CNN) which are designed to perform feature extraction and classification in the same step.
The other tries to learn the latent features through deep representative models (\eg, DBN-RBM) and then send the learned representations into a powerful classifier (such as HMM and MLP). It can be observed that the former models (\cite{zhang2018spatial,hiroyasu2014gender}) seem outperform the latter methods (\cite{zheng2015investigating}) with a small margin on the SEED dataset. 

Drive fatigue detection can be easily integrated in the platforms such as self-driving vehicles. Nevertheless, there are only a few publications in this area due to the expensive experimental cost and the lack of accessible dataset. Moreover, there are a lot of interesting applications (\eg, guilty knowledge test and gender detection) have been explored by deep learning models.

\section{Open Issues} 
\label{sec:future_directions}


Although deep learning has lifted the performance of brain signal systems, technical and usability challenges remain. The technical challenges concern the classification ability in complex scenarios, and the usability challenges refer to limitations in large scale real-world deployment. In this section, we introduce these challenges and point out the possible solutions.

\subsection{Explainable General Framework} 
\label{sub:general_framework}
Until now, we have introduced several types of brain signals (\eg, spontaneous EEG, ERP, fMRI) and deep learning models that have been applied for each type. One promising research direction for deep learning-based brain signal research is to develop a general framework that can handle various brain signals regardless of the number of channels used for signal collection, the sample dimensions (\eg, 1-D or 2-D sample), and stimulation types (\eg, visual or audio stimuli), etc. 
The general framework would require two key capabilities: the attention mechanism and the ability to capture latent feature. The former guarantees the framework can focus on the most valuable parts of input signals, and the latter enables the framework to capture the distinctive and informative features. 

The attention mechanism can be implemented based on attention scores or by various machine learning algorithms such as reinforcement learning. The attention scores can be inferred from the input data and work as a weight to help the framework to pay attention to the parts with high attention scores. Reinforcement learning has shown to be able to find the most valuable part through a policy search \cite{zhang2018multimodality}. CNN is the most suitable structure for capturing features at various levels and ranges. In the future, CNN could be used as a fundamental feature learning tool and be integrated with suitable attention mechanisms to form a general classification framework.

One additional direction we may consider is how to interpret the feature representation derived by the deep neural network, what is the intrinsic relationship between the learned features and the task-related neural pattern, or neuropathology of mental disorders. More and more people are realizing that interpretation could be even more important than prediction performance, since we usually just treat deep learning as a black box.

\subsection{Subject-Independent Classification} 
\label{sub:person_independent_classification}
Until now, most brain signal classification tasks focus on person-dependent scenarios, where the training samples and testing samples are collected from the identical individual. The future direction is to realize person-independent classification so that the testing data will never appear in the training set. High-performance person-independent classification is compulsory for the wide application of brain signals in the real world. 

One possible solution to achieving this goal is to build a personalized model with transfer learning. A personalized affective model can adopt a transductive parameter transfer approach to construct individual classifiers and to learn a regression function that maps the relationship between data distribution and classifier parameters \cite{zheng2016personalizing}.
Another potential solution is mining the subject-independent component from the input data. The input data can be decomposed into two parts: a subject-dependent component, which depends on the subject and a subject-independent component, which is common for all subjects. A hybrid multi-task model can work on two tasks simultaneously, one focusing on person identification and the other on class recognition. A well-trained and converged model is supposed to extract the subject-independent features in the class recognition task. 

\subsection{Semi-supervised and Unsupervised Classification} 
\label{sub:unsupervised_classification}
The performance of deep learning highly depends on the size of training data, which, however, requires expensive and time-consuming manual labeling to collect abundant class labels for a wide range of scenarios such as sleep EEG. While supervised learning requires both observations and labels for the training, unsupervised learning requires no labels, and semi-supervised learning only requires partial labels \cite{jia2014novel}. They are, therefore, more suitable for problems with little ground truth.

Zhang et al. proposed an Adversarial Variational Embedding (AVAE) framework that combines a VAE++ model (as a high-quality generative model) and semi-supervised GAN (as a posterior distribution learner) \cite{zhang2019adversarial} for robust and effective semi-supervised learning. Jia et al. proposed a semi-supervised framework by leveraging the data distribution of unlabelled data to prompt the representation learning of labelled data \cite{jia2014novel}. 

Two methods may enhance the unsupervised learning: one is to employ crowd-sourcing to label the unlabeled observations; the other is to leverage unsupervised domain adaption learning to align the distribution of source brain signals and the distribution of target signals with a linear transformation.

\subsection{Online Implementation} 
\label{sub:online_implementation}
Most of the existing brain signal systems focus on offline procedure which means that the training and testing dataset are pre-collected and evaluated offline. However, in the real-world scenarios, the brain signal systems are supposed to receive live data stream and produce classification results in real time, which is still very challenging. 

For EEG signals, in the online system, compared to the offline procedure, the gathered live signals are more noisy and unstable due to lots of factors such as the less-concentrating of the subject \cite{koyama2010comparison} and the inherent destabilization of the equipment (\eg, fluctuating sampling rate). Through our empirical experiments, online brain signal systems generally perform a lower accuracy of 10\% than their counterparts.
One future scope of online implementation is to develop a batch of robust algorithms in order to handle the influence factors and discover the latent distinctive patterns underlying the noisy live brain signals. 
\cite{aliakbaryhosseinabadi2019online} implemented an EEG-based online system that achieves comparable performance, however, this work only investigates a very high-level target (\ie, human attention). 
Discovering the latent invariant representations through covariance matrices of EEG signals can help to mitigate the influence of extinct perturbations \cite{kalunga2016online}.
Some post-processing methods (\eg, voting and aggregating) \cite{cecotti2011convolutional,chu2017individual} can help to improve the decoding performance by averaging the results from multiple continues samples. However, these methods will inevitably bring higher latency. Thus, the post-processing requires a trade-off between the high-accuracy and low-latency.

For fNIRS and fMRI, the online evaluation is relatively less challenging since they have a rather low temporal resolution. The online images with less dynamic can be regarded as static images to some extent, which makes the online system approximating to the offline system. Furthermore, most fMRI and MEG signals are used to evaluate the user's neurological status (\eg, detect the effects of tumor) which does not require an instantaneous response. Thus, they have less demand for a real-time monitoring system.

\subsection{Hardware Portability} 
\label{sub:portable_hardware}
Poor portability of hardware has been preventing brain signals from wide application in the real world. In most scenarios, users would like to use small, comfortable, or even wearable brain signal hardware to collect brain signals and to control appliances and assistant robots.

Currently, there are three types of EEG collection equipment: the unportable, the portable headset, and ear-EEG sensors. The unportable equipment has high sampling frequency, channel numbers, and signal quality but is expensive. It is suitable for physical examination in a hospital.
The portable headsets (\eg, Neurosky, Emotiv EPOC) have 1 $\sim$ 14 channels and 128$\sim$ 256 sampling rate but has inaccuracy readings and cause discomfort after long-time use.
The ear-EEG sensors, which are attached to the outer eat, have gained increasing attention recently but remain mostly at the laboratory stage \cite{pacharra2017concealed}.
 The ear-EEG sensors contain a series of electrodes which are placed in each ear canal and concha \cite{mikkelsen2015eeg}. 
The EEGrids, to the best of our knowledge, is the only commercial ear-EEG. It has multi-channel sensor arrays placed around the ear using an adhesive \footnote{http://ceegrid.com/home/concept/} and is even more expensive. A promising future direction is to improve the usability by developing a cheaper (\eg, lower than 200\$) and more comfortable (\eg, can last longer than 3 hours without feeling uncomfortable) wireless ear-EEG equipment. 




\section{Conclusion} 
\label{sec:conclusion}
In this paper, we thoroughly summarize the recent advances in deep learning models for non-invasive brain signal analysis. Compared with traditional machine learning methods, deep learning not only enables to learn high-level features automatically from brain signals but also have less dependency on domain knowledge. We organize brain signals and dominant deep learning models, followed by discussing state-of-the-art deep learning techniques for brain signals. Moreover, we provide guidelines to help researchers to find the suitable deep learning algorithms for each category of brain signals. Finally, we overview deep learning-based brain signal applications and point out the open challenges and future directions.

\section*{References}
\bibliographystyle{IEEEtran}
\bibliography{survey.bib}
\balance

\newpage

\begin{appendices}

\section{Non-invasive Brain Signals}
\label{sec:brain_signals}

Here, we present a detailed introduction of brain signals as shown in Figure~\ref{fig:BCI_signals}.
Non-invasive brain imaging technique can be collected using electrical, magnetic or metabolic methods, which mainly include Electroencephalogram (EEG), Functional near-infrared spectroscopy (fNIRS), Functional magnetic resonance imaging (fMRI), and Magnetoencephalography (MEG). 


\subsection{Electroencephalography (EEG)} 
\label{sub:electroencephalography_eeg_}
\begin{figure}[b]
    \centering
    \hspace{3mm}
    \begin{subfigure}[t]{0.24\textwidth}
        \centering
        \includegraphics[width=0.8\textwidth]{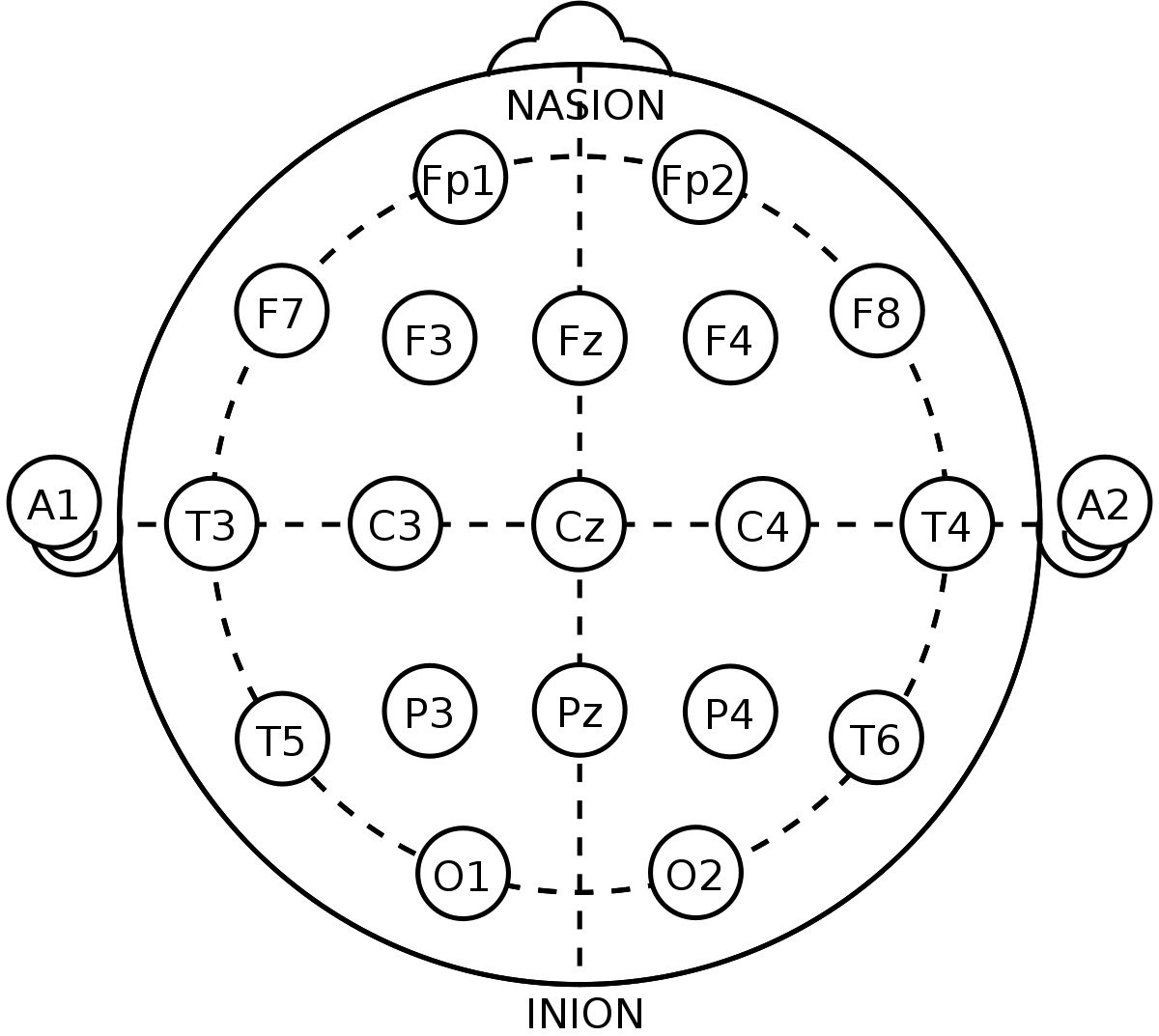}
        \caption{EEG electrode locations}
    \end{subfigure}%
    ~ 
    \begin{subfigure}[t]{0.24\textwidth}
        \centering
        \includegraphics[width=\textwidth]{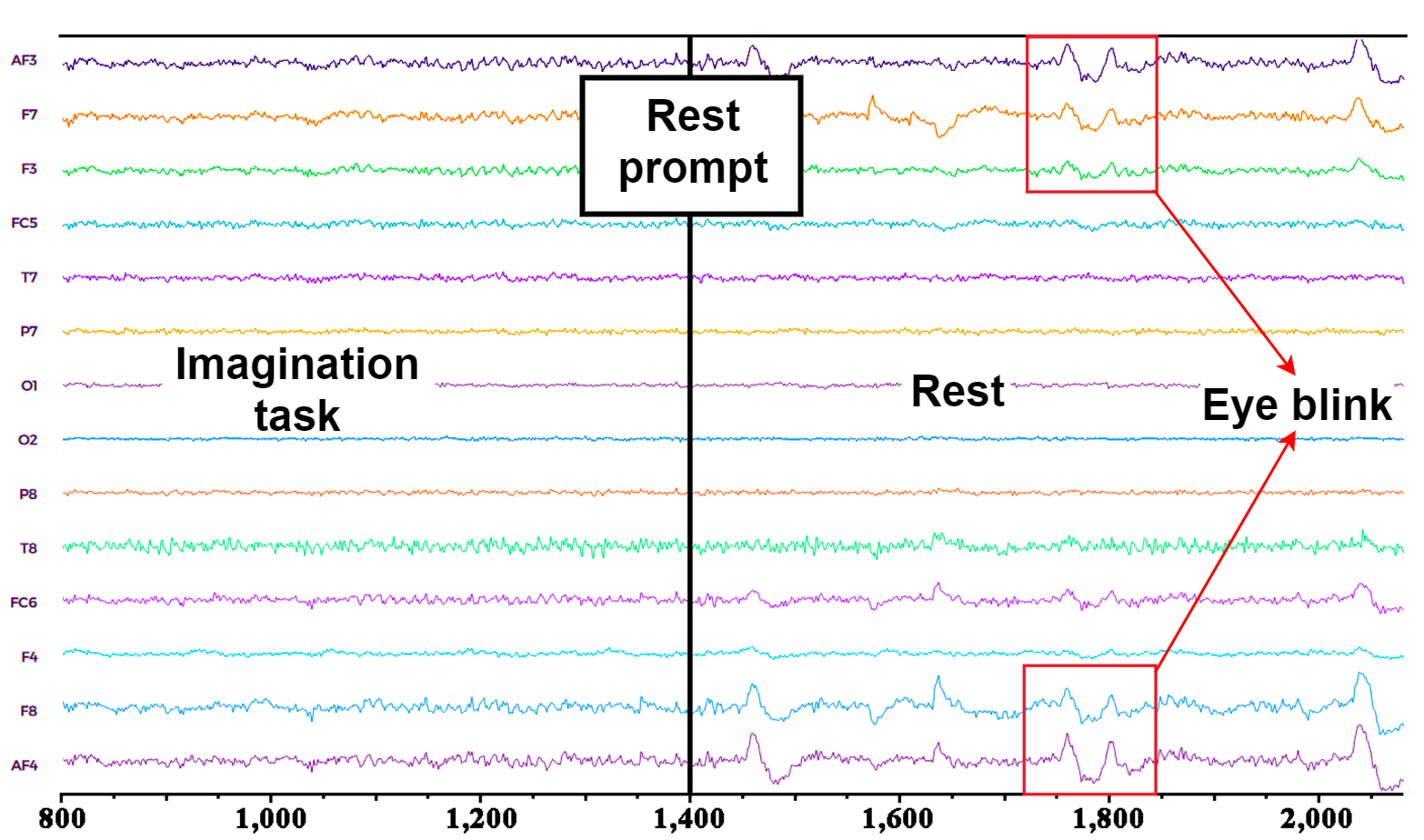}
        \caption{EEG signals}
    \end{subfigure}
 \caption{EEG electrode locations on scalp (10-20 system) and the gathered EEG signals \cite{zhang2018converting}. The electrodes' names are marked by their position: Fp (pre-frontal), F (frontal), T (temporal), P (parietal), O (occipital), and C ( central).
 }
\label{fig:EEG_signal}
\end{figure}

Electroencephalography (EEG) is the most commonly used non-invasive technique for measuring brain activities. EEG monitors the voltage fluctuations generated by an electrical current within human neurons. 
Electrodes placed on the scalp measure the amplitude of EEG signals. EEG signals have a low spatial resolution due to the effect of volume conduction which refers to the complex effects of measuring electrical potentials a distance from the source generators \cite{rutkove2007introduction,burle2015spatial}. 
EEG electrode locations generally follow the international 10-20 system \cite{malmivuo1995bioelectromagnetism}. The specific placement of electrodes is presented in Figure~\ref{fig:EEG_signal} \cite{zhang2018converting}. The EEG signals are collected while the subject is undertaking imagination task. Each line represents the signal stream collected from a single EEG electrode (also called `channel`) over time.

The temporal resolution of EEG signals is much better than the spatial resolution. The ionic current changes rapidly, which offers a temporal resolution higher than $1000$ Hz. The SNR of EEG is generally very poor due to both objective and subjective factors. Objective factors include environmental noises, the obstruction of the skull and other tissues between cortex and scalp, and different stimulations. Subjective factors contain the subject's mental stage, fatigue status, the variance among different subjects, and so on.

EEG recording equipment can be installed in a cap-like headset.
The EEG headset can be mounted on the user's head to gather signals. Compared to other equipment used to measure brain signals, EEG headsets are portable and more accessible for most applications.

The EEG signals collected from any typical EEG hardware have several non-overlapping frequency bands (Delta, Theta, Alpha, Beta, and Gamma) based on the strong intra-band correlation with a distinct behavioral state \cite{zhang2018converting}. Each EEG pattern contains signals associated with particular brain information. Table~\ref{tab:bands} shows EEG frequency patterns and the corresponding characteristics. Here, the degree of awareness denotes the perception of individuals when presented with external stimuli. 

Compared to other signals (e.g., fMRI, fNIRS, MEG), EEG has several important advantages: 1) the hardware has higher portability with much lower price; 2) the temporal resolution is very high (milliseconds level). Among other non-invasive techniques, only MEG has the same level of temporal resolution; 3) EEG is relatively tolerant of subject movement and artifacts, which can be minimized by existing signal processing methods; 4) the subject doesn't need to be exposed to high-intensity ($>$1 Tesla) magnetic fields, therefore, EEG can serve subjects that have metal implants in their body (such as metal-containing pacemakers). 

As the most commonly used signals, there are a large number of sub-classes of EEG signals.
In this section, we present a methodical introduction of EEG sub-class signals. As shown in Figure~\ref{fig:BCI_signals}, we divided EEG signals into spontaneous EEG and evoked potentials. Evoked potentials can be split into event-related potentials and steady-state evoked potentials based on the frequency of the external stimuli \cite{abdulkader2015brain}. Each potential contains visual-, auditory-, and somatosensory- potentials based on the external stimuli types. The dashed quadrilaterals in Figure~\ref{fig:BCI_signals}, such as Intracortical, SEP, SSAEP, SSSEP, and RSAP, are not included in this survey because there are very few existing studies working on them with deep learning algorithms. We list these signals for systematic completeness.

\begin{table*}[t]
\centering
\caption{EEG patterns and corresponding characters. Awareness Degree denotes the degree of being aware of an external world. 
The awareness degree mentioned here is mainly defined in physiology instead of psychology.
}
\label{tab:bands}
\resizebox{\textwidth}{!}{
\begin{tabular}{llllll}
\hline
\textbf{Patterns} & \textbf{Frequency ($Hz$)} & \textbf{Amplitude} & \textbf{Brain State} & \textbf{Awareness Degree} & \textbf{Produced Location} \\ \hline
\textbf{Delta} & 0.5-4 & Higher & Deep sleep pattern & Lower & Frontally and posteriorly \\
\textbf{Theta} & 4-8 & High & Light sleep pattern & Low &  Entorhinal cortex, hippocampus  \\
\textbf{Alpha} & 8-12 & Medium & Closing the eyes, relax state & Intermediate & Posterior regions of head \\
\textbf{Beta} & 12-30 & Low & Active thinking, focus, high alert, anxious & High & Most evident frontally, motor areas \\
\textbf{Gamma} & 30-100 & Lower & During cross-modal sensory processing & Higher & Somatosensory, auditory cortices\\ \hline
\end{tabular}
}
\end{table*}

\subsubsection{Spontaneous EEG} 
\label{sub:eeg}
Typically, when we talk about the term `EEG,' we refer to \textit{spontaneous} EEG which measures the brain signals under a specific state without external stimulation \cite{tortella2014spontaneous,salek2003studying,ikeda2019spontaneous}. In particular, spontaneous EEG includes the EEG signals while the individual is sleeping, undertaking a mental task (e.g., counting), suffering brain disorders, undertaking motor imagery tasks, in a certain emotion, etc. 

The EEG signals recorded while a user stares at a color/shape/image belong to this category. While the subject is gazing at a specific image, the visual stimuli are steady without any change. This scenario differs from the visual stimuli in evoked potential, where the visual stimuli are changing at a specific frequency. Thus, we regard the image stimulation as a particular state and regard it as spontaneous EEG. Spontaneous EEG-based systems are challenging to train, due to the lower SNR and the larger variation across subjects \cite{pfurtscheller2001motor}.

According to the gathering scenarios, the spontaneous EEG contains several subordinates: sleeping, motor imagery, emotional, mental disease and others.


\subsubsection{Evoked Potential (EP)} 
\label{sub:evoked_potential}
Evoked Potentials (EP) or evoked responses refers to the EEG signals which are evoked by an external stimulus instead of spontaneously. An EP is time-locked to the external stimulus while the aforementioned spontaneous EEG is non-time-locked. In contrast to spontaneous EEG, EP generally has higher amplitude and lower frequency. As a result, the EP signals are more robust across subjects. According to the stimulation method, there exist two categories of EP: the Event-Related Potential (ERP) 
and the Steady State Evoked Potential (SSEP).
ERP records the EEG signals in response to an isolated discrete stimulus event (or event change). To achieve this isolation, stimuli in an ERP experiment are typically separated from each other by a long inter-stimulus interval, allowing for the estimation of a stimulus-independent baseline reference \cite{norcia2015steady}. The stimuli frequency of ERP is generally lower than 2 Hz.
In contrast, SSEP is generated in response to a periodic stimulus at a fixed rate. The stimuli frequency of SSEP generally ranges within 3.5-75 Hz.

\noindent\textbf{Event-related potential (ERP).} 
There are three kinds of evoked potentials in extensive research and clinical use: Visual Evoked Potentials (VEP); Auditory Evoked Potentials (AEP); and Somatosensory Evoked Potentials (SEP) \cite{cecotti2017best}. The VEP signals are mainly on the occipital lobe, and the highest signal amplitudes are collected at the Calcarine sulcus. 

1) Visual Evoked Potentials (VEP).
Visual Evoked Potentials are a specific category of ERP which is caused by visual stimulus (e.g., an alternating checkerboard pattern on a computer screen). VEP signals are hidden within the normal spontaneous EEG. To separate VEP signals from the background EEG readings, repetitive stimulation and time-locked signal-averaging techniques are generally employed. 

Rapid Serial Visual Presentation (RSVP) \cite{lees2018review} can be regarded as one kind of VEP. An RSVP diagram is commonly used to examine the temporal characteristics of attention. The subject is required to stare at a screen where a series of items (e.g., images) are presented one-by-one. There is a specific item (called the target) separates from the rest of the other items (called distracters). The subject knows which is the target before the RSVP experiment. For instance, the distracters can be a color change or letters among numbers. RSVP contains a static mode (the items appear on the screen and then disappear without moving) and a moving mode (the items appear on the screen, move to another place, and finally disappear). Nowadays, brain signal research mainly focuses on the static mode RSVP. Usually, the frequency of RSVP is 10Hz which means that each item will stay on the screen for 0.1 seconds.

2) Auditory Evoked Potentials (AEP).
Auditory Evoked Potentials are a specific subclass of ERP in which responses to auditory (sound) stimuli are recorded. AEP is mainly recorded from the scalp but originates at the brainstem or cortex. The most common AEP measured is the auditory brainstem response (ABR) which is generally employed to test the hearing ability of newborns and infants. In the brain signal area, AEP is mainly used in clinical tests for its accuracy and reliability in detecting unilateral loss \cite{chiappa1997evoked}. Similar to RSVP, Rapid Serial Auditory Presentation (RSAP) refers to the experiments with rapid serial presentation of sound stimuli. The task for the subject is to recognize the target audio among the distracters.

3) Somatosensory Evoked Potentials (SEP).\footnote{Generally, Somatosensory Evoked Potentials is abbreviated as SSEP or SEP. In this paper, we choose SEP as the abbreviation in case of the conflict with Steady-State Evoked Potentials (SSEP).}
Somatosensory Evoked Potentials are another commonly used subcategory of ERP which is elicited by electrical stimulation of the peripheral nerves. SEP signals conclude a series of amplitude deflection that can be elicited by virtually any sensory stimuli.


\begin{figure}[]
    \centering
    \begin{subfigure}[t]{0.24\textwidth}
        \centering
        \includegraphics[width=\textwidth]{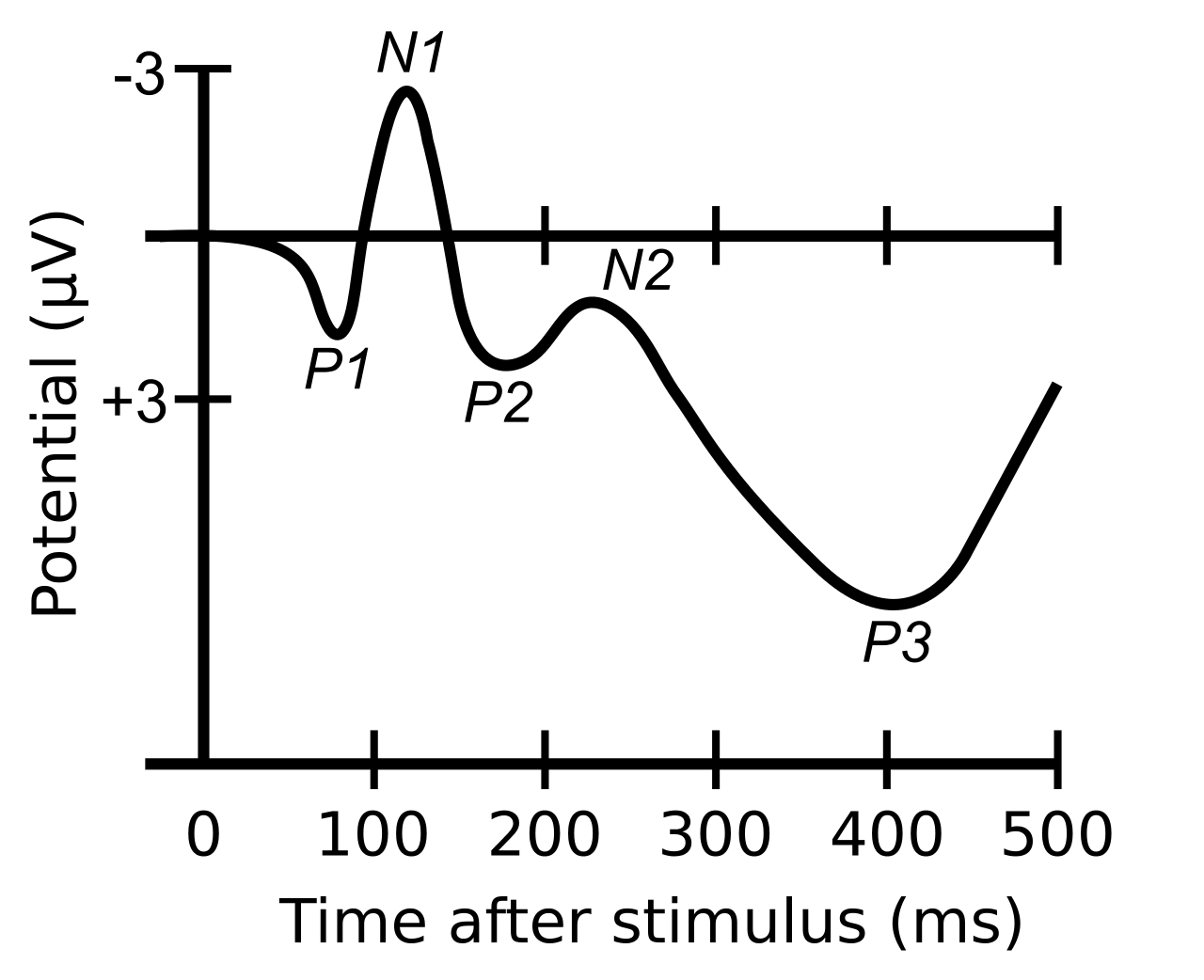}
        \caption{ERP components
         }
        \label{fig:ERP_component}
    \end{subfigure}%
    ~
    \begin{subfigure}[t]{0.24\textwidth}
        \centering
        \includegraphics[width=\textwidth]{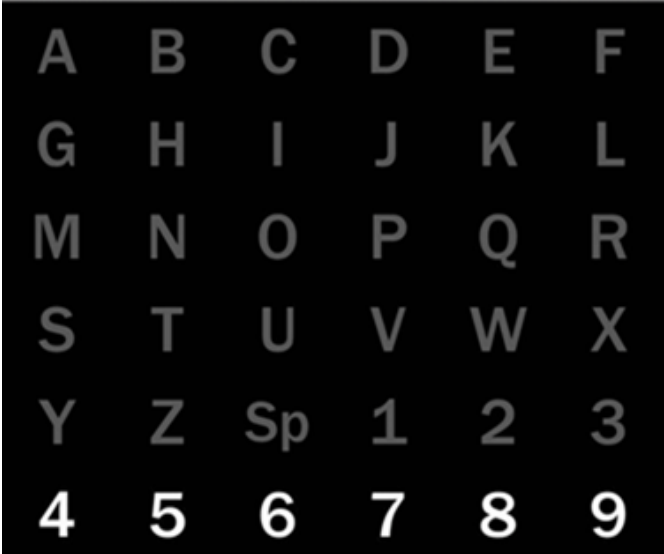}
        \caption{P300 speller}
        \label{fig:P300_speller}
        \end{subfigure}
 \caption{P300 waves and visual-based P300 speller \cite{mayya2017information}.}
\end{figure}

\noindent\textbf{P300.}
P300 (also called P3) is an important component in ERP \cite{guger2009many}. Here we introduce P300 signal separately since it is widely-used in brain signal analysis. 
Figure~\ref{fig:ERP_component} shows the ERP signal fluctuation in the 500 ms after the stimuli onset.
The waveform mainly concludes five components, P1, N1, P2, N2, and P3. The capital character P/N represents positive/negative electrical potentials. The following number refers to the occurrence time of the specific potential. Thus, P300 denotes the positive potential of ERP waveform at approximately 300 ms after the presented stimuli. Compared to other components, P300 has the highest amplitude and is easiest to detect. Thus, a large number of brain signal studies focus on P300 analysis. P300 is more of an informative feature instead of a type of brain signal (e.g., VEP). Therefore, we do no list P300 in Figure~\ref{fig:BCI_signals}. P300 can be analyzed in most of ERP signals such as VEP, AEP, SEP.

In practice, P300 can be elicited by rare, task-relevant events in an `oddball' paradigm (e.g., P300 speaker). In the oddball paradigm, the subject receives a series of stimuli where low-probability target items are mixed with high-probability non-target items. Visual and auditory stimuli are the most commonly used in the oddball paradigm. 
Figure~\ref{fig:P300_speller} shows an example of visual-based P300 speller which enables the subject the spell letters/numbers directly through brain signals \cite{mayya2017information}. The 26 letters of the alphabet and the Arabic numbers are displayed on a computer screen which serves as the keyboard. The subject focuses attention successively on the characters they wish to spell. The computer detects the chosen character online in real time. This detection is achieved by repeatedly flashing rows and columns of the matrix. When the elements containing the selected characters are flashing, a P300 fluctuation is elicited. In the $6 \times 6$ matrix screen, the rows and columns flash in mixed random order.  The flash duration and interval among adjacent flashes are generally set as 100 ms \cite{belitski2011p300}. The columns and rows flash separately. First, the columns flash six times with each column flashing one time. Second, the rows will flash for six times. After that, this paradigm repeats for several times (e.g., $N$ times). The P300 signals of the total $12N$ flash will be analyzed to output a single outcome (i.e., one letter/number). 

\noindent\textbf{Steady State Evoked Potentials (SSEP).}
Steady State Evoked Potential is another subcategory of evoked potentials, which are periodic cortical responses evoked by certain repetitive stimuli with a constant frequency. It has been demonstrated that the brain oscillations generally maintain a steady level over time while the potentials are evoked by steady state stimuli (e.g., a flickering light with fixed frequency). Technically, SSEP is defined as a form of response to repetitive sensory stimulation in which the constituent frequency components of the response remain constant over time in both amplitude and phase \cite{regan1977steady}. Depending on the type of stimuli, SSEP divides into three subcategories: Steady-State Visually Evoked Potentials (SSVEP), Steady-State Auditory Evoked Potentials (SSAEP), and Steady-State Somatosensory Evoked Potentials (SSSEP). In the brain signal area, most studies are focused on visual evoked steady potentials, and only rarely do papers focus on auditory and somatosensory stimuli. Therefore, in this survey, we mainly introduce SSVEP rather than SSAEP and SSSEP.

\noindent\textbf{Commonly Used Visual-Related Potentials.}
Visual evoked potentials are the most common used potentials. Therefore, it is essential to distinguish the three different visual evoked potential paradigms: VEP, RSVP, SSVEP. Here, we theoretically introduce the characteristics of each paradigm and then give three demonstration videos to provide a better understanding. First, the frequencies are different: the frequency of VEP is less than 2Hz while the frequency of RSVP is around 10Hz, and the frequency of SSVEP ranges from 3.5 to 75Hz.
Second, they have various presentation protocols. In the VEP paradigm, different visual patterns will be presented on the screen to check the user's brain signals changes. For instance, in this video\footnote{https://www.youtube.com/watch?v=iUW\_l5YAEEM}, the image pattern is full of the screen and changes dramatically.
In RSVP diagram, several items will be presented on a screen one-by-one. All the items are shown in the same place and share the same frequency. For example, the video\footnote{https://www.youtube.com/watch?v=5yddeRrd0hA\&t=36s} shows an RSVP scenario which is called speed reading.
In SSVEP paradigm, several items will be presented on a screen at the same time while the items are shown at \textit{variant positions} with different frequencies. 
For example, in this demonstration video\footnote{https://www.youtube.com/watch?v=t96rl1SFHlI}, there are four circles distributed on the up, down, left, and right sides of a screen and the frequency of each item differs from each other.

\subsection{Functional Near-infrared Spectroscopy (fNIRS)} 
\label{sub:fNIRS}


Functional near-infrared spectroscopy (fNIRS) is a non-invasive functional neuro-imaging technology using near-infrared (NIR) light \cite{naseer2016analysis}. In specific, fNIRS employs NIR light to measure the aggregation degree of oxygenated hemoglobin (Hb) and deoxygenated-hemoglobin (deoxy-Hb) because Hb and deoxy-Hb have higher absorbence of light than other head components such as the skull and scalp. fNIRS relies on blood-oxygen-level-dependent (BOLD) response or hemodynamic response to form a functional neuro-image. The BOLD response can detect the oxygenated or deoxygenated blood level in the brain blood. The relative levels reflect the blood flow and neural activation, where increased blood flow implies a higher metabolic demand caused by active neurons. For example, when the user is concentrating on a mental task, the prefrontal cortex neurons will be activated, and the BOLD response in the prefrontal cortex area will be stronger \cite{hennrich2015investigating}.

Single or multiple emitter-detector pairs measure the Hb and deoxy-Hb: the emitter transmits NIR light through the blood vessels to the detector. Most existing studies use fNIRS technologies to measure the status of prefrontal and motor cortex. The former response to mental tasks and music/image imagery while the latter is a response to motor-related tasks (e.g., motor imagery). The monitored Hb and deoxy-Hb change slowly since the blood speed varies in a relatively slow ratio compared to electrical signals. Temporal resolution refers to the smallest time of neural activity reliably separated by the signal. The fNIRS has lower temporal resolution compared with electrical or magnetic signals. The spatial resolution depends on the number of emitter-detector pairs. In current studies, three emitters and eight detectors would suffice for adequately acquiring the prefrontal cortex signals; and six emitters and six detectors would suffice for covering the motor cortex area \cite{naseer2015fnirs}. fNIRS has a drawback in that it cannot be used to measure cortical activity occurring deeper than 4cm in the brain, due to the limitations in light emitter power and spatial resolution.

\subsection{Functional Magnetic Resonance Imaging (fMRI)} 
\label{sub:fMRI}


Functional magnetic resonance imaging (fMRI) monitors brain activities by detecting changes associated with blood flow in brain areas \cite{wen2018deeplearning}. Similar to fNIRS, fMRI relies on the BOLD response. The main differences between fNIRS and fMRI are as follows \cite{liu2018applications}. First, as the name implies, fMRI measures BOLD response through magnetic instead of optical methods. Hemoglobin differs in how it responds to magnetic fields, depending on whether it has a bound oxygen molecule. The magnetic fields are more sensitive to and are more easily distorted by deoxy-Hb than Hb molecules. 
Second, the magnetic fields have higher penetration than NIR light, which gives fMRI greater ability to capture information from deep parts of the brain than fNIRS. Third, fMRI has a higher spatial resolution than fNIRS since the latter's spatial resolution is limited by the emitter-detector pairs. However, the temporal resolutions of fMRI and fNIRS are at an equal level because they both constrained by the blood flow speed. 

fMRI has several flaws compared to fNIRS: 1) fMRI requires an expensive scanner to generate magnetic fields; 2) the scanner is heavy and has poor portability.
In order to measure the signal of interest, CNR (Contrast-to-Noise Ratio) has been investigated to measure the image quality of fMRI because researchers are more interested in the contrast between images rather than the raw images.
So for fMRI data, using the CNR of the time series instead of (t)SNR is more preferred because CNR compares a measure of the activation fluctuations to the noise \cite{welvaert2013definition}.

\subsection{Magnetoencephalography (MEG)} 
\label{sub:MEG}
Magnetoencephalography (MEG) is a functional neuroimaging technique for mapping brain activity by recording magnetic fields produced by electrical currents occurring naturally in the brain, using very sensitive magnetometers \cite{cichy2017dynamics}. The ionic currents of active neurons will create weak magnetic fields. The generated magnetic fields can be measured by magnetometers like SQUIDs (superconducting quantum interference devices). However, producing a detectable magnetic field requires massive (e.g., 50,000) active neurons with similar orientation. The source of the magnetic field measured by MEG is the pyramidal cells which are perpendicular to the cortex surface. 

MEG has a relatively low spatial resolution since the signal quality highly depends on the measurement factors (e.g., brain area, neuron orientations, neuron depth). However, MEG can provide very high temporal resolution ($\geq$1000Hz) since MEG directly monitors the brain activity from the neuron level, which is in the same level of intracortical signals. 
MEG equipment is expensive and not portable which limits its real-world deployment. 


\newpage
\section{Basic Deep Learning in Brain Signal Analysis}
\label{sec:deep_learning_models_appendix}
In this part, we will give relative detail introduction of various deep learning models for the reason that a part of the potential readers who are from non-computer area (e.g., biomedical) are not familiar to deep learning. 

For simplification, we first define an operation $\mathcal{T}(\cdot)$ as 
\begin{equation} \mathcal{T}(\bm{x}) = \bm{w}*\bm{x} + \bm{b} \end{equation}
\begin{equation} \mathcal{T}(\bm{x}, \bm{x'}) = \bm{w}*\bm{x} + \bm{b} + \bm{w'}*\bm{x'} + \bm{b'} \end{equation}
where $\bm{x}$ and $\bm{x'}$ denote two variables while $\bm{w}$, $\bm{w'}$, $\bm{b}$, and $\bm{b'}$ denote the corresponding weights and basis.

\subsection{Discriminative Deep Learning Models} 
\label{sub:discriminative_deep_learning_models}
Since the main task of brain signal analysis is brain signal recognition, the discriminative deep learning models are the most popular and powerful algorithms. Suppose we have a dataset of brain signal samples $\{\mathbb{X}, \mathbb{Y}\}$ where $\mathbb{X}$ denotes the set of brain signal observations and $\mathbb{Y}$ denotes the set of sample ground truth (i.e., labels). Suppose an specific sample-label pair $\{\bm{x} \in \mathbb{R}^N, \bm{y} \in \mathbb{R}^M \}$ where $N$ and $M$ denote the dimension of observations and the number of sample categories, respectively. The aim of discriminative deep learning models is to learn a function with the mapping: $\bm{x} \rightarrow \bm{y}$. In short, the discriminative models receive the input data and output the corresponding category or label. All the discriminative models introduced in this section are supervised learning techniques which require the information of both the observations and the ground truth.

\subsubsection{Multi-Layer Perceptron (MLP)} 
\label{sub:MLP}
\begin{figure}[]
    \centering
    \begin{subfigure}[t]{0.25\textwidth}
        \centering
        \includegraphics[width=\textwidth]{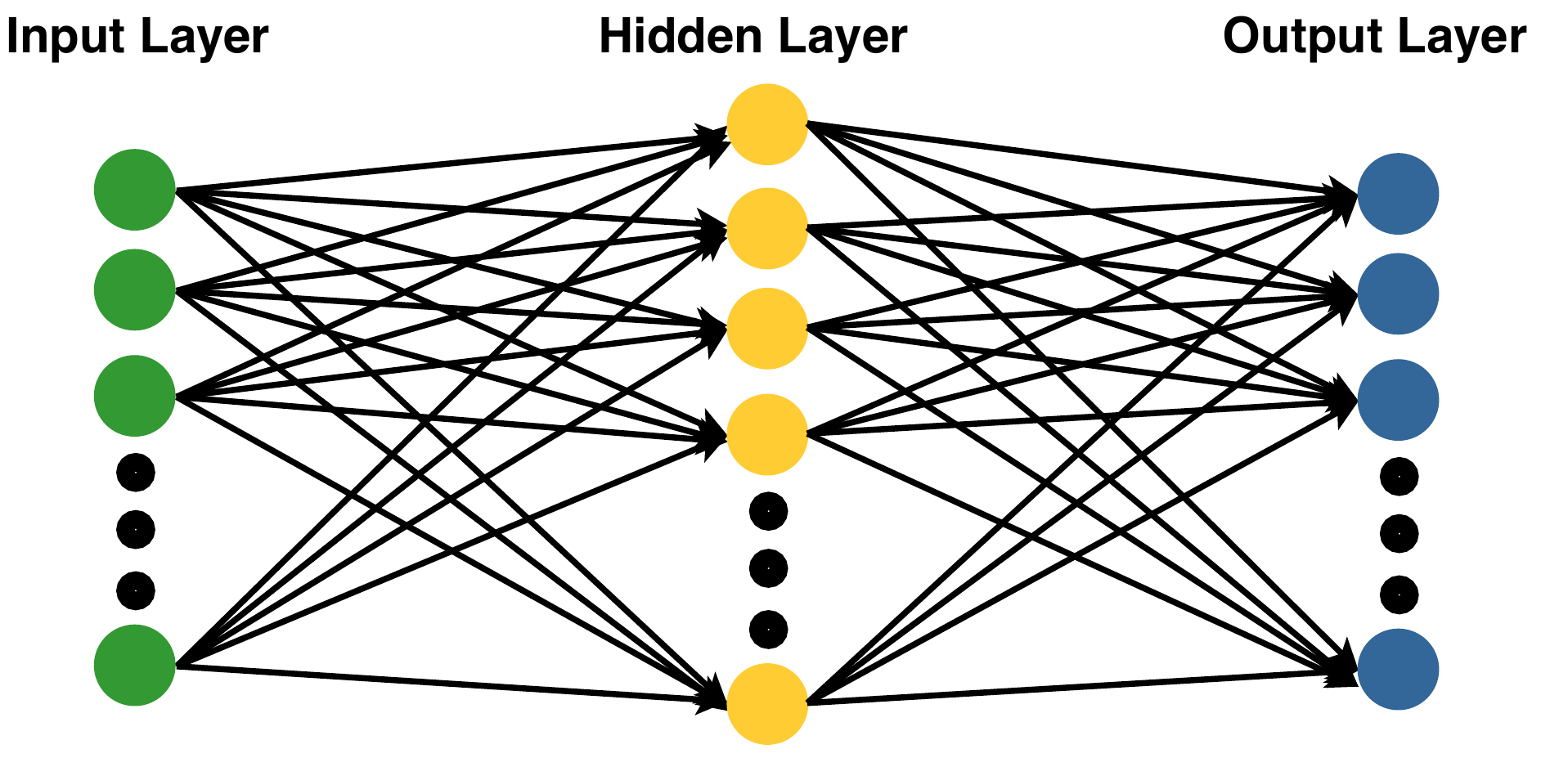}
        \caption{Fully-connected neural network}
        \label{fig:basic_neural_network}
    \end{subfigure}%
    ~
    \begin{subfigure}[t]{0.25\textwidth}
        \centering
        \includegraphics[width=\textwidth]{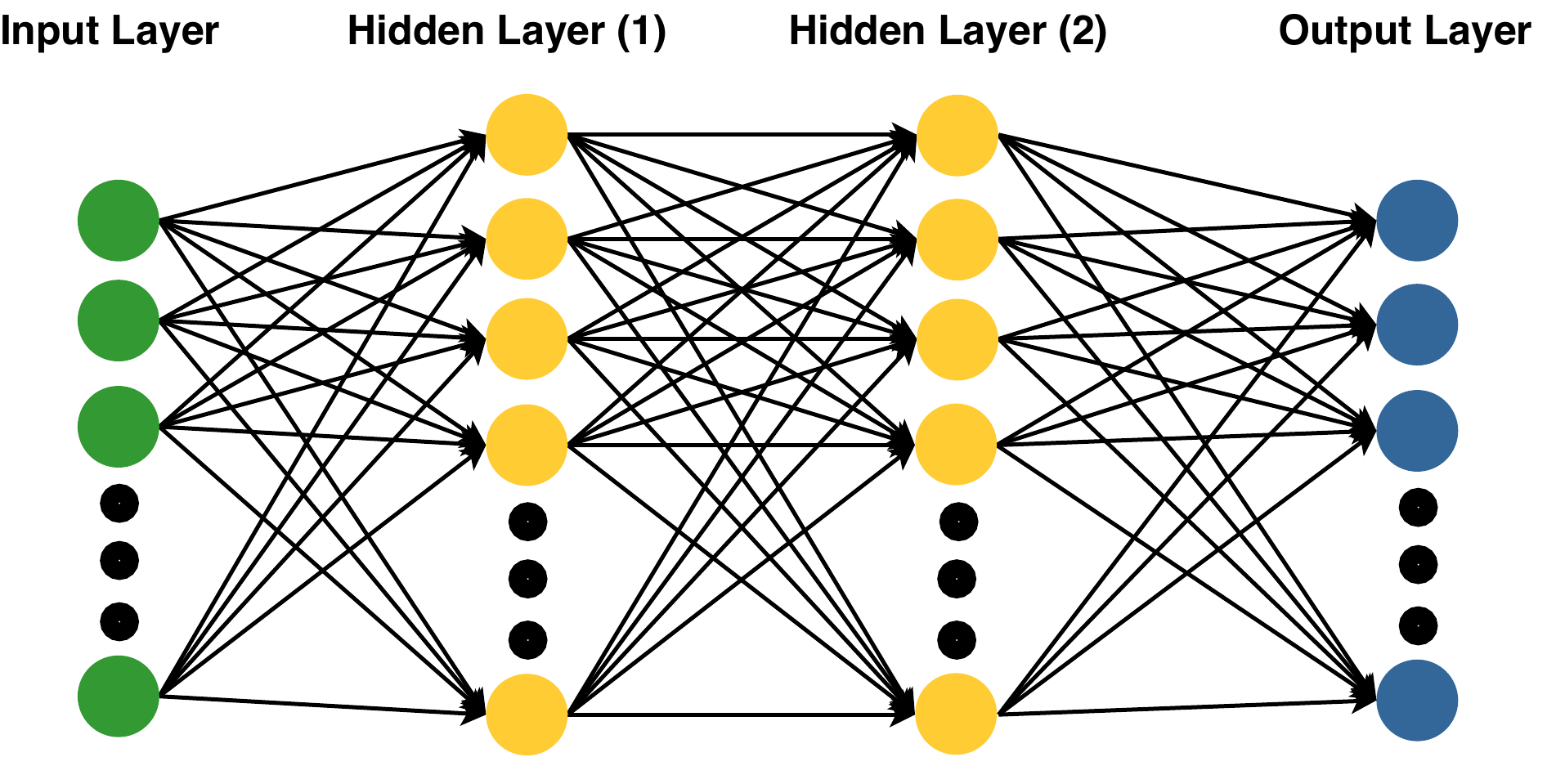}
        \caption{MLP}
        \label{fig:MLP}
        \end{subfigure}
 \caption{Illustration of standard neural network and multilayer perceptron. (a) The basic structure of the fully-connected neural network. The input layer receives the raw data or extracted features of brain signals while the output layer shows the classification results. The term `fully-connected' denotes each node in a specific layer is connected with all the nodes in the previous and next layer. (b) MLP could have multiple hidden layers, the more, the deeper. This is an example of MLP with two hidden layers, which is the simplest MLP model.}
\end{figure}

The most basic neural network is fully-connected neural networks (Figure~\ref{fig:basic_neural_network}) which only contains one hidden layer. The input layer receives the raw data or extracted features of brain signals while the output layer shows the classification results. The term `fully-connected' denotes each node in a specific layer is connected with all the nodes in the previous and next layer. This network is too `shallow` and generally not regarded as `deep` neural networks.

Multilayer Perceptron is the simplest and the most basic deep learning model. The key difference between MLP and the fully-connected neural network is that MLP has more than one hidden layers. All the nodes are fully-connected with the nodes of the adjacent layers but without connection with the other nodes of the same layer. MLP includes multiple hidden layers. As shown in Figure~\ref{fig:MLP}, we take a structure with two hidden layers as an example to describe the data flow in MLP.

The input layer receives the observation $\bm{x}$ and feeds forward to the first hidden layer,
\begin{equation} \bm{x^{h1}} = \sigma(\mathcal{T}(\bm{x})) \end{equation}
where $\bm{x^{h1}}$ denotes the data flow in the first hidden layer and $\sigma$ represents the non-linear activation function. There are several commonly used activation functions such as sigmoid/Logistic, Tanh, ReLU, we choose sigmoid activation function as an example in this section. Then, the data flow to the second hidden layer and the output layer,
\begin{equation} \bm{x^{h2}} = \sigma(\mathcal{T}(\bm{x^{h1}})) \end{equation}
\begin{equation} \bm{y'} = \sigma(\mathcal{T}(\bm{x^{h2}})) \end{equation}
where $\bm{y'}$ denotes the predict results in one-hot format.
The error (i.e., loss) could be calculated based on the distance between $\bm{y'}$ and the ground truth $\bm{y}$. For instance, the Euclidean-distance based error can be calculated by 

\begin{equation}
\label{eq:1}
error =  \left \| \bm{y'}- \bm{y} \right \|_2    
\end{equation}
where $\left \| \cdot \right \|_2$ denotes the Euclidean norm. 
 Afterward, the error will be back-propagated and optimized by a suitable optimizer. The optimizer will adjust all the weights and basis in the model until the error converges. The most widely used loss functions includes cross-entropy, negative log likelihood, mean square estimation, etc. The most widely used optimizers include Adaptive moment estimation (Adam), Stochastic Gradient Descent (SGD), Adagrad (Adaptive subgradient method), etc. 

Several terms may be easily confused with each other: Artificial Neural Network (ANN), Deep Neural Network (DNN), and MLP. These terms have no strict difference and often mixed in literature and commonly used as synonyms. Generally, ANN and DNN can be used to describe deep learning models overall, including not only fully-connected networks but also other networks (e.g., recurrent, convolutional networks), but MLP can only refer to fully-connected network. 
Additionally, ANN contains all the models of neural networks, can be either shallow (one hidden layer) or deep (multiple hidden layers) while DNN doesn't cover shallow neural network \cite{schmidhuber2015deep,deng2014tutorial}.

\subsubsection{Recurrent Neural Networks (RNN)} 
\label{sub:RNN}
\begin{figure}[]
    \centering
    \begin{subfigure}[t]{0.25\textwidth}
        \centering
        \includegraphics[width=\textwidth]{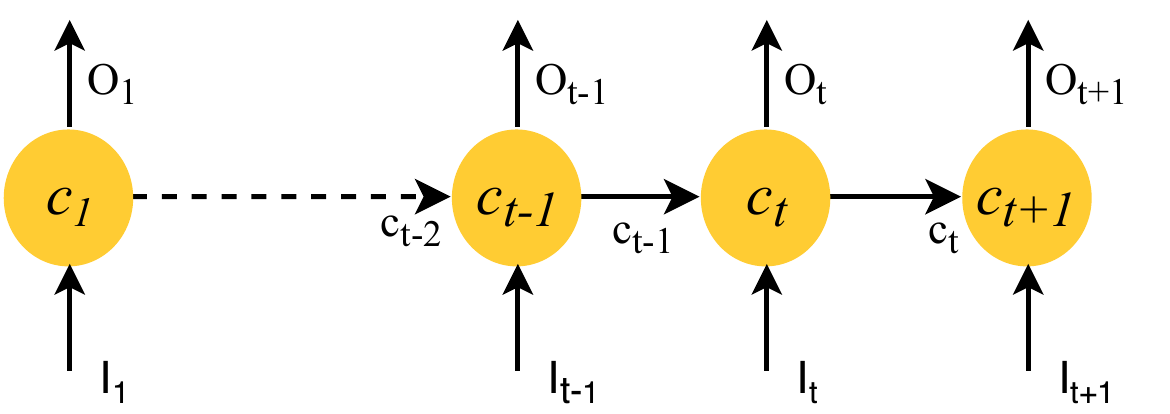}
        \caption{RNN}
        \label{fig:RNN}
    \end{subfigure}%
    ~
    \begin{subfigure}[t]{0.25\textwidth}
        \centering
        \includegraphics[width=\textwidth]{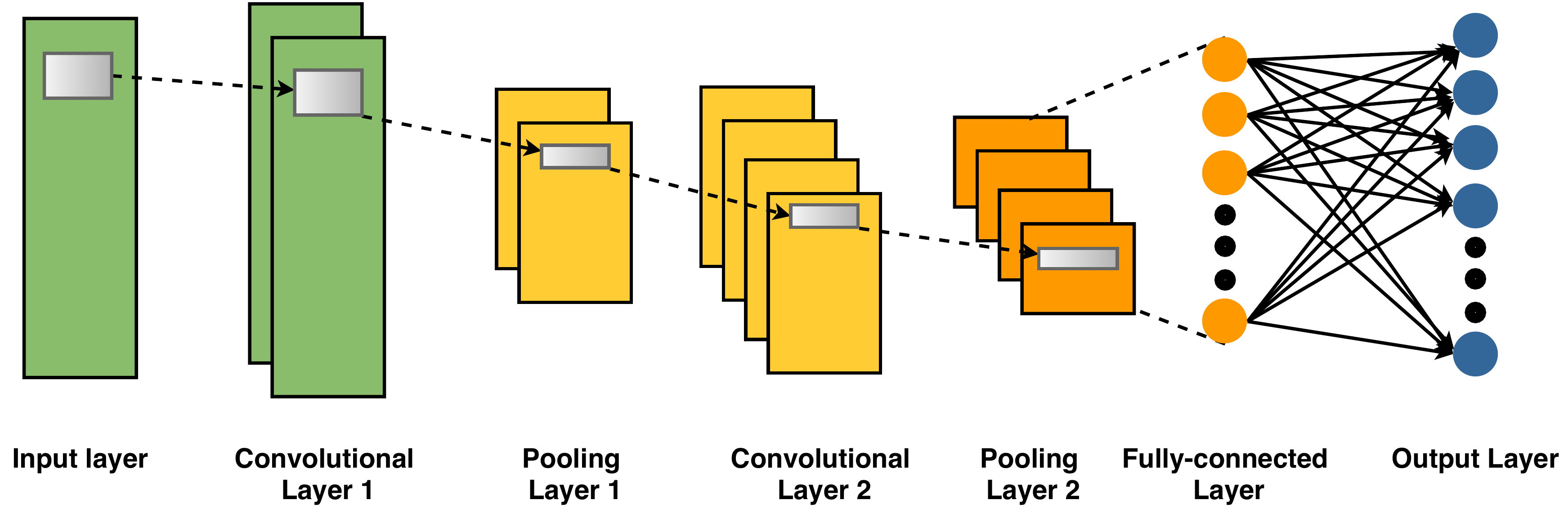}
        \caption{CNN}
        \label{fig:CNN}
        \end{subfigure}
 \caption{Illustration of RNN and CNN models. (a) The recurrent procedure of the RNN model. This procedure describes the recurrent procedure of a specific node in time range $[1, t+1]$. The node at time $t$ receives two inputs variables ($I_t$ denotes the input at time $t$ and $c_{t-1}$ denotes the hidden state at time $t-1$) and exports two variables (the output $O_t$ and the hidden state $c_t$ at time $t$). (b) The paradigm of CNN model which includes two convolutional layers, two pooling layers, and one fully-connected layer.}
\label{fig:}
\end{figure}

Recurrent Neural Network is a specific subclass of discriminative deep learning model which are designed to capture temporal dependencies among input data \cite{mikolov2010recurrent}. 
Figure~\ref{fig:RNN} describes the activity of a specific RNN node in the time domain. At each time ranges from $[1, t+1]$, the node receives an input $I$ (the subscript represents the specific time) and a hidden state $c$ from the previous time (except the first time). 
For instance, at time $t$ it receives not only the input $I_t$ but also the hidden state of the previous node $c_{t-1}$. The hidden state can be regarded as the `memory' of the nodes which can help the RNN `remember' the historical input. 

Next, we will report two typical RNN architectures which have attracted much attention and achieved great success: long short-term memory and gated recurrent units. They both follow the basic principles of RNN, and we will pay our attention to the complicated internal structures in each node. Since the structure is much more complicated than general neural nodes, we call it a `cell.' Cells in RNN are equivalent to nodes in MLP.

\begin{figure}[]
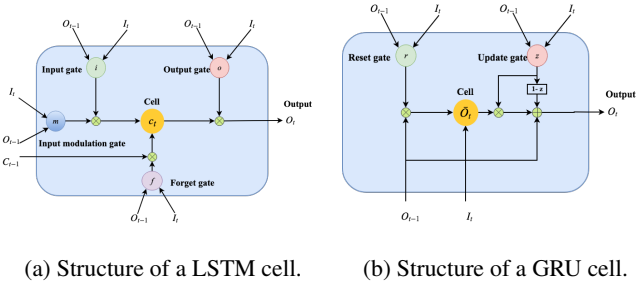

    \centering
    \begin{subfigure}[t]{0.25\textwidth}
        \centering
        \includegraphics[width=\textwidth]{figure/DL_models/LSTM.png}
        \caption{Structure of a LSTM cell.}
        \label{fig:LSTM}
    \end{subfigure}%
    ~
    \begin{subfigure}[t]{0.25\textwidth}
        \centering
        \includegraphics[width=0.95\textwidth]{figure/DL_models/GRU.png}
        \caption{Structure of a GRU cell.}
        \label{fig:GRU}
        \end{subfigure}
 \caption{Illustration of detailed LSTM and GRU cell structures. (a) LSTM cell receives three inputs ($I_t$ denotes the input at time $t$, $O_{t-1}$ denotes the output of previous time, and $c_{t-1}$ denotes the hidden state of the previous time) and exports two outputs (the output of this time $O_t$ and the hidden state of this time $c_t$). LSTM cell contains four gates in order to control the data flow, which are the input gate, output gate, forget gate, and input modulation gate. (b) GRU cell receives two inputs (the input of this time $I_t$ and the output of the previous time $O_{t-1}$) and exports its output $O_t$. GRU cell only contains two gates which are the reset gate and the update gate. Unlike the hidden state $c_t$ in LSTM cell, there is no transmittable hidden state in GRU cell except one intermediate variable $\bar{O_t}$.}
\end{figure}

\noindent\textbf{Long Short-Term Memory (LSTM).}
Figure~\ref{fig:LSTM} shows the structure of a single LSTM cell at time $t$ \cite{hochreiter1997long}. The LSTM cell has three inputs ($I_t$, $O_{t-1}$, and $c_{t-1}$) and two outputs ($c_t$ and $O_t$). The operation is as follows:
\begin{equation} I_t, O_{t-1}, c_{t-1} \rightarrow c_t, O_t \end{equation}
$I_t$ denotes the input value at time $t$, $O_{t-1}$ denotes the output at the previous time (i.e., time $t-1$), and $c_{t-1}$ denotes the hidden state at the previous time. $c_t$ and $O_t$ separately denote the hidden state and the output at time $t$. Therefore, we can observe that the output $O_t$ at time $t$ not only related to the input $I_t$ but also related to the information at the previous time. In this way, LSTM is empowered to remember the important information in the time domain.
Moreover, the essential idea of LSTM is to control the memory of specific information. For this aim, LSTM cell adopts four gates: the input gate, forget gate, output gate, and input modulation gate. Each gate is a weight to control how much information can flow through this gate. For example, if the weight of the forget gate is zero, the LSTM cell would remember all the information passed from the previous time $t-1$; if the weight is one, the LSTM cell would remember nothing. The corresponding activation function determines the weight.
The detailed data flow as follows:
\begin{equation} f = \sigma(\mathcal{T}(I_t, O_{t-1})) \end{equation}
\begin{equation} i = \sigma(\mathcal{T}(I_t, O_{t-1})) \end{equation}
\begin{equation} o = \sigma(\mathcal{T}(I_t, O_{t-1})) \end{equation}
\begin{equation} m = tanh(\mathcal{T}(I_t, O_{t-1})) \end{equation}
\begin{equation} c_t = f*c_{t-1} + i*m \end{equation}
\begin{equation} h_t = o*tanh(c_t) \end{equation}
where $i$, $f$, $o$ and $m$ represent the input gate, forget gate, output gate and input modulation gate, respectively.

\noindent\textbf{Gated Recurrent Units (GRU).}
Another widely used RNN architecture is GRU \cite{cho2014learning}. Similar to LSTM, GRU attempts to exploit the information from the past. GRU does not require hidden states, however, it receives temporal information only from the output of time $t-1$. Thus, as shown in Figure~\ref{fig:GRU}, GRU has two inputs ($I_t$ and $O_{t-1}$) and one output ($O_t$). The mapping can be described as:
\begin{equation} I_t, O_{t-1} \rightarrow O_t \end{equation}
GRU contains two gates: reset gate $r$ and update gate $z$. The former decides how to combine the input with previous memory. The latter decides how much of previous memory to keep around, which is similar to the forget gate of LSTM. The data flow as follows:
\begin{equation} z = \sigma(\mathcal{T}(I_t, O_{t-1})) \end{equation}
\begin{equation} r = \sigma(\mathcal{T}(I_t, O_{t-1})) \end{equation}
\begin{equation} \bar{O_t} = tanh(\mathcal{T}(I_t, r*O_{t-1})) \end{equation}
\begin{equation} O_t = (1-z)*O_{t-1} + z*\bar{O_t} \end{equation}
It can be observed that there's an intermediate variable $\bar{O_t}$ which is similar to the hidden state of LSTM. However, $\bar{O_t}$ only works on this time point and unable to pass to the next time point. 

We here give a brief comparison between LSTM and GRU since they are very similar. First, LSTM and GRU have comparable performance as studied by literature. For any specific task, it is recommended to try both of them to determine which provides better performance. Second, GRU is lightweight since it only has two gates and without the hidden state. Therefore, GRU is faster to train and requires few data for generalization. Third, in contrast, LSTM generally works better if the training dataset is big enough. The reason is that LSTM has better non-linearity than GRU since LSTM has two more control gates (input modulation gate and forget gate). As a result, LSTM, compared with GRU, is more powerful to discover the latent distinct information from large-level training dataset.

\subsubsection{Convolutional Neural Networks (CNN)} 
\label{sub:CNN}
Convolutional Neural Networks is one of the most popular deep learning models specialized in spatial information exploration \cite{sainath2013deep}. This section will briefly introduce the working mechanism of CNN. 
CNN is widely used to discover the latent spatial information in applications such as image recognition, ubiquitous, and object searching due to their salient features such as regularized structure, good spatial locality, and translation invariance. In the area of brain signal, specifically, CNN is supposed to capture the distinctive dependencies among the patterns associated with different brain signals. 

We present a standard CNN architecture as shown in Figure~\ref{fig:CNN}. The CNN contains one input layer, two convolutional layers with each followed by a pooling layer, one fully-connected layer, and one output layer. The square patch in each layer shows the processing progress of a specific batch of input values. The key to the CNN is to reduce the input data into a form which is easier to recognize, with as little information loss as possible. CNN has three stacked layers: the convolutional Layer, pooling Layer, and fully-connected Layer.

The convolutional layer is the core block of CNN, which contains a set of filters to convolve the input data followed by a nonlinear transformation to extract the geographical features. In the deep learning implementation, there are several key hyper-parameters should be set in the convolutional layer, like the number of filters, the size of each filter, etc. 
The pooling layer generally follows the convolutional layer. The pooling layer aims to reduce the spatial size of the features progressively. In this way, it can help to decrease the number of parameters (e.g., weights and basis) and the computing burden. There are three kinds of pooling operation: max, min, average. Take max pooling for example. The pooling operation outputs the maximum value of the pooling area as a result. The hyper-parameters in the pooling layer includes the pooling operation, the size of the pooling area, the strides, etc. In the fully-connected layer, as in the basic neural network, the nodes have full connections to all activations in the previous layer.

The CNN is the most popular deep learning model in brain signal research, which can be used to exploit the latent spatial dependencies among the input brain signals like fMRI image, spontaneous EEG, and so on. More details will be reported in Section~\ref{sec:state_of_the_art_dl_techniques_for_bci}.

\subsection{Representative Deep Learning Models} 
\label{sub:representative_deep_learning_models}
The term of representative deep learning refers to use deep neural network for representation learning. It aims to learn representations of input data that makes it easier to perform a downstream task (e.g., classification, generation, and clustering) \cite{bengio2013representation}.

The essential blocks of representative deep learning models are autoencoders, and restricted Boltzmann machines\footnote{AE and RBM are generally regarded as kind of deep learning although they only have three and two layers, respectively.}. Deep Belief Networks are composed of AE or RBM. The representative models including AE, RBM\footnote{We regard AE, and RBMas representative methods as most researches in brain researches adopt them for feature representation.}, and DBN, are unsupervised learning methods. Thus, they can learn the representative features from only the input observations $\bm{x}$ without the ground truth $\bm{y}$. In short, representative models receive the input data and output a dense representation of the data.
There are various definitions in different studies for several models (such as DBN, Deep RBM, and Deep AE), in this survey, we choose the most understandable definitions and will present them in detail in this section.

\begin{figure}[]
    \centering
    \begin{subfigure}[t]{0.24\textwidth}
        \centering
        \includegraphics[width=\textwidth]{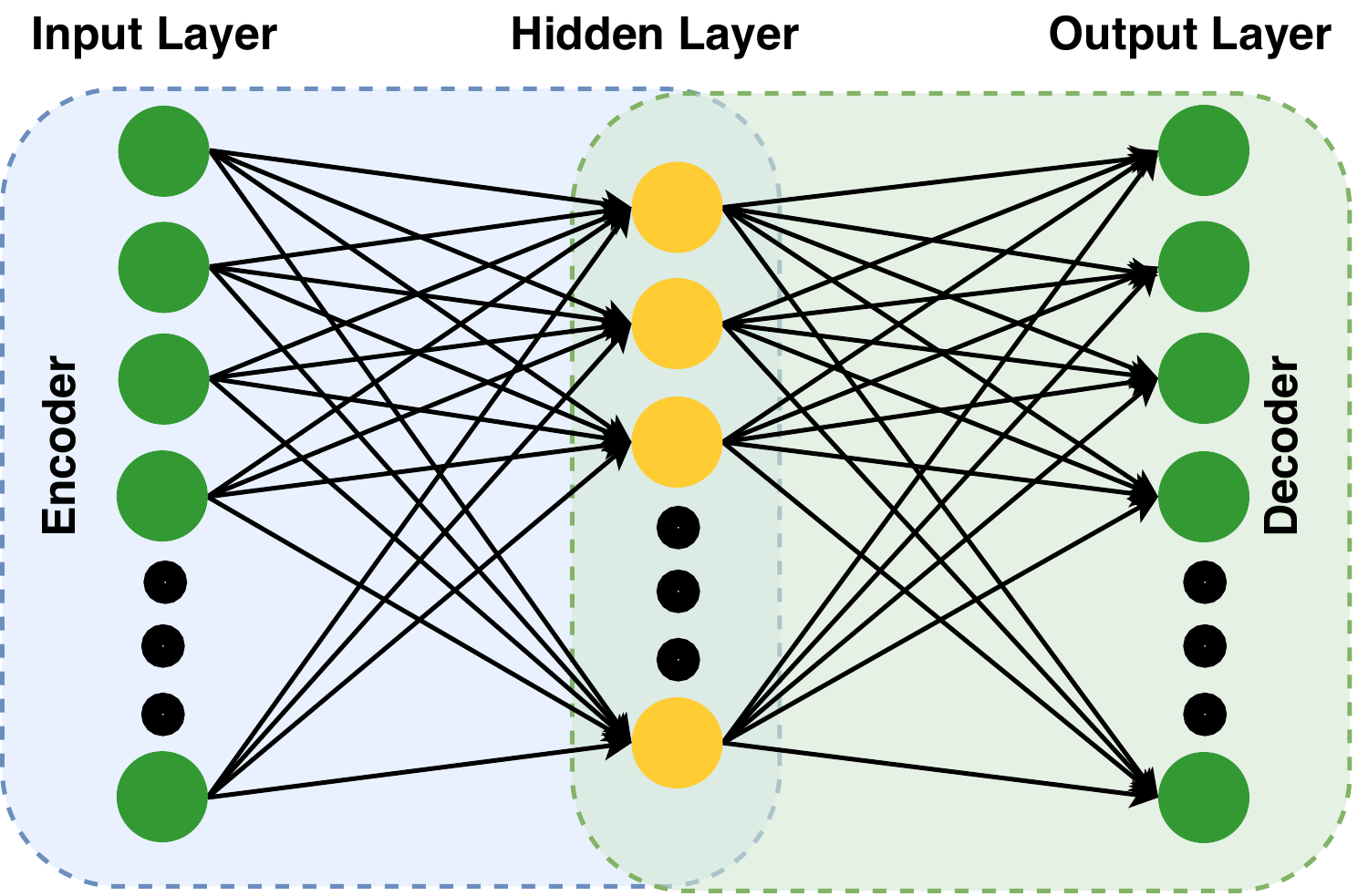}
        \caption{Autoencoder}
        \label{fig:AE}
    \end{subfigure}%
    \hfill
    \begin{subfigure}[t]{0.24\textwidth}
        \centering
        \includegraphics[width=0.6\textwidth]{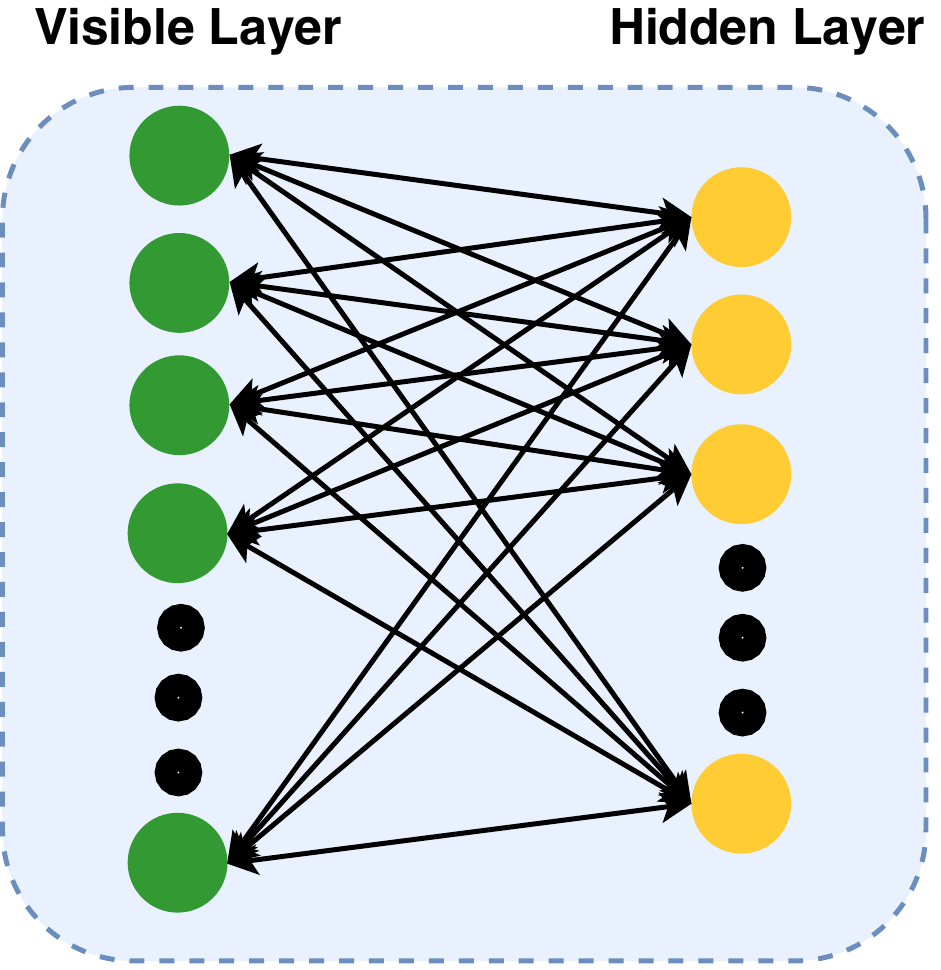}
        \caption{RBM}
        \label{fig:RBM}
        \end{subfigure}
\vfill
    \begin{subfigure}[t]{0.24\textwidth}
        \centering
        \includegraphics[width=\textwidth]{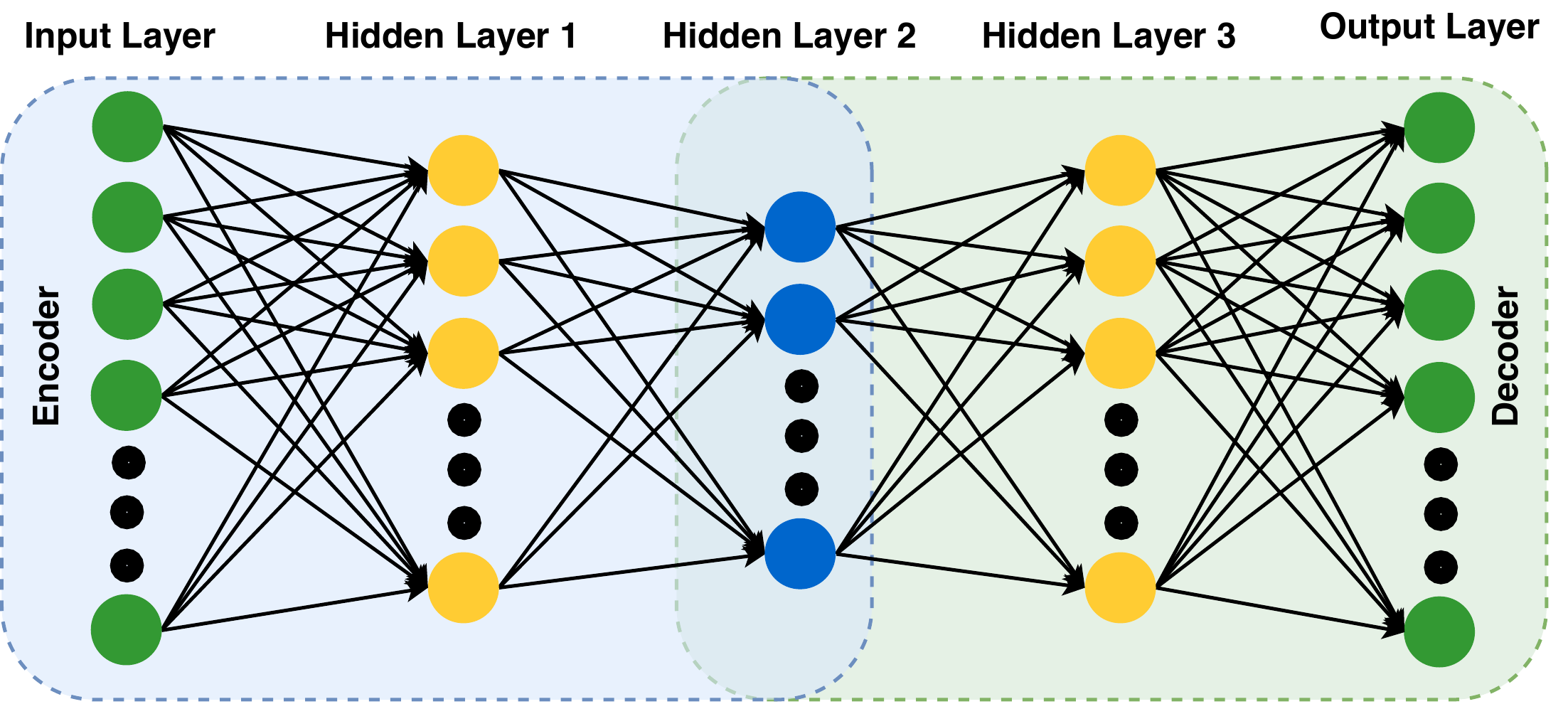}
        \caption{Deep AE}
        \label{fig:S_AE}
    \end{subfigure}%
    \begin{subfigure}[t]{0.24\textwidth}
        \centering
        \includegraphics[width=0.6\textwidth]{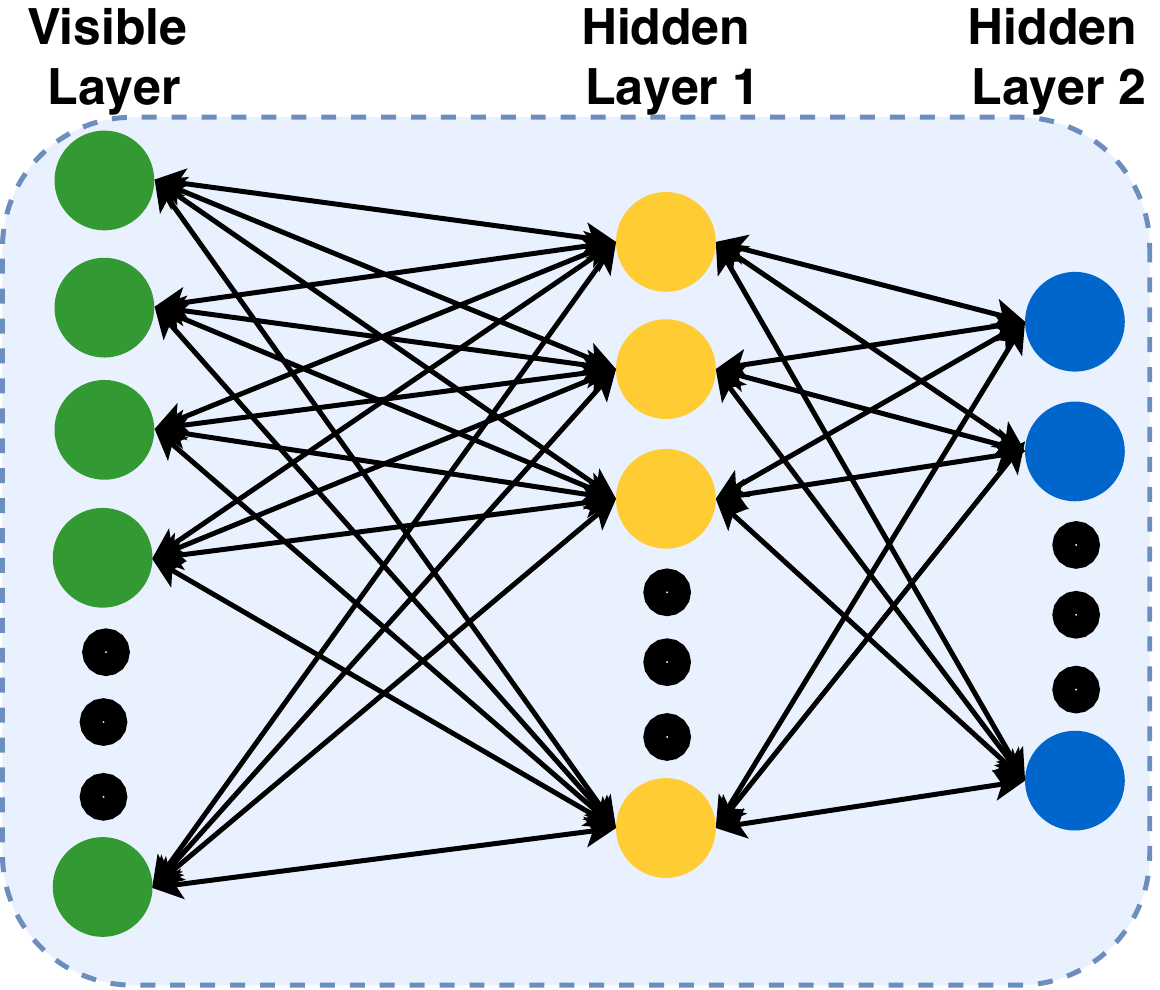}
        \caption{Deep RBM}
        \label{fig:S_RBM}
    \end{subfigure}
 \caption{Illustration of several standard representative deep learning models. (a) A basic autoencoder contains one hidden layer.
 The process from the input layer to the hidden layer is an encoder while the process from the hidden layer to the output layer is a decoder. 
 (b) Restricted Boltzmann Machine, the encoder and the decoder share the same transformation weights. The input layer and the output layer are merged into the visible layer.
 (c) Deep AE with hidden layers. Generally, the number of hidden layers is odd, and the middle layer is the learned representative features. (d) Deep RBM has one visible layer and multiple hidden layers, the last layer is the encoded representation.
 }
\end{figure}

\subsubsection{Autoencoder (AE)} 
\label{sub:AE}
As shown in Figure~\ref{fig:AE}, A autoencoder is a neural network that has three layers: the input layer, the hidden layer, and the output layer \cite{kramer1991nonlinear}. It differs from the standard neural network,
in that the AE is trained to reconstruct its inputs, which forces the hidden layer to try to learn good representations of the inputs. 

The structure of AE contains two blocks. The first block is called the encoder, which embeds the observation to a latent representation (also called `code'), 
\begin{equation} \bm{x^{h}} = \sigma(\mathcal{T}(\bm{x})) \end{equation}
where $\bm{x^{h}}$ represents the hidden layer. The second block is called the decoder, which decodes the representation into the original space,
\begin{equation} \bm{y'} = \sigma(\mathcal{T}(\bm{x^{h}})) \end{equation}
where $\bm{y'}$ represents the output. 

AE forces $\bm{y'}$ to be equal to the input $\bm{x}$ and calculates the error based on the distance between them. Thus, AE can compute the loss function only by $\bm{x}$ without the ground truth $\bm{y}$ 
\begin{equation}\label{eq:2}
error = \left \| \bm{y'}- \bm{x} \right \|_2
\end{equation}
Compared to Equation~\ref{eq:1}, this equation does not involve the variable $\bm{y}$ because it takes the input $\bm{x}$ as the ground truth. This is why AE is able to perform unsupervised learning. 

Naturally, one variant of AE is Deep-AE (D-AE) which has more than one hidden layer. We present the structure of D-AE with three hidden layers in Figure~\ref{fig:S_AE}. From the figure, we can observe that there is one more hidden layer in both the encoder and the decoder. The symmetrical structure ensures the smoothness of encoding and decoding procedure. Thus, D-AE generally has an odd number of hidden layers (e.g., $2n+1$) where the first $n$ layers belong to the encoder, the $(n+1)$-th layer works as the code which belongs to both encoder and decoder, and the last $n$ layers belong to the decoder. The data flow of D-AE (Figure~\ref{fig:S_AE}) can be represented as
\begin{equation} \bm{x^{h1}} = \sigma(\mathcal{T}(\bm{x})) \end{equation}
\begin{equation} \bm{x^{h2}} = \sigma(\mathcal{T}(\bm{x^{h2}})) \end{equation}
where $\bm{x^{h2}}$ denotes the median hidden layer (the code). Then decode the hidden layer, we can get 
\begin{equation} \bm{x^{h3}} = \sigma(\mathcal{T}(\bm{x^{h2}})) \end{equation}
\begin{equation} \bm{y'} = \sigma(\mathcal{T}(\bm{x^{h3}})) \end{equation}
It is almost the same as AE except that D-AE has more hidden layers. Apart from D-AE, AE has many other variants like denoising autoencoder, sparse autoencoder, contractive AE, etc. Here we only introduce the D-AE because it is easily confused with the AE-based deep belief network. The key difference between them will be provided in Section~\ref{sub:DBN}.

The core idea of AE and its variants is simple, which is that condensing the input data $\bm{x}$ into a code $\bm{x^{h}}$ (generally the code layer has lower dimension) and then reconstructing the data based on the code. If the reconstructed $\bm{y'}$ can approximate to the input data $\bm{x}$, it can be demonstrated that the condensed code $\bm{x^{h}}$ carries enough information about $\bm{x}$, thus, we can regard $\bm{x^{h}}$ as a representation of the input data for future operation (e.g., classification). 

\subsubsection{Restricted Boltzmann Machine (RBM)}  
\label{sub:RBM}
Restricted Boltzmann Machine is a stochastic artificial neural network that can learn a probability distribution over its set of inputs \cite{hinton2006reducing}. It contains two layers including one visible layer (input layer) and one hidden layer, as shown in Figure~\ref{fig:RBM}. From the figure, we can see that the connection lines between the two layers are bidirectional. RBM is a variant of Boltzmann Machine with stronger restriction of being without intra-layer connections. In a general Boltzmann machine, the nodes in the same hidden layer will connect.
Similar to AE, the procedure of RBM also includes two steps. The first step condenses the input data from the original space to the hidden layer in a latent space. After that, the hidden layer is used to reconstruct the input data in an identical way. Compared to AE, RBM has a stronger constraint which is that the encoder weights and the decoder weights should be equal. We have 
\begin{equation} \bm{x^{h}} = \sigma(\mathcal{T}(\bm{x})) \end{equation}
\begin{equation} \bm{x'} = \sigma(\mathcal{T}(\bm{x^h})) \end{equation}
In the above two equations, the weights of $\mathcal{T}(\cdot)$ are the same. Then, the error for training can be calculated by
\begin{equation} error = \left \| \bm{x'}- \bm{x} \right \|_2 \end{equation}
We can observe from the Figure~\ref{fig:S_RBM} that the Deep-RBM (D-RBM) is an RBM with multiple hidden layers. The input data from the visible layer firstly flow to the first hidden layer and then the second hidden layer. Then, the code will flow backward into the visible layer for reconstruction. 

\begin{figure}[]
    \centering
    \begin{subfigure}[t]{0.25\textwidth}
        \centering
        \includegraphics[width=\textwidth]{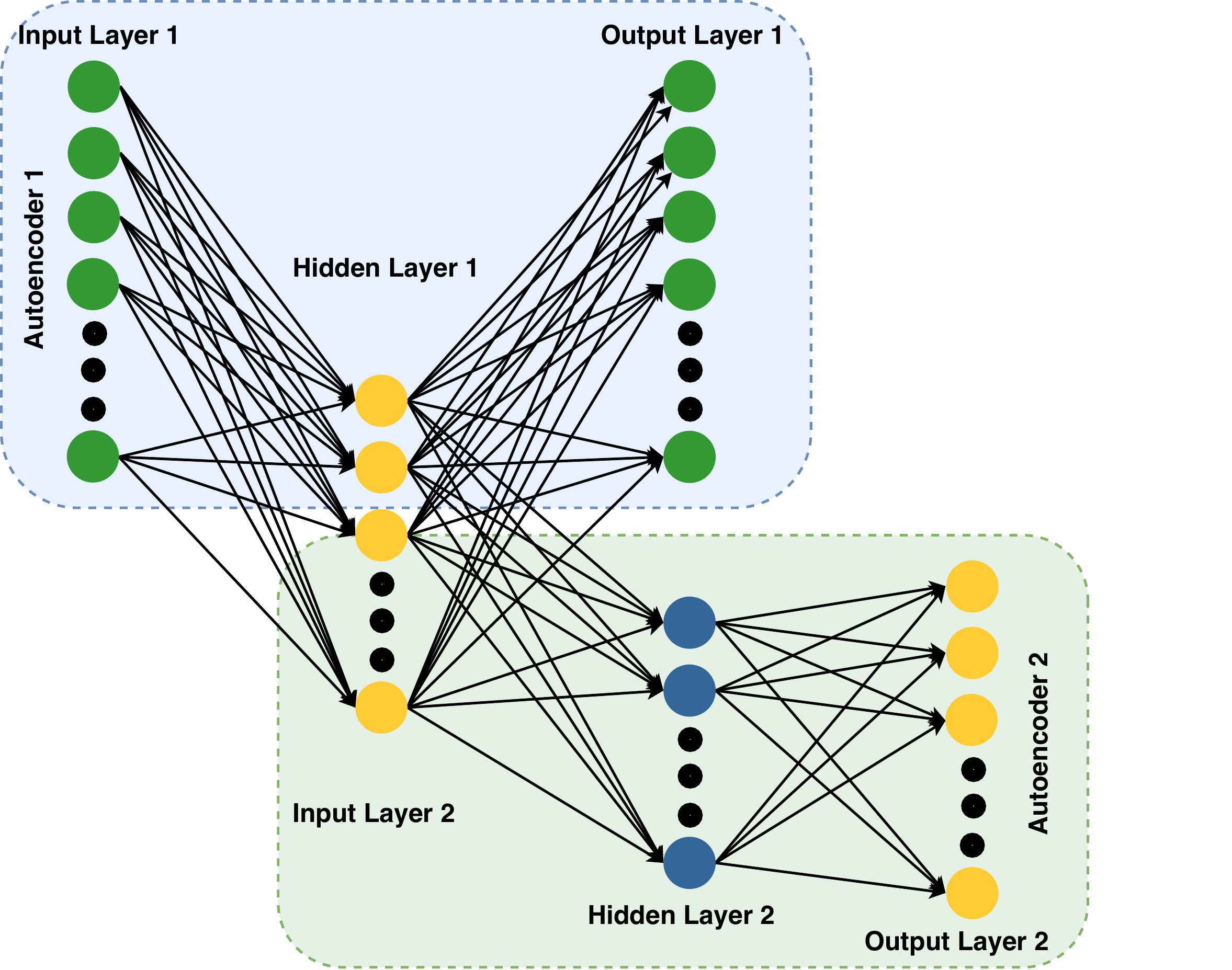}
        \caption{DBN-AE}
        \label{fig:DBN_AE}
    \end{subfigure}%
    ~
    \begin{subfigure}[t]{0.25\textwidth}
        \centering
        \includegraphics[width=\textwidth]{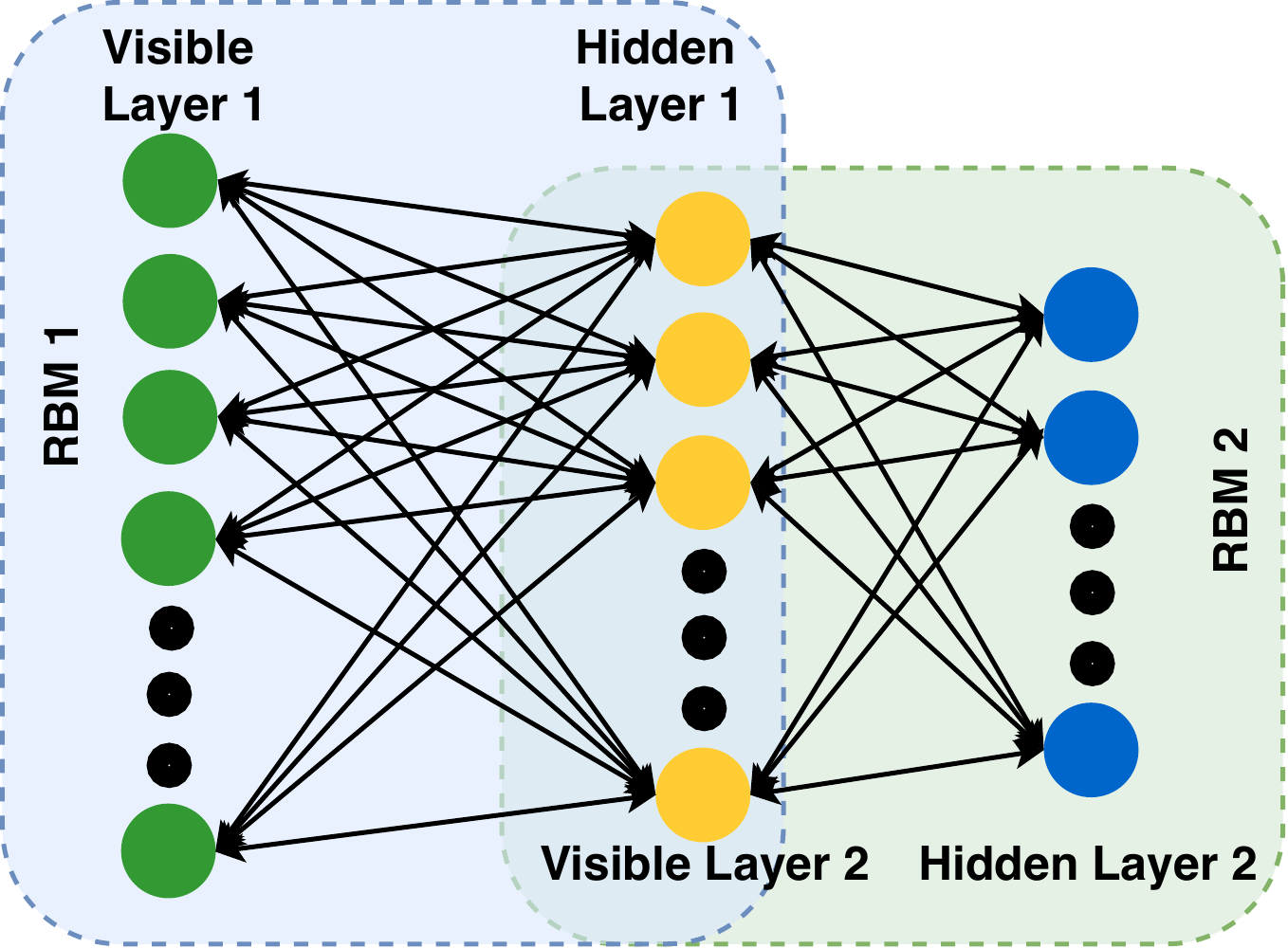}
        \caption{DBN-RBM}
        \label{fig:DBN_RBM}
        \end{subfigure}
 \caption{Illustration of deep belief networks. (a) DBN composed of autoencoders. DBN-AE contains multiple AE components (in this case, two AE), with the hidden layer of the previous AE working as the input layer of the next AE. The hidden layer of the last AE is the learned representation. (b) DBN composed of RBM. In this illustration, there are two RBM components with the hidden layer of the first RBM working as the visible layer of the second RBM. The last hidden layer is the encoded representation. While DBN-RBM and D-RBM (Figure~\ref{fig:S_RBM}) have similar architecture, the former is trained greedily while the latter is trained jointly .}
\label{fig:DBN}
\end{figure}

\subsubsection{Deep Belief Networks (DBN)} 
\label{sub:DBN}
A Deep Belief Network (DBN) is a stack of simple networks, such as AEs or RBMs \cite{glauner2015comparison}. Thus, we divided DBN into DBN-AE (also called stacked AE) which is composed of AE and DBN-RBM (also called stacked RBM) which is composed of RBM. 

As shown in Figure~\ref{fig:DBN_AE}, the DBN-AE contains two AE structures while the hidden layer of the first AE works as the input layer of the second AE. This diagram has two stages. In the first stage, the input data feed into the first AE follows the rules introduced in Section~\ref{sub:AE}. The reconstruction error is calculated and back propagated to adjust the corresponding weights and basis. This iteration continues until the AE converges. We get the mapping,
 \begin{equation} \bm{x^1} \rightarrow \bm{x^{h1}} \end{equation}

Then, we move on to the second stage where the learned representative code in the hidden layer $\bm{x^{h1}}$ will be used as the input layer of the second AE, which is 
\begin{equation} \bm{x^{2}} = \bm{x^{h1}} \end{equation}
and then, after the second AE converges, we have
\begin{equation} \bm{x^2} \rightarrow \bm{x^{h2}} \end{equation}
where $\bm{x^{h2}}$ denotes the hidden layer of the second AE, meanwhile, it is the final outcome of the DBN-AE. 

The core idea of AE is that of learning a representative code with lower dimensionality but containing most information of the input data. The idea behind DBN-AE is to learn a more representative and purer code. 

Similarly, the DBN-RBM is composed of several single RBM structures. Figure~\ref{fig:DBN_RBM} shows a DBN with two RBMs where the hidden layer of the first RBM is used as the visible layer of the second RBM. 

Compare the DBN-RBM (Figure~\ref{fig:DBN_RBM}) and D-RBM (Figure~\ref{fig:S_RBM}). They almost have the same architecture.  Moreover, DBN-AE (Figure~\ref{fig:DBN_AE}) and D-AE (Figure~\ref{fig:S_AE}) have similar architecture. The most important difference between the DBN and the deep AE/RBM is that the former is trained greedily while the latter is trained jointly. In particular, for the DBN, the first AE/RBM is trained first, 
after it converges, the second AE/RBM is trained\cite{hinton2006reducing}. For the deep AE/RBM, jointly training means that the whole structure is trained together, no matter how layers it has.

\begin{figure}[]
    \centering
    \begin{subfigure}[t]{0.24\textwidth}
        \centering
        \includegraphics[width=0.9\textwidth]{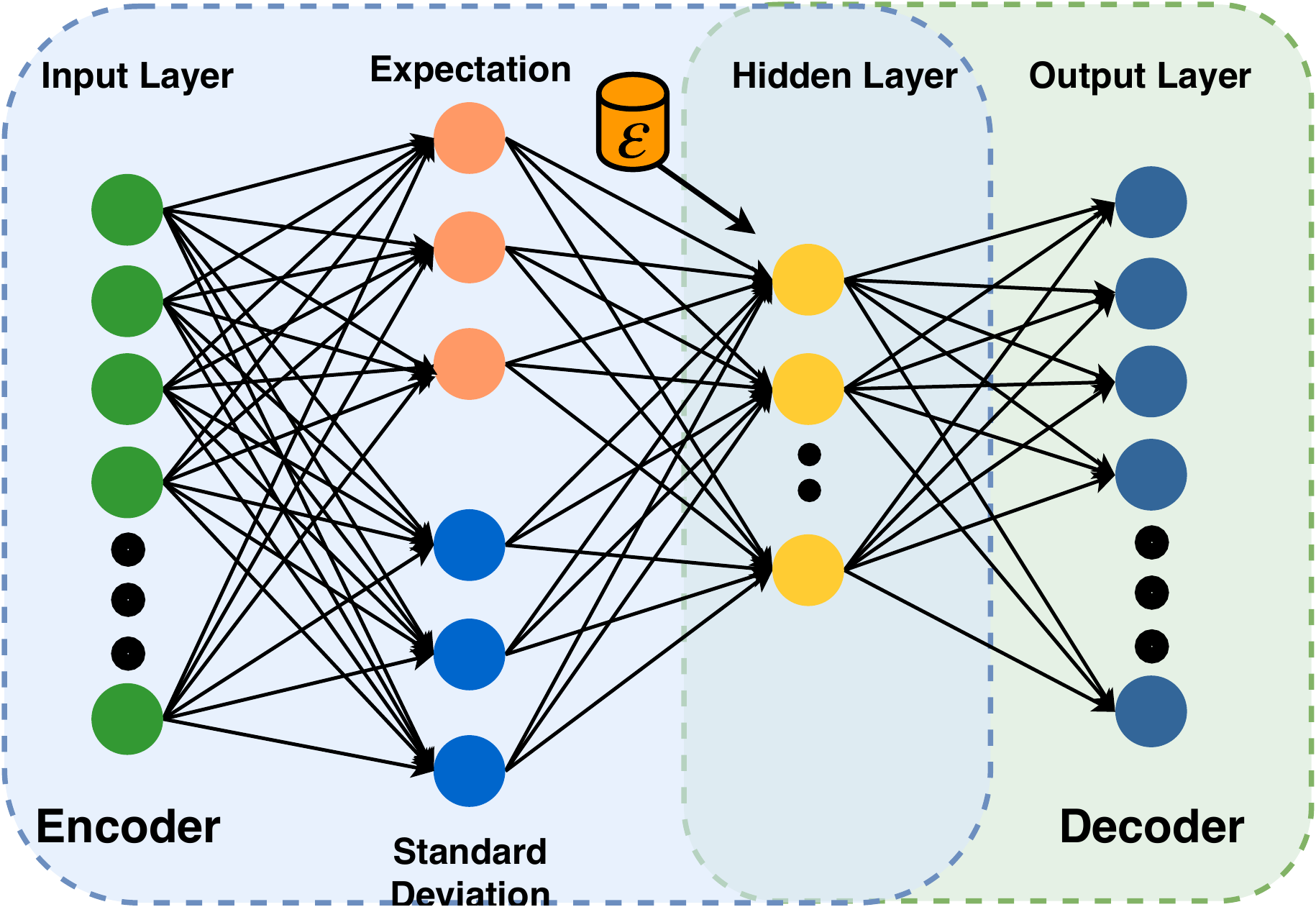}
        \caption{Variational Autoencoder}
        \label{fig:VAE}
    \end{subfigure}%
    \hfill
    \begin{subfigure}[t]{0.24\textwidth}
        \centering
        \includegraphics[width=\textwidth]{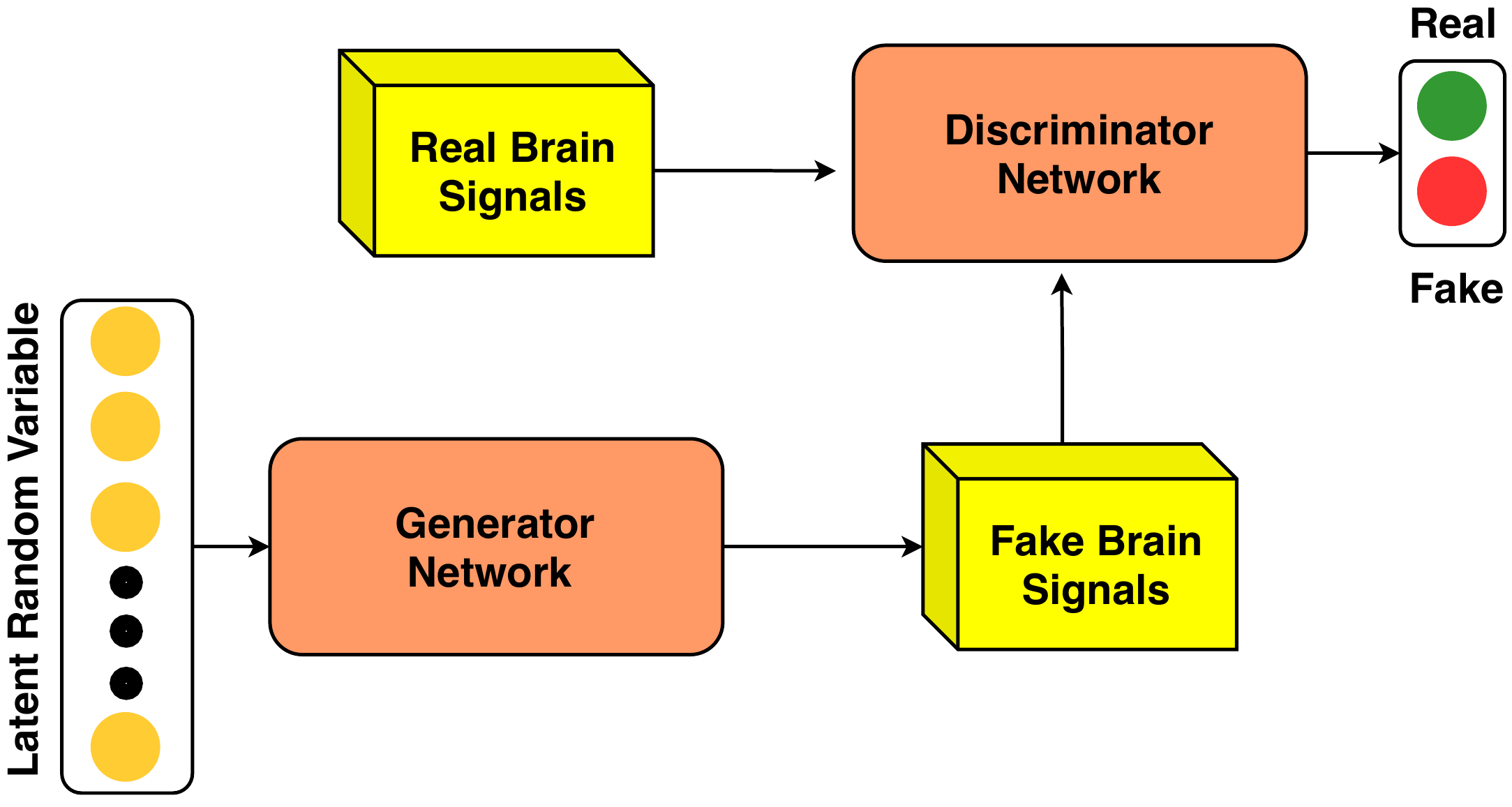}
        \caption{GAN}
        \label{fig:GAN}
        \end{subfigure}
 \caption{Illustration of generative deep learning models. (a) VAE contains two hidden layers. The first hidden layer is composed of two components: the expectation and the standard deviation, which are learned separately from the input layer. The second hidden layer represents the encoded information. $\epsilon$ denotes the standard normal distribution. (b) GAN mainly contain two crucial components: the generator and the discriminator network. The former receives a latent random variable to generate a fake brain signal while the latter receives both the real and the generated brain signals and attempts to determine if its generated or not. In the are of brain signals, GAN reconstructs or augments data instead of classification.}
\label{fig:generative}
\end{figure}

\subsection{Generative Deep Learning Models} 
\label{sub:generative_deep_learning_models}
Generative deep learning models are mainly used to generate training samples or data augmentation. In other words, generative deep learning models play a supporting role in the brain signal field to enhance the training data quality and quantity. After the data augmentation, the discriminative models will be employed for the classification. This procedure is created to improve the robustness and effectiveness of the trained deep learning networks, especially when the training data is limited. In short, the generative models receive the input data and output a batch of similar data. In this section, we will introduce two typical generative deep learning models: variational Autoencoder (VAE) and Generative Adversarial Networks (GAN).

\subsubsection{Variational Autoencoder (VAE)}  
\label{sub:variational_autoencoder}
Variational Autoencoder, proposed in 2013 \cite{kingma2013auto}, is an important variant of AE, and one of the most powerful generative algorithms. The standard AE and its other variants can be used for representation but fail in generation for the reason that the learned code (or representation) may not be continuous. Therefore, we cannot generate a random sample which is similar to the input sample. In other words, the standard AE does not allow interpolation. Thus, we can replicate the input sample but cannot generate a similar one. 
VAE has one fundamentally unique property that separates it from other AEs, and it is this property that makes VAE so useful for generative modeling: the latent spaces are designed to be continuous which allows easy random sampling and interpolation. Next, we will introduce how VAE works.

Similar to the standard AE, VAE can be divided into an encoder and decoder where the former embeds the input data to a latent space and the latter transfers the data from the latent space to the original space. However, the learned representation in the latent space is forced to approximate a prior distribution $\bm{\bar{p(z)}}$ which is generally set as Standard Gaussian distribution. Based on the reparameterization trick \cite{kingma2013auto}, the first hidden layer of VAE is designed to have two parts where one denotes the expectation $\bm{\mu}$ and another denotes the standard deviation $\bm{\sigma}$, thus we have
\begin{equation} \bm{\mu} = \sigma(\mathcal{T}(\bm{x})) \end{equation}
\begin{equation} \bm{\sigma} = \sigma(\mathcal{T}(\bm{x})) \end{equation}
Then, the latent code in the hidden layer is not directly calculated but sampled from a Gaussian distribution $\mathcal{N}(\bm{\mu}, \bm{\sigma}^2)$. The statistic code
\begin{equation} \label{eq:eq3}
\bm{z} = \bm{\mu} + \bm{\sigma}*\bm{\varepsilon}
\end{equation}
where $\bm{\varepsilon} \sim \mathcal{N}(\bm{0}, \bm{I})$. The representation $\bm{z}$ is forced to a prior distribution, and the distance $error_{KL}$ is measured by Kullback–Leibler divergence,
\begin{equation} error_{KL} = D_{KL}(z, \bm{\bar{p(z)}}) \end{equation}
where $\bm{\bar{p(z)}}$ denotes the prior distribution. In the decoder, $\bm{z}$ is decoded into the output $\bm{y}'$,
\begin{equation} \bm{y'} = \sigma(\mathcal{T}(\bm{z})) \end{equation}
and the reconstruction error is
\begin{equation} error_{recon} = \left \| \bm{y'}- \bm{x} \right \|_2 \end{equation}
The overall error for VAE is combined by the DL divergence and the reconstruction error,
\begin{equation} error = error_{KL} + error_{recon} \end{equation}

The key point of VAE is that all the latent representations $\bm{z}$ are forced to obey the normal distribution. Thus, we can randomly sample a representation $\bm{z'} \in \bm{\bar{p(z)}}$ from the prior distribution and then reconstruct a sample based on $\bm{z'}$. This is why VAE is so powerful in generation.

\subsubsection{Generative Adversarial Networks (GAN)} 
\label{sub:GAN}
Generative Adversarial Networks \cite{goodfellow2014generative} is proposed in 2014 and achieved great success in a wide range of research areas (e.g., computer vision and natural language processing). GAN is composed of two simultaneously trained neural networks with a generator and a discriminator. The generator captures the distribution of the input data, and the discriminator is used to estimate the probability that a sample came from the training data. The generator aims to generate fake samples while the discriminator aims to distinguish whether the sample is genuine. The functions of the generator and the discriminator are opposite; that's why GAN is called `adversarial.' After the convergence of both the generator and the discriminator, the discriminator ought to be unable to recognize the generated samples. Thus, the pre-trained generator can be used to create a batch of samples and use them for further operations such as as classification.

Figure~\ref{fig:GAN} shows the procedure of a standard GAN. The generator receives a noise signal $\bm{s}$ which is randomly sampled from a multimodal Gaussian distribution and outputs the fake brain signals $\bm{x}_F$. The distributor receives the real brain signals $\bm{x}_R$ and the generated fake sample $\bm{x}_F$, and then it predicts whether the received sample is real or fake. The internal architecture of the generator and discriminator are designed depending on the data types and scenarios. For instance, we can build the GAN by convolutional layers on fMRI images since CNN has an excellent ability to extract spatial features. The discriminator and the generator are trained jointly. After the convergence, numerous brain signals $\bm{x}_G$ can be created by the generator. Thus, the training set is enlarged from $\bm{x}_R$ to $\{\bm{x}_R, \bm{x}_G\}$ to train a more effective and robust classifier.

\subsection{Hybrid Model} 
\label{sub:hybrid_model}
Hybrid deep learning models refers to models which are composed of at least two deep basic learning models where the basic model is a discriminative, representative, or generative deep learning model. Hybrid models comprise two subcategories based on their targets: classification-aimed (CA) hybrid models and the non-classification-aimed (NCA) hybrid models. 

Most of the deep learning related studies in brain signal area are focused on the first category. Based on the existing literature, the representative and generative models are employed to enhance the discriminative models. The representative models can provide more informative and low dimensional features for the discrimination while the generative models can help to augment the training data quality and quantity which supply more information for the classification. The CA hybrid models can be further subdivided into: 1) several discriminative models combined to extract more distinctive and robust features (e.g., CNN+RNN); 2) representative model followed by a discriminative model (e.g., DBN+MLP); 3) generative + representative model followed by a discriminative model; 4) generative + representative model followed by a non-deep learning classifier. In which, a representative model followed by a non-deep learning classifier is regarded as a representative deep learning model.

A few NCA hybrid models aim for brain signal reconstruction. For example, St-yves et al. \cite{st2018generative} adopted GAN to reconstruct visual stimuli based on fMRI images. 
\end{appendices}

\end{document}